\documentclass[a4paper,onecolumn,11pt,accepted=2022-02-02]{quantumarticle}
\pdfoutput=1
\usepackage[utf8]{inputenc}
\usepackage[english]{babel}
\usepackage[T1]{fontenc}
\usepackage{amsmath}
\usepackage{dsfont}
\usepackage{amssymb}
\usepackage{amsthm}

\usepackage{tikz}
\usepackage{lipsum}
\usepackage[numbers,sort&compress]{natbib}

\usepackage{tikz}
\usepackage{ifthen}
\usepackage{pgfplots}
\pgfplotsset{compat=1.9}
\usetikzlibrary{shapes,arrows}
\usetikzlibrary{positioning}
\usetikzlibrary{shapes.geometric}

\RequirePackage[framemethod=default]{mdframed}

\usepackage[margin=1in]{geometry}
\usepackage[pagebackref, colorlinks = true, linkcolor = blue, urlcolor  = blue, citecolor = red]{hyperref}

\def\hs{\mathcal{H}}

\newcommand{\tr}{\operatorname{tr}}
\newcommand{\Tr}{\operatorname{Tr}}
\newcommand{\identity}{\ensuremath{\mathds{1}}}
\newcommand{\norm}[1]{\left\| #1 \right\|}
\newcommand{\abs}[1]{\left| #1 \right|}

\definecolor{grey}{RGB}{220,220,220}

\def\Block[#1,#2,#3,#4]{

\def\r{0.3};

\ifthenelse{\NOT #4=0}{
\fill [#2] (-0.5,-0.5) rectangle ({#1-0.5},0.5);
}

\foreach \n in {1,...,#1}{ 

\shade[shading=ball, ball color=red] ({\n-1},0) circle (\r);

}

\begin{scope}[decoration={brace,mirror,amplitude=7}]

\ifthenelse{#4=1}{

\draw [decorate] (-0.5,-0.6) --node[below=3mm]{$#3$} ({#1-0.5},-0.6);

}

\ifthenelse{#4=2}{
\draw [decorate] ({#1-0.5},0.6) --node[above=3mm]{$#3$} (-0.5,0.6);
}

\end{scope}

}

\newmdenv[skipabove=7pt,
skipbelow=7pt,
backgroundcolor=grey!10,
innerleftmargin=5pt,
innerrightmargin=5pt,
innertopmargin=5pt,
leftmargin=0cm,
rightmargin=0cm,
innerbottommargin=5pt,
linewidth=1pt]{sBox}

\newmdenv[skipabove=7pt,
skipbelow=7pt,
backgroundcolor=grey!50,
innerleftmargin=5pt,
innerrightmargin=5pt,
innertopmargin=5pt,
leftmargin=0cm,
rightmargin=0cm,
innerbottommargin=5pt,
linewidth=1pt]{tBox}

\newtheorem{thm}{Theorem}[section]
\newtheorem{lem}[thm]{Lemma}
\newtheorem{cor}[thm]{Corollary}

\newtheorem{defi}[thm]{Definition}
\newtheorem{prop}[thm]{Proposition}
\newtheorem{remark}[thm]{Remark}

\newtheorem{stp}{Step}

\newenvironment{theo}{\begin{tBox}\begin{thm}}{\end{thm}\end{tBox}}
\newenvironment{step}{\begin{sBox}\begin{stp}}{\end{stp}\end{sBox}}

\begin{document}

\title{Exponential Decay of Mutual Information for Gibbs states of local Hamiltonians}

\author{Andreas Bluhm}
\email{bluhm@math.ku.dk}
\affiliation{QMATH, Department of Mathematical Sciences, University of Copenhagen, Universitetsparken 5, 2100 Copenhagen, Denmark}
\orcid{0000-0003-4796-7633}
\author{Ángela Capel}
\email{angela.capel@uni-tuebingen.de}
\affiliation{Fachbereich Mathematik, Universit\"at T\"ubingen, 72076 T\"ubingen, Germany}
\affiliation{Zentrum Mathematik, Technische Universit\"at M\"unchen, Boltzmannstrasse 3, 85748 Garching, Germany}
\affiliation{Munich Center for Quantum Science and Technology (MCQST), M\"unchen, Germany}
\orcid{0000-0001-6713-6760}
\author{Antonio Pérez-Hernández}
\email{antperez@ind.uned.es}
\affiliation{Departamento de Matem\'{a}tica Aplicada I, Escuela T\'{e}cnica Superior de Ingenieros Industriales, Universidad Nacional de Educación a Distancia, calle Juan del Rosal 12, 28040 Madrid (Ciudad Universitaria), Spain}
\affiliation{Departamento de An\'{a}lisis Matemático y Matemática Aplicada, Universidad Complutense de Madrid, 28040 Madrid, Spain}
\orcid{0000-0001-8600-7083}

\maketitle

\begin{abstract}
The thermal equilibrium properties of physical systems can be described using Gibbs states. It is therefore of great interest to know when such states allow for an easy description. In particular, this is the case if correlations between distant regions are small. In this work, we consider 1D quantum spin systems with local, finite-range, translation-invariant interactions at any temperature. In this setting, we show that Gibbs states satisfy uniform exponential decay of correlations and, moreover, the mutual information between two regions decays exponentially with their distance, irrespective of the temperature. In order to prove the latter, we show that exponential decay of correlations of the infinite-chain thermal states, exponential uniform clustering and exponential decay of the mutual information are equivalent for 1D quantum spin systems with local, finite-range interactions at any temperature.  In particular, Araki's seminal results yields that the three conditions hold in the translation-invariant case. The methods we use are based on the Belavkin-Staszewski relative entropy and on techniques developed by Araki. Moreover, we find that the Gibbs states of the systems we consider are superexponentially close to saturating the data-processing inequality for the Belavkin-Staszewski relative entropy.

\end{abstract}

\tableofcontents

\section{Introduction}\label{sec:intro}

The thermal equilibrium properties of quantum systems can be described by \textit{quantum Gibbs states}. Such states inherit the locality properties of their associate Hamiltonian, featuring locality of correlations and allowing for an efficient description in certain cases. A typical measure of correlations in many-body systems is the \emph{operator} or \emph{covariance correlation}, defined for a quantum state \mbox{$\rho_{ABC} \in \mathcal{B}(\mathcal{H}_{A} \otimes \mathcal{H}_{B} \otimes \mathcal{H}_{C})$}:
$$  \text{Corr}_\rho (A:C) := \sup_{O_{A}, O_{C}}{\big|\text{Tr}[O_A \otimes O_C\, (\rho_{AC} - \rho_{A} \otimes \rho_{C})] \big|} \, , $$
where the supremum is taken over all operator norm-one operators $O_A$ and $O_C$ supported on  subsystems $A$ and $C$,  respectively, and $\rho_X$ is the reduced state of $\rho_{ABC}$ on $X$ for $X \in \{ A, C, AC\}$. When these correlations are spatially localized, there are efficient methods to approximate Gibbs states of local Hamiltonians using tensor network methods \cite{Molnar2015, Cirac2020}.

In his seminal work in 1969, Araki showed that for any infinite 1D quantum spin system with finite-range and translation-invariant interactions, bipartite correlation functions decay exponentially with the distance between $A$ and $C$ \cite{Araki1969}. This opened a new field of research that has been exceptionally active in the last years. Remarkably, Araki's result was extended for different correlation functions to higher dimensional systems above a threshold temperature in a series of papers \cite{Gross1979,Park1995,Kliesch2014,Ueltschi2004}, both for classical and quantum systems. In a recent work \cite{Harrow2020}, the exponential decay of correlations property was related to the absence of complex zeroes of the partition function close to the real axis. Moreover, it is known that systems with a positive \textit{spectral gap} exhibit exponential decay of correlations \cite{Kastoryano2013}.

From an information-theoretic point of view, the most prominent measure of correlations is however the
 \textit{mutual information}, defined for $\rho_{AC} \in \mathcal B(\mathcal{H}_{A}  \otimes \mathcal{H}_{C})$ by 
\begin{equation} \label{eq:mutu-info-intro}
    I_{\rho}(A:C):=D(\rho_{AC}||\rho_{A} \otimes \rho_{C}) \, ,
\end{equation}
where $D(\rho||\sigma) := \Tr[ \rho \, (\log \rho - \log \sigma)]$ is the Umegaki relative entropy between $\rho$ and $\sigma$ \cite{Umegaki-RelativeEntropy-1962}. Indeed, it is shown in \cite{Groisman2005} that this measure quantifies the total amount of correlation between subsystems $A$ and $C$ via an operational interpretation. It moreover has several desirable properties, e.g., it inherits positivity and data processing from the relative entropy. Note that estimates on the decay of the mutual information are stronger than those on the decay of the operator correlation function, since
$$ I_{\rho}(A:C) \geq \frac{1}{2} \| \rho_{AC} - \rho_{A} \otimes \rho_{C} \|_{1}^{2} \geq \frac{1}{2}\, \text{Corr}_\rho (A:C)^{2}\,,   $$
by Pinsker's inequality \cite{Pinsker-Information-1964} and the duality of Schatten $p$-norms \cite{Bhatia1997}. Furthermore, it is well known that, in the context of data hiding, it is possible to find a state whose operator correlation is arbitrarily small, whereas it is still highly correlated in terms of the mutual information \cite{Hastings2007b, Hayden2004}. The mutual information has  attracted significant attention in the past years, because it satisfies an \textit{area law} \cite{Wolf2008}, i.e., the correlations between adjacent systems scale only as the boundary between them. The connection between area laws and exponential decay of correlations was subsequently explored in \cite{Brandao2013, Brandao2015b}. See also the recent papers \cite{Kuwahara2020, Scalet2021} for related results on the area law in more general settings and for different quantities. Furthermore, it is known that systems with the so-called \textit{rapid mixing} property exhibit exponential decay of mutual information \cite{Kastoryano2013}.

Another information-theoretic quantity that has been thoroughly studied in the context of decay of correlations is the \textit{conditional mutual information} (CMI for short), given for a state \mbox{$\rho_{ABC} \in \mathcal B(\mathcal{H}_{A} \otimes \mathcal{H}_{B} \otimes \mathcal{H}_{C})$} by
\begin{equation*}
    I_\rho(A:C | B) := S(\rho_{AB})+S(\rho_{BC}) - S(\rho_{B}) - S(\rho_{ABC}) \, ,
\end{equation*}
where $S(\rho):= - \Tr[\rho \, \log \rho]$ is the von Neumann entropy. As shown in \cite{FawziRenner-ConditionalMutualInformation-2015, Sutter2018}, this quantity is related to the approximate recoverability of a state, and thus it is widely used as a measure of independence in quantum systems.  Building on Araki's results, Kato and Brand{\~a}o showed in  \cite{Kato2019} that the CMI of any Gibbs state in the setting of \cite{Araki1969} decays subexponentially. This was subsequently improved in a joint work with Kuwahara to exponential decay and extended to generic graphs in \cite{Kuwahara2019}, provided the system is at high-enough temperature and the interactions are short ranged. Additionally, in \cite{Aragones2020}, an exponential decay of the CMI for injective matrix product states (MPS) in finite 1D lattices with finite bond dimension and open boundary conditions was obtained. See also \cite{Chen2020}, in which the authors {presented} the conjecture that the CMI for any matrix product density operator (MPDO) should decay exponentially. To support the conjecture, the authors {were} able to prove it in some cases of interest. In particular, the conjecture implies that the parent Hamiltonians of such MPDOs are (quasi-)local.

In this work, we build on Araki's result to prove an exponential decay of the (unconditional) mutual information at any temperature for quantum Gibbs states in 1D.  This should be compared to the previous work by Kuwahara et al.\ \cite{Kuwahara2019}, in which an exponential decay of the conditional mutual information was shown above a threshold temperature for arbitrary lattices.

The main novelty in the approach  we follow is the use of another of the possible extensions of the Kullback-Leibler divergence to the quantum setting, namely the \textit{Belavkin-Staszewski relative entropy} (BS-entropy for short) \cite{BelavkinStaszewski-BSentropy-1982}. The interest in this quantity has increased in the past few years,  with a comprehensive study of the fundamental properties of the maximal $f$-divergences  \cite{Matsumoto2018, Hiai2017}, of which the BS-entropy is a prominent example. Recently, a recovery condition and a strengthened data-processing inequality for such divergences has been obtained \cite{Bluhm2020}, as well as some weak quasi-factorization results for the BS-entropy \cite{Bluhm2021}. Remarkably, a subclass of maximal $f$-divergences, namely the \textit{geometric Rényi divergences}, has found applications in estimating channel capacities \cite{FangFawzi-GeometricRenyiDivergences-2019}. In the current paper, we present other important applications of geometric Rényi divergences and the BS-entropy in the fields of quantum many-body systems and quantum information theory. In particular, we define a BS-mutual information in the same spirit as \eqref{eq:mutu-info-intro} and show that it exhibits an exponential decay under some reasonable assumptions.

The rest of the paper is organized as follows: In Section~\ref{sec:main_results}, we list the main results of the present article. Some basic notions and preliminary results are reviewed in Section~\ref{sec:preliminaries}. In Section \ref{sec:locality_observables}, we provide norm estimates on some local observables that are necessary for the rest of the article. They are in particular used in Section \ref{sec:approximate_factorization} to prove one of the main results of the paper, namely a faster-than-exponential (i.e.~\emph{superexponential}) decay for the distance of a quantum Gibbs state from being BS-recoverable. In Section~\ref{sec:infinite-chain}, we prove another main result regarding the exponential decay of correlations (on a finite interval) of a Gibbs state with the distance between two spatially separated regions in 1D. This condition is further used in Section~\ref{sec:clustering} to show the local indistinguishability of such states. Subsequently, in Section \ref{sec:decay-of-mutual-info} we prove our third main result: an exponential decay for the BS-mutual information of a Gibbs state with the distance between two spatially separated regions in 1D. Finally, we conclude our manuscript with {a discussion of our results} in Section~\ref{sec:conclusions}.

\section{Main results}\label{sec:main_results}

The main results of this paper concern the BS-entropy and some other quantities derived from it. The BS-entropy for two strictly positive states $\rho$ and $\sigma$ is given by
$$ \widehat{D}(\rho \| \sigma) := \text{Tr}[\rho \, \text{log} (\rho^{1/2} \sigma^{-1} \rho^{1/2})] \, . $$
It relates to the (Umegaki) relative entropy by means of the following inequality \cite{Ohya1993}
$$  \widehat{D}(\rho \| \sigma) \geq D(\rho \| \sigma)\, ,$$
which is strict if, and only if, $[\rho, \sigma] \neq 0$ \cite{Hiai2017}. In particular, for a bipartite Hilbert space $\mathcal H_{AC}=\mathcal H_A \otimes \mathcal H_C$ and a positive state $\rho_{AC} \in \mathcal B(\hs_{AC})$,
we can define the \textit{BS-mutual information} in an analogous way to the usual mutual information, and clearly the following inequality holds:
$$  I_\rho (A:C) \leq \widehat{I}_\rho (A:C) := \widehat{D}(\rho_{AC} \| \rho_A \otimes \rho_C) \, . $$
This basic inequality allows us to obtain an estimate for the mutual information via estimating the BS-mutual information. 

 The first main result of the paper is the so-called \textit{exponential uniform clustering} for Gibbs states.
 
 \vspace{0.2cm}

\noindent \textbf{Theorem I.}
\textit{We consider a quantum spin system on $\mathbb{Z}$ with local, finite-range, translation-invariant (non-commuting) interactions. In  this setting, there exists a positive function  $\ell \mapsto \varepsilon(\ell)$ exhibiting exponential decay such that for every finite interval $I \subset \mathbb{Z}$ split into three subintervals $I=ABC$, where $B$ shields $A$ from $C$, the Gibbs state  $\rho=e^{-H_{I}}/\Tr(e^{-H_{I}})$ satisfies that the associated  covariance correlation decays exponentially with the size of $B$, namely
\begin{equation}\label{eq:uniform_clustering2}
    \operatorname{Corr}_\rho (A:C) \leq \varepsilon (|B|) \, .
\end{equation}
}

We remark that the previous condition, which appears later in the main text as Theorem~\ref{Theo:ArakiImpliesUnifCluster}, holds at any inverse temperature $\beta >0$. As far as we know, there is no proof for this condition in this setting in any previous paper (although it was shown to hold in any dimension for high-enough temperature in \cite{Kliesch2014}). It was a necessary condition for some of the main results in \cite{Kato2019} and \cite{Brandao2019}, where the authors claimed that it directly follows from Araki's result on infinite-chain exponential decay of correlations. However, it is not obvious how the former result follows from the latter, and that is the reason for including a proof of such a result in the current manuscript.

Next, building on the fact that any Gibbs state of a local, finite-range, translation-invariant Hamiltonian in 1D satisfies exponential uniform clustering, we prove exponential decay of the mutual information for such states. 

 \vspace{0.2cm}

\noindent \textbf{Theorem II.}
\textit{
Under the conditions of Theorem I, there is a positive function $\ell \mapsto \delta_{1}(\ell)$, depending on the local interactions and $\varepsilon(\ell)$ in~\eqref{eq:uniform_clustering2}, that exhibits exponential decay and satisfying
$$ I_\rho(A:C) \, \, \leq \, \, \widehat I_\rho(A:C) \,\, \leq \,\, \delta_{1}(|B|) \, .$$}

This result appears in the main text as Theorem \ref{Theo:MIdecay-main-result} and builds heavily on the well-known methods  of Araki for complex-time evolutions and finite-range interactions that constitute a generalization of the classical Lieb-Robinson bounds \cite{Araki1969, Perez2020}.  Throughout the paper, we will absorb $\beta$ in the Hamiltonian so that it is an implicit parameter of the interaction strength $J$. Given an arbitrary finite-range Hamiltonian $H$, all our results hold for the Gibbs state \mbox{$\rho = e^{-H}/\mathrm{Tr}(e^{-H})$} associated to it; thus, they actually hold for every finite temperature $\beta>0$. The  previous theorem should be compared to \cite[Corollary 5]{Kato2019}, which shows that the CMI of a Gibbs state in the above setting decays subexponentially in the distance $|B|$. Considering the  (unconditional) mutual information instead, we can improve the decay to being exponential.

Let us remark, that our results allow to formally complete the equivalence between several different notions of decay of correlations for Gibbs states of local, finite-range Hamiltonians in 1D (not necessarily translation-invariant). More specifically, we show for such states that the following three conditions are equivalent:

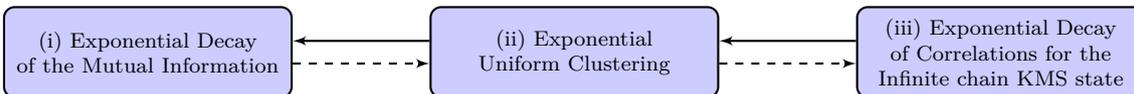
\begin{figure}[h]

\begin{center}
    
\begin{tikzpicture}[node distance = 2cm, thick, scale=0.5]
\scriptsize

 \tikzstyle{block} = [rectangle, aspect=2, draw, fill=blue!20, 
     text width=12em, text centered, rounded corners, minimum height=4em]
\tikzstyle{line} = [draw, -latex']    
    
    
\node [block] (UnifCluster) {(ii) Exponential \\ Uniform Clustering};       

\node [block, left=1.8cm of UnifCluster] (ExpDecayMI) {(i) Exponential Decay \\ of the Mutual Information};      
  
\node [block, right=1.8cm of UnifCluster] (ExpDecay) {(iii) Exponential Decay\\ of Correlations for the \\ Infinite chain KMS state};   
    

 \path [line] ([yshift=0.3cm]UnifCluster.west) -- ([yshift=0.3cm]ExpDecayMI.east);  

 \path [line, dashed] ([yshift=-0.3cm]ExpDecayMI.east) -- ([yshift=-0.3cm]UnifCluster.west) ; 
    
\path [line, dashed] ([yshift=-0.3cm]UnifCluster.east) -- ([yshift=-0.3cm]ExpDecay.west) ;
    
\path [line] ([yshift=0.3cm]ExpDecay.west) -- ([yshift=0.3cm]UnifCluster.east);    
    
\end{tikzpicture}    
\end{center}

  \caption{Equivalence between three different forms of decay of correlations in 1D. }
  \label{fig:0}
\end{figure}

 \noindent Dashed lines indicate implications that were previously known or straightforward consequences of well-known results, while the reverse implications are formally proved in Theorems \ref{Theo:MIdecay-main-result} and \ref{Theo:ArakiImpliesUnifCluster}, respectively. The \emph{translation-invariant} condition is applied to deduce from Araki's result \cite{Araki1969} that (i)-(iii) hold. For further detail, we refer the reader to Section~\ref{sec:conclusions} and the diagram in Figure \ref{fig:scheme} therein.

The use of the BS-entropy in our methods yields another result of independent interest, namely that the distance of a Gibbs state from being BS-recoverable (i.e.~from saturating the data-processing inequality for the BS-entropy) decays superexponentially with the size of the middle system. More specifically, by the data-processing inequality, the following holds for any quantum channel $\mathcal{T}$ and any positive states $\rho$ and $\sigma$:
\begin{equation*}
    \widehat D(\rho || \sigma ) \geq \widehat D(\mathcal{T} (\rho) || \mathcal{T} (\sigma) ) \, ,
\end{equation*}
and by the results of \cite{Bluhm2020}, equality in the previous inequality is equivalent to 
\begin{equation*}
    \rho = \sigma \,  \mathcal{T}^* \left( \mathcal{T}(\sigma)^{-1} \mathcal{T}(\rho) \right)  \, ,
\end{equation*}
where the map $ \mathcal{B}_\mathcal{T}^\sigma (\cdot) := \sigma \,  \mathcal{T}^* \left( \mathcal{T}(\sigma)^{-1} (\cdot) \right)$ is called \textit{BS-recovery condition}. This map represents the analogue to the Petz recovery map for the 
saturation in the data-processing inequality for the Umegaki relative entropy, although the former is not a quantum channel. Furthermore, another one of the main results of \cite{Bluhm2020} is a strengthened version of the data-processing inequality for the BS-entropy in terms of the distance of a state to be BS-recoverable, i.e., to satisfying the latter equality. That result in particular shows that if there is almost-saturation of the DPI for the BS-entropy, then the state is close to being BS-recoverable. The converse implication, though, is as far as we know an open problem. The result below might shed some light on this question for a particular class of states.

If we consider $\hs_{ABC}= \hs_A \otimes \hs_B \otimes \hs_C$, the quantum channel 
$\Tr_A(\cdot)\mathds{1}_A/d_A \otimes \operatorname{id}_{BC}$, a positive state $\rho_{ABC} \in 
\mathcal B(\hs_{ABC})$, and if we define $\sigma_{ABC}= \rho_{AB} \otimes \identity_C / d_C$, we say that $\rho_{ABC}$ is a \textit{BS-recoverable state} if it is a fixed point of the BS-recovery condition composed with this quantum channel, i.e., if
\begin{equation*}
     \rho_{ABC} = \rho_{AB} \rho_{B}^{-1} \rho_{BC} \, .
\end{equation*}
We are now in the position to informally state our third main result. 

 \vspace{0.2cm}

\noindent \textbf{Theorem III.}
  \textit{
 Under the conditions of Theorem I, there is a positive function  $\ell~\longmapsto~\delta_{2}(\ell)$, depending only on the local interactions  and exhibiting superexponential decay, such that for every three adjacent and finite intervals $ABC$, where $B$ shields $A$ from $C$,  the associated Gibbs state $\rho_{ABC}$ satisfies
$$ \| \rho_{ABC} - \rho_{AB} \rho_{B}^{-1} \rho_{BC} \|_{1}\, \leq \, \delta_{2}(|B|)\,. $$
}

This result appears in the main text as Theorem \ref{thm:main_BS_recoverable} and its proof also follows from the aforementioned Araki's methods. It should be compared to \cite[Theorem 4]{Kato2019}, which shows that Gibbs states in the above setting are subexponentially close to being recoverable. Replacing the Petz recovery map by the BS-recovery condition, we find that the Gibbs states are superexponentially close to fulfilling the latter. A natural question that arises from this result is, since Gibbs states in the conditions above are close to being BS-recoverable, whether there is almost saturation on the DPI of the BS-entropy for such states. This is left for future work. Additionally to the theorems mentioned above, we include some other results of independent interest in this text.

\vspace{0.2cm}

We  expect that the techniques we developed, based on the BS-entropy, can also be useful in other contexts, such as for deriving area laws in quantum many-body systems (see e.g.\ \cite{Scalet2021}).  Indeed, the main technique in the proof of exponential decay of the mutual information, namely Proposition \ref{Theo:MIdecay}, is to show that tracing out $B$ in a Gibbs state on an interval $ABC$ leads to the reduced state being exponentially close in operator norm to a product state, where the decay is with $|B|$. This finding is one of the key ingredients in the proof that Davies generators in 1D which converge to a Gibbs state of a local, finite-range, translation-invariant, commuting Hamiltonian, at any temperature, satisfy a positive modified logarithmic Sobolev inequality, and hence exhibit rapid mixing \cite{BardetCapelGaoLuciaPerezGarciaRouze-Davies1DMLSIlong-2021,BardetCapelGaoLuciaPerezGarciaRouze-Davies1DMLSIshort-2021}. 

\section{Preliminaries}\label{sec:preliminaries}

\subsection{Mutual information and relative entropy}\label{subsec:mutualinfo}

A class of important tools for quantum information theory which we will use in this work are the generalizations of the classical Kullback-Leiber (KL) divergence \cite{KullbackLeibler-KLD-1951} to the quantum setting. The generalization is not unique and therefore the KL-divergence can be extended to the quantum case in a wide variety of measures of distinguishability for states. Let $\mathcal{H}_{AC}:=\mathcal{H}_{A} \otimes \mathcal{H}_{C}$ be a finite-dimensional Hilbert space and let $\rho_{AC}, \sigma_{AC}$ be full-rank states. Their (Umegaki) relative entropy \cite{Ohya1993} is given by
\[ D(\rho_{AC} 	\, || \, \sigma_{AC}) \, := \, \operatorname{Tr}[ \, \rho_{AC} \, (\log{\rho_{AC}} - \log{\sigma_{AC}}) \, ]\, , \]
where by $\operatorname{Tr}$ we are denoting the usual (unnormalized) trace, and their Belavkin-Staszewski relative entropy (shortened BS-entropy) \cite{BelavkinStaszewski-BSentropy-1982} is
\[ \widehat{D}(\rho_{AC} 	\, || \, \sigma_{AC}) \, := \, \operatorname{Tr}\left[\, \rho_{AC} \, \log{\left(\rho_{AC}^{1/2} \, \sigma_{AC}^{-1} \, \rho_{AC}^{1/2}\right)} \,\right]\,. \]
An equivalent way to write the BS-entropy, which follows easily from the power series of the matrix logarithm, is
\begin{equation*}
    \widehat{D}(\rho_{AC} 	\, || \, \sigma_{AC}) \, = \, \operatorname{Tr}\left[\, \rho_{AC} \, \log{\left( \sigma_{AC}^{-1}  \, \rho_{AC}\right)} \,\right]\,.
\end{equation*}
Let us recall that the two quantum relative entropies are related through
\begin{equation}
D(\rho_{AC} 	\, || \, \sigma_{AC}) \,\, \leq \,\, \widehat{D}(\rho_{AC} 	\, || \, \sigma_{AC}), \label{eq:rel-BS-entropy} 
\end{equation}
which is indeed an equality if and only if $\rho_{AC}$ and $\sigma_{AC}$ commute \cite[Theorem 4.3]{Hiai2017}. Additionally, throughout the text we make use of the following family of maximal $f$-divergences which converge to the BS-entropy: For $1 <q < \infty$,  and $\rho_{AC} , \, \sigma_{AC}$ as above, their \emph{\mbox{$q$-geometric} Rényi divergence} \cite{Matsumoto2018} is given by
\begin{equation*}
    \widehat{D}_q (\rho_{AC} \| \sigma_{AC}) := \frac{1}{q - 1} \log \Tr\left[ \sigma_{AC}^{1/2} \left( \sigma_{AC}^{-1/2} \rho_{AC} \sigma_{AC}^{-1/2} \right)^q \sigma_{AC}^{1/2}\right] \, .
\end{equation*}
In the following result, given $q >1$, we prove upper bounds for the $q$-geometric Rényi divergence between two full-rank states in terms of an operator norm, denoted by $\norm{\cdot}$ throughout the text.
\begin{lem}\label{lem:boundBSrelativeEntropy}
For any $q >1$ and any $\rho_{AC}, \sigma_{AC}$ full-rank states, we have
\begin{equation}\label{equa:boundGeometricRenyiDivergence}
 \widehat{D}_q(\rho_{AC} 	\, || \, \sigma_{AC}) \,\, \leq  \,\,  \norm{ \sigma_{AC}^{-1} \, \rho_{AC} - \identity_{AC} }  \, .
\end{equation}
In particular,
\begin{equation}\label{equa:boundBSrelativeEntropy2}
 \widehat{D}(\rho_{AC} 	\, || \, \sigma_{AC}) \,\, \leq  \,\, \norm{ \sigma_{AC}^{-1} \, \rho_{AC} - \identity_{AC} } \, .
\end{equation}
\end{lem}

\noindent \textit{Proof.}
For $q >1$, note that the expression inside the logarithm in the definition of the $q$-geometric Rényi divergences can be written as 
\begin{align*}
\Tr \left[ \sigma_{AC}^{1/2} \left(  \sigma_{AC}^{-1/2}  \, \rho_{AC} \, \sigma_{AC}^{-1/2}\right)^q  \sigma_{AC}^{1/2} \right] & = \Tr \left[  \rho_{AC} \, \sigma_{AC}^{-1/2} \left( \sigma_{AC}^{-1/2} \, \rho_{AC} \, \sigma_{AC}^{-1/2} \right)^{q -1} \sigma_{AC}^{1/2} \right] \\
& = \Tr \left[  \rho_{AC}^{1/2} \left( \rho_{AC}^{1/2} \, \sigma_{AC}^{-1} \, \rho_{AC}^{1/2} \right)^{q -1} \rho_{AC}^{1/2} \right] 
\end{align*}
where in the last equality we have used that $f(X^{\dagger}X) = X^{-1}f(XX^{\dagger}) X$ whenever $f$ is continuous and $X$ is invertible. Thus,
\begin{align*}
 \Tr \left[ \sigma_{AC}^{1/2} \left(  \sigma_{AC}^{-1/2}  \, \rho_{AC} \, \sigma_{AC}^{-1/2}\right)^q  \sigma_{AC}^{1/2} \right] \leq  \norm{ \left( \, \rho_{AC}^{1/2} \,  \sigma_{AC}^{-1} \, \rho_{AC}^{1/2} \right)^{q - 1}} \leq  \norm{ \rho_{AC}^{1/2} \,  \sigma_{AC}^{-1} \, \rho_{AC}^{1/2} }^{q - 1} \, .
\end{align*}
By definition of the $q$-geometric Rényi divergence, monotonicity of the logarithm and the inequality $\log (x + 1) \leq x$, we obtain
\begin{align*}
  \widehat{D}_q (\rho_{AC} \| \sigma_{AC})  \, \leq \,  \frac{1}{q - 1} \log  \norm{ \rho_{AC}^{1/2} \,  \sigma_{AC}^{-1} \, \rho_{AC}^{1/2} }^{q - 1}  \, & = \; \log \norm{ \rho_{AC}^{1/2}  \, \sigma_{AC}^{-1} \, \rho_{AC}^{1/2} } \\[1mm]
       & \, \leq \; \log \left( \norm{ \rho_{AC}^{1/2} \, \sigma_{AC}^{-1} \, \rho_{AC}^{1/2} - \identity } + 1 \right) \\[1mm]
    &  \, \leq \; \norm{ \rho_{AC}^{1/2} \, \sigma_{AC}^{-1} \, \rho_{AC}^{1/2} - \identity } \\[1mm]
    & \,\leq  \; \norm{ \sigma_{AC}^{-1}  \rho_{AC} - \identity } \, ,
\end{align*}
where we have used triangle inequality and \cite[Proposition IX.1.1]{Bhatia1997}, which  implies, for $P > 0$, that
\begin{equation}\label{equa:boundBSrelativeEntropyAux1} 
\left\| P^{1/2} Q P^{1/2} - \mathbbm{1}\right\| \,\, \leq \,\, \left\| Q P - \mathbbm{1} \right\|\, . 
\end{equation}
 In particular, taking limit when $q \rightarrow 1^{+}$, we deduce that
\[ \widehat{D}(\rho_{AC} \| \sigma_{AC}) = \lim_{q \rightarrow 1^{+}} \widehat{D}_q (\rho_{AC} \| \sigma_{AC}) \leq \norm{ \sigma_{AC}^{-1} \, \rho_{AC} - \identity_{AC} } \,.\]
 \qed
\vspace{0.2cm}

Let $\rho_{AC}$ be a full-rank state on $\mathcal{H}_{AC}$. Let us recall that the \emph{mutual information} of $\rho$ between two regions $A$ and $C$ is defined as
\[ I_{\rho}(A:C) := D(\rho_{AC} 	\, || \, \rho_{A}  \otimes \rho_{C}) \]
The analogous notion for the BS-entropy, which we call \textit{BS-mutual information}, can be defined as
\[ \widehat{I}_{\rho}(A:C) := \widehat{D}(\rho_{AC} 	\, || \, \rho_{A}  \otimes \rho_{C}) \]
According to \eqref{eq:rel-BS-entropy},
\[ I_{\rho}(A:C) \,\, \leq \,\, \widehat{I}_{\rho}(A:C)\,. \]
Moreover, we can define the Rényi mutual information arising from the $q$-geometric Rényi divergence by
\begin{equation*}
     \widehat{I}^q_\rho (A : C) := \widehat{D}_q (\rho_{AC} \| \rho_A \otimes \rho_C) \, ,
\end{equation*}
as done in \cite{Scalet2021}. Then, we conclude from \eqref{equa:boundGeometricRenyiDivergence}, for any $q >1$,
\begin{equation*}
     \widehat{I}^q_\rho (A : C) \leq \norm{ \rho_{A}^{-1} \otimes \rho_C^{-1}  \rho_{AC} - \identity_{AC} } \, . 
\end{equation*}

\noindent Furthermore, as a consequence of \eqref{equa:boundBSrelativeEntropy2} it holds that
\begin{equation}\label{equa:BasicBoundMutualInfo2}
\widehat{I}_{\rho}(A:C) \,\, \leq \,\, \left\| \rho_{A}^{-1} \otimes \rho_{C}^{-1} \rho_{AC} - \identity_{AC} \right\| \, . \\
\end{equation}
To conclude this subsection, let us recap and emphasize that the following chain of relations, which will appear frequently throughout the text, holds true for any $q>1$ and any full-rank state $\rho_{AC}$:
\begin{align*}
    \frac{1}{2}\operatorname{Corr}_{\rho}(A:C)^{2} & \leq \frac{1}{2} \| \rho_{AC} - \rho_{A} \otimes \rho_{C} \|_{1}^{2} \\
    & \leq I_{\rho}(A:C) 
    \leq \, \widehat{I}_\rho(A:C) \, \leq \, \widehat{I}^q_\rho(A:C) \, \leq \, \| \rho_{A}^{-1} \otimes \rho_{C}^{-1} \rho_{AC} - \identity_{AC}  \|. 
\end{align*}

\vspace{0.2cm}

\subsection{Quantum spin chains} \label{sec:quantum-spin-chains}
In this work, we will consider a quantum spin system over $\mathbb{Z}$.  For each finite subinterval $I \subset \mathbb{Z}$ we denote by $|I|$ its cardinal.  At every site $x \in \mathbb{Z}$ we have a finite-dimensional local Hilbert space $\mathcal{H}_{x} \equiv \mathbb{C}^{d}$ and the corresponding space of operators $\mathfrak{A}_{x}:= \mathcal B(\mathbb{C}^{d})$.  For each finite subset $\Lambda \subset \mathbb{Z}$ let
\[ \mathcal{H}_{\Lambda} := \otimes_{x \in \Lambda}{\mathcal{H}_{x}} \,, \quad \mathfrak{A}_{\Lambda}:= \mathcal B(\mathcal{H}_{\Lambda}) = \otimes_{x \in \Lambda}{\mathfrak{A}_{x}}\,. \] 
If $\Lambda' \subset \Lambda$ we have a canonical linear isometry
\[ \mathfrak{A}_{\Lambda'} \longrightarrow \mathfrak{A}_{\Lambda} = \mathfrak{A}_{\Lambda'} \otimes \mathfrak{A}_{\Lambda \setminus \Lambda'}\,, \quad Q \longmapsto Q \otimes \mathbbm{1}_{\Lambda \setminus \Lambda'}\,. \]
The (normed) algebra of local observables $\mathfrak{A}_{loc}$ is then formally defined as the inductive limit of the family $(\mathfrak{A}_{\Lambda})_{\Lambda}$, and its completion with respect to the operator norm is the algebra of quasi-local observables $\mathfrak{A}_{\mathbb{Z}}$. In the following, $\norm{A}$ will always be the operator norm of $A \in \mathfrak{A}_{\mathbb Z}$ if we do not specify the norm. The norms $\norm{\cdot}_p$ for $p \in [1,\infty]$ will be the Schatten $p$-norms. Notably, $p = 1$ corresponds to the trace norm.

\subsection{Locality estimates in 1D}

Let us also fix a local interaction $\Phi = (\Phi_{X})_{X}$, where $X$ runs over all finite subsets of $\mathbb{Z}$, and where $\Phi_X \in \mathfrak{A}_X$, $\Phi_X = \Phi_X^\ast$. Recall that the diameter of $X$ is defined as $\operatorname{diam}(X) = \max_{x,y \in X}|x-y|$. We assume that $\Phi$ has finite range $r > 0$ and strength $J > 0$, namely $\| \Phi_{X}\| = 0$ if $X$ has diameter greater than $r$ and $\| \Phi_{X}\| \leq J$ for all $X$. As usual, we denote for every finite subset $\Lambda \subset \mathbb{Z}$ the corresponding Hamiltonian by
\[ H_{\Lambda} := \sum_{X \subset \Lambda}{\Phi_{X}} \, , \]
and the (possibly complex) time evolution of an observable $Q \in \mathfrak{A}_{\mathbb{Z}}$ by
\[ \Gamma_{H_{\Lambda}}^{s}(Q) := e^{isH_\Lambda}Qe^{-isH_\Lambda}\,, \quad s \in \mathbb{C}\,. \]

The following proposition gathers several results by Araki \cite{Araki1969} for finite-range interactions, see also \cite{Perez2020}  for a more recent result applying to a more general type of interactions.

\begin{prop}\label{Theo:mainLocality}
Let $\Phi$ be a local interaction with range $r$ and strength $J$ as above. Then, for every $\beta > 0$ there is a constant   $\mathcal{G}=\mathcal{G}(\beta, J,r) > 1$ satisfying the following properties:\\
\begin{enumerate}
\item[(i)] For every pair $a,b \in \mathbb{Z}$ with $a \leq b$, every observable $Q$ in $\mathfrak{A}_{[a,b]}$\,, every complex $s \in \mathbb{C}$ with $|s| \leq \beta$ and every pair $n$, $m \in \mathbb{Z}$ with $0 \leq n \leq m$
\begin{align*}
\| \Gamma_{H_{[a-m,b+m]}}^{s}(Q) - \Gamma_{H_{[a-n,b+n]}}^{s}(Q) \| & \leq \mathcal{G}^{b-a + 1} \, \frac{\mathcal{G}^{n}}{\left( \lfloor n/r \rfloor + 1\right)!} \, \| Q\| \, ,\\[2mm]
\| \Gamma_{H_{[a-n,b+n]}}^{s}(Q) \| & \leq  \mathcal{G}^{b-a+1} \, \| Q\| \, .
\end{align*}
\item[(ii)] Denoting
\[ E_{n}(s) := e^{-s H_{[1-n,n]}} e^{s H_{[1-n,0]} + s H_{[1,n]}}\,,\quad n \in \mathbb{N}\,,\, s \in \mathbb{C}, \]
we have for every $1 \leq n \leq m$ and every $s \in \mathbb{C}$ with $|s| \leq \beta$
\[ \| E_{n}(s) \| \,, \, \| E_{n}^{-1}(s) \| \, \leq \,  \mathcal{G}\, \, , \]
\[ \| E_{n}(s) - E_{m}(s) \| \,, \, \| E_{n}(s)^{-1} - E_{m}(s)^{-1}\| \, \leq  \, \frac{\mathcal{G}^{n}}{\left( \lfloor n/r \rfloor + 1\right)!}\,.   \]
\end{enumerate}
\end{prop}

The first part of the previous proposition can be interpreted as an extended version of the celebrated Lieb-Robinson bounds \cite{Lieb1972} for complex-time evolution (in the particular setting of 1D and finite-range interactions). These estimates are one of the main ingredients in Araki's argument regarding the absence of (thermal) phase transitions in 1D.

\begin{remark}
To simplify notation, we will absorb $\beta$ in the Hamiltonian so that it will be an implicit parameter of the interaction strength $J$. Since all our results hold for Gibbs states \mbox{$\rho = e^{-H}/\mathrm{Tr}(e^{-H})$} associated to arbitrary finite-range Hamiltonians, they actually hold for every positive temperature. 
\end{remark}

Let $I \subset \mathbb{Z}$ be a finite interval. Let us split $I$ into two subintervals $X$ and $Y$ so that $I=XY$. Without loss of generality, we can assume that they correspond to $X=[1-a,0]$ and $Y = [1, b]$ for some $a, b \in \mathbb{N}$. We present now an alternative formulation of Araki's results that will be useful.

\begin{cor}\label{Coro:mainLocality}
Let $\Phi$ be a local interaction with range $r$ and strength $J$ as above.  For a finite interval $I=XY \subset \mathbb{Z}$ split into two subintervals $X$ and $Y$, let us write
\[ E_{X,Y}(s) \, := \, e^{-s \, H_{XY}} \, e^{s \, H_{X} + s \, H_{Y}} \quad , \quad s \in \mathbb{C}\,. \]
\[ E_{X,Y} \, := \, E_{X,Y}(1) \, = \, e^{-\, H_{XY}} \, e^{\, H_{X} \, + \, H_{Y}} \,. \]
Then, there is an absolute constant $\mathcal{G}>1$ depending only on $J$ and $r$ such that:
\begin{enumerate}
\item[(i)] For every $s \in \mathbb{C}$ with $|s| \leq 1$
\begin{equation*}\label{equa:AuxiliarBound1} 
\| E_{X,Y}(s)\| \, , \, \| E_{X,Y}(s)^{-1}\| \,\, \leq \,\, \mathcal{G} \, .
\end{equation*}
\item[(ii)] If we add two intervals $\widetilde{X}$ and $\widetilde{Y}$ adjacent to $X$ and $Y$, respectively, so that we get a larger interval $\widetilde{J}:=\widetilde{X}XY\widetilde{Y}$, then 
\begin{equation*}\label{equa:AuxiliarBound2} 
 \left\| E_{X,Y}^{-1}(s) - E^{-1}_{\widetilde{X}X,Y\widetilde{Y}}(s) \right\|, \left\| E_{X,Y}(s) - E_{\widetilde{X}X,Y\widetilde{Y}}(s) \right\| \, \leq \, \frac{\mathcal{G}^{\ell}}{(\lfloor \ell/r \rfloor + 1)!} \quad , \quad |s| \leq 1 \,.
 \end{equation*}
 for any $\ell \in \mathbb{N}$ such that $\ell \, \leq \, |X| \, , \, |Y|$.\\
 \item[(iii)] Let us identify $X=[1-a,0]$ and $Y=[1,b]$ for some $a,b \in \mathbb{N}$, and denote $X_{n}:=X \cap [1-n,0]$ and $Y_{n}:=Y \cap [1,n]$ for each $n \in \mathbb{N}$. Then, we can decompose
 \[ E_{X,Y} = \sum_{n=1}^{\infty} 
\widetilde{E}^{(n)}\quad \text{where} \quad  \widetilde{E}^{(1)}:=E_{X_{1}, Y_{1}} \,\, , \,\, \widetilde{E}^{(n)} := E_{X_{n}, Y_{n}} - E_{X_{n-1}, Y_{n-1}} \quad (n \geq 2)\]  
so that this series is absolutely convergent with
\[ \widetilde{E}^{(n)} \in \mathfrak{A}_{[1-n, n]},\quad \big\| \widetilde{E}^{(1)} \big\| \leq \mathcal G   \quad \text{and} \quad \big\| \widetilde{E}^{(n)} \big\| \leq \frac{\mathcal{G}^{n-1}}{(\lfloor (n-1)/r \rfloor + 1)!}  \quad \text{ for every } n \geq 2\,.\]
 \end{enumerate}
\end{cor}

\begin{remark}\label{rem:alternative_corollary}
One-sided versions of Corollary \ref{Coro:mainLocality}.(ii) can be given when adding an adjacent interval to only one side of $I=XY$:
\begin{enumerate}
\item[(ii)$'$] If we add an interval $\widetilde{Y}$ adjacent to $Y$ so that we get a larger interval $\widetilde{J}:=XY\widetilde{Y}$, then 
\begin{equation*}
 \Big\| E_{X,Y}^{-1}(s) - E^{-1}_{X,Y\widetilde{Y}}(s) \Big\| \, , \quad \Big\| E_{X,Y}(s) - E_{X,Y\widetilde{Y}}(s) \Big\| \, \leq \, \frac{\mathcal{G}^{\ell}}{(\lfloor \ell/r \rfloor + 1)!} \quad , \quad |s| \leq 1 \,.
 \end{equation*}
 for any $\ell \in \mathbb{N}$ such that $\ell \, \leq  \, |Y|$.\\
\item[(ii)$''$]  If we add an interval $\widetilde{X}$ adjacent to $X$ so that we get a larger interval $\widetilde{J}:=\widetilde{X}XY$, then 
\begin{equation*}
 \Big\| E_{X,Y}^{-1}(s) - E^{-1}_{\widetilde{X}X,Y}(s) \Big\| \, , \quad \Big\| E_{X,Y}(s) - E_{\widetilde{X}X,Y}(s) \Big\| \, \leq \, \frac{\mathcal{G}^{\ell}}{(\lfloor \ell/r \rfloor + 1)!} \quad , \quad |s| \leq 1 \,.
 \end{equation*}
 for any $\ell \in \mathbb{N}$ such that $\ell \, \leq  \, |X|$.\\
 \end{enumerate}
 Both statements $(ii)'$ and $(ii)''$ are actually a consequence of Corollary \ref{Coro:mainLocality}.(ii). Indeed, to deduce $(ii)'$ consider a new set of interactions $\widehat{\Phi}$ obtained from $\Phi$ by setting \mbox{$\widehat{\Phi}_{\Lambda}=\Phi_{\Lambda}$} whenever $\Lambda$ is contained in $XY\widetilde{Y}$ and $\widehat{\Phi}_{\Lambda}=0$ otherwise. Then, the corresponding expansionals $\widehat{E}$ for this new interaction satisfy $\widehat{E}_{X,Y} = E_{X,Y}$ and $\widehat{E}_{\widetilde{X}X,Y\widetilde{Y}} = E_{X,Y\widetilde{Y}}$. Since $\widehat{\Phi}$ has the same bounds for interaction range and strength than $\Phi$, we can apply Corollary \ref{Coro:mainLocality}.(ii) to get the desired estimate. A similar argument shows that $(ii)''$ holds.
\end{remark}

\vspace{0.2cm}

\subsection{Partial trace} \label{sec:partial-trace}

We will mainly deal with a finite one-dimensional spin system on a finite interval $I \subset \mathbb{Z}$. Let us split $I$ into two parts, namely $ \Lambda$ and $\Lambda^{c}:=I \setminus \Lambda$. Thus, $\mathfrak{A}=\mathfrak{A}_{I}=\mathfrak{A}_{\Lambda} \otimes \mathfrak{A}_{\Lambda^{c}}$. The (unnormalized) trace over $I$ will be written as $\mathrm{Tr}_I: \mathfrak{A}_{I} \to \mathbb C$. Moreover, we will consider the operator  (unnormalized) partial trace on $\Lambda$,  given for $\Lambda  \subset I$  by  the range-extended map
\begin{equation}\label{eq:partial-trace}
\operatorname{tr}_{\Lambda}: \mathfrak{A}_I \longrightarrow \mathfrak{A}_I \quad, \quad R \otimes S \longmapsto \operatorname{Tr}_{\Lambda}(R) (\mathbbm{1}_{\Lambda } \otimes S)\, ,  \; \; \forall R \in  \mathfrak{A}_{\Lambda}  \, , \, \, \forall S \in  \mathfrak{A}_{ \Lambda^c}\,.     
\end{equation}
Note that for every $R \in \mathfrak{A}_\Lambda$, $S \in \mathfrak{A}_{\Lambda^c}$ as above
\[ \| \operatorname{tr}_{\Lambda}(R \otimes S)\| = |\operatorname{Tr}_\Lambda(R)| \, \| S\| \, .  \]
Sometimes, it will be more convenient to work with the conditional expectation in the sense of \cite[Proposition 1.12]{Ohya1993}, which is a unital completely positive map \cite[Proposition 5.2.2]{Benatti2009}. We will write $\mathbb E_{\Lambda^c}:\mathfrak A \to \mathfrak A_{\Lambda^c}$ for this conditional expectation. It is defined as in \eqref{eq:partial-trace}, but using the normalized trace instead of the unnormalized one. 

With this notation, if $H_{\Lambda}$ is the Hamiltonian of the system on $\Lambda$ with local interaction $\Phi$, the local Gibbs state on $\Lambda$ is given by
\[ \rho^{\Lambda}:= e^{-H_{\Lambda}}/\operatorname{Tr}_{\Lambda}(e^{-H_{\Lambda}})\,. \]
 We might omit the superscript when it is clear from the context. The reduced operator of $\rho=\rho^{I}$ on $\Lambda$ will be denoted by
\[ \rho_{\Lambda}^{I} = \rho_{\Lambda} := \operatorname{tr}_{\Lambda^c}(\rho)\,. \]
Note that $H_{\Lambda}$ has support in $\Lambda$, and so the map
\[ \mathfrak{A} \longrightarrow \mathfrak{A}\,, \quad Q \longmapsto \operatorname{tr}_{\Lambda}(e^{-H_{\Lambda}} Q) = \operatorname{tr}_{\Lambda}(e^{-\frac{1}{2}H_{\Lambda}} Q e^{-\frac{1}{2}H_{\Lambda}})  \]
is positive. The well-known Russo-Dye theorem yields
\[ \| \operatorname{tr}_{\Lambda}(e^{-H_{\Lambda}} Q) \| \leq \|\operatorname{tr}_{\Lambda}(e^{-H_{\Lambda}} )\| \, \| Q\|=\operatorname{Tr}_{\Lambda}(e^{-H_{\Lambda}} ) \, \| Q\|\,. \]
 As a consequence, the map
\[ \mathfrak{A} \longrightarrow \mathfrak{A}\,, \quad Q \longmapsto \operatorname{tr}_{\Lambda}( \rho^{\Lambda}Q) \]
is positive, unital and hence contractive.  This observation will be repeatedly used in the following results. Note that $\operatorname{tr}_{\Lambda}$ is an operator valued map, whereas $\operatorname{Tr}_{\Lambda}$ maps to scalars.

\section{Locality of observables}\label{sec:locality_observables}
In this section, we will prove some norm estimates on local observables which will be used repeatedly in the rest of the article. 

Let $I \subset \mathbb{Z}$ be a finite interval. For each local observable $Q \in \mathfrak A_{\mathrm{loc}}$, let us define by
\[ \|Q\|_{I}:= \inf{\{ \|Q-P\| \colon P \in \mathfrak{A}_{I} \}} \]
the distance from $Q$ to the subspace $\mathfrak{A}_{I}$. Beware that this is clearly not a norm. 

\begin{lem}\label{lem:localityInverse}
Let $Q,Q'$ be two local observables. Then
\[ \| Q \, Q' \|_{I} \, \leq \, 2\| Q\|_{I} \, \| Q'\| \, + \, 2\| Q\| \, \| Q'\|_{I}\,. \]
Moreover, if $Q$ is positive and invertible, then
\[ \| Q^{-1}\|_{I} \, \leq \, 2 \, \| Q^{-1}\|^{2} \, \| Q\|_{I} \, .\]
\end{lem}

\noindent \textit{Proof.}
We start with an observation: By compactness, there exists an element $P_{I} \in \mathfrak{A}_{I}$ satisfying $\| Q\|_{I} = \| Q -P_{I}\|$.
But using the conditional expectation, we have that the element
\[ Q_{I}:= \mathbb E_I(Q)\in \mathfrak{A}_{I} \]
satisfies
\begin{equation}\label{equa:equivLocalApprox}
\| Q\|_{I} \leq \| Q - Q_{I}\| \leq 2 \| Q\|_{I}\,.
\end{equation}
Indeed, the left-hand side is clear, and the right-hand side follows simply from
\begin{align*} 
\| Q - Q_{I} \| & \leq \| Q - P_{I} \| + \| P_{I} - Q_{I}\|\\ 
& = \| Q -P_{I}\| + \| \mathbb E_I(Q - P_{I})\|\\ 
& \leq 2 \, \| Q - P_{I}\| = 2 \| Q\|_{I}\,. 
\end{align*}
Here, we have used that the conditional expectation is a unital completely positive map and thus contractive. Hence, we can bound
\begin{align*}
\| Q Q'\|_{I} & \, \leq \, \| Q Q' - Q_{I} Q'_{I}\|\\[2mm]
& \, \leq \, \| Q\| \, \| Q' - Q'_{I}\| \, + \, \| Q - Q_{I}\| \, \| Q'_{I}\| \\[2mm]
& \, \leq \, 2\| Q\|  \,  \|Q'\|_{I}  +  2\| Q\|_{I} \, \| Q'\|
\end{align*}

\noindent If we assume that $Q$ is positive and invertible, that is
\begin{equation}\label{equa:PositiveInvertibleCondition} 
\| Q^{-1}\|^{-1} \mathbbm{1} \leq Q \leq \| Q\| \mathbbm{1} \,
\end{equation} 
we can then apply the conditional expectation to get from \eqref{equa:PositiveInvertibleCondition} that $Q_{I}$ is also positive and invertible with
\begin{equation}\label{equa:PositiveInvertibleCondition2} 
\| Q^{-1} \|^{-1} \mathbbm{1} \leq Q_{I} \leq \| Q\| \mathbbm{1}\,.  
\end{equation}
In particular, it easily follows that $\| Q_{I}^{-1}\| \leq \| Q^{-1}\|$. Combining this and \eqref{equa:equivLocalApprox} we conclude
\begin{align*}
\| Q^{-1}\|_{I} \leq \| Q^{-1} - Q_{I}^{ - 1}\| & \leq \| Q^{-1}\|  \, \| Q - Q_{I}\| \, \| Q_{I}^{- 1}\| \\ 
& \leq \| Q^{-1}\|^{2} \, \| Q - Q_{I} \| \\ 
& \leq 2 \| Q^{-1}\|^{2} \, \| Q\|_{I}.
\end{align*}
This finishes the proof.\qed
\vspace{0.2cm}

The following result appears in the proof of \cite[Theorem 4.2]{Araki1969}, but we will anyway prove it here for the sake of completeness as an immediate consequence of Proposition \ref{Theo:mainLocality}.

\begin{lem}\label{lem:localityDynamics}
Let $n \in \mathbb N$ and let $I_{n}:=[1-n,n]$. 
Then, for every observable $Q \in \mathfrak{A}_{\mathbb{Z}}$ with finite support, every $s \in \mathbb{C}$ with $|s| \leq 1$ and every finite subset $\Lambda \subset \mathbb{Z}$
\[ \| \Gamma_{H_{\Lambda}}^{s}(Q) \|  \leq \mathcal{G}^{2} \|Q\| + (1+ \mathcal{G}^{2})  \sum_{n \geq 1} \mathcal{G}^{2n} \| Q\|_{I_{n}}\, , \, \]
where $\mathcal{G} = \mathcal{G}(1)$ is the constant from Proposition \ref{Theo:mainLocality}.
\end{lem}

\noindent \textit{Proof.}
  For each $n \in \mathbb{N}$, compactness allows us to fix $P_{n} \in \mathfrak{A}_{I_{n}}$ such that $\| Q\|_{I_{n}} = \| Q - P_{n}\|$. We can then write $Q = \sum_{n \geq 1}{Q_{n}}$, where $Q_{1}:=P_{1}$ and $Q_{n} :=P_{n} - P_{n-1}$ for each $n>1$. Note that this series is actually a finite sum, since $Q$ has finite support. On the one hand, we can easily bound
\[ \| Q_{1}\| \leq \|Q \|_{I_{1}} + \|Q \| \quad , \quad  \| Q_{n}\| \leq \| Q\|_{I_{n}} + \| Q\|_{I_{n-1}} \quad (n \geq 2)\,. \]
On the other hand, using Proposition \ref{Theo:mainLocality} and noticing that $Q_{n}$ has support in $I_{n}$, we have
\[ \| \Gamma_{H_{\Lambda}}^{s}(Q_{n}) \| \,  \leq \, \mathcal{G}^{2n} \, \| Q_{n}\|\,. \]
Combining all these inequalities, we conclude
\begin{align*}
\| \Gamma_{H_{\Lambda}}^{s}(Q)\| \leq \sum_{n \geq 1} \| \Gamma_{H_{\Lambda}}^{s}(Q_{n})\| &  \leq \sum_{n \geq 1} \mathcal{G}^{2n}\| Q_{n}\|\\[2mm] 
& \leq \, \mathcal{G}^{2} \|Q\| + (1+ \mathcal{G}^{2})  \sum_{n \geq 1} \mathcal{G}^{2n} \| Q\|_{I_{n}}\,.
\end{align*}\qed
\vspace{0.2cm}

The next proposition combines the two previous lemmas to give an estimate on the norm of the complex time evolution of observables of a certain kind.  
\begin{prop}\label{Theo:auxiliarLocalityPartialTrace}
Let $I=ABC$ be a finite interval split into three adjacent intervals so that $B$ shields $A$ from $C$, and let $Q \in \mathfrak{A}_{ABC}$ be a positive and invertible observable.  Define 
\[ F(s) \, := \, e^{sH_{AC}} \, \tr_{B}(\rho^{B}Q) \, e^{-sH_{AC}} \quad , \quad s \in \mathbb{C} \,. \]
Let us identify  $B=[1,k]$ for some $k \geq 0$, where $k=0$ corresponds to the case $B=\emptyset$, and denote $I_{n} = [1-n, k+n]$ for each $n \geq 1$.
Then, for any $s \in \mathbb C$ such that $|s|\leq 1$, \\
\begin{align*}
\| F(s)\| & \leq (1+\mathcal{G})^{2} \bigg( \| Q\| + \sum_{n \geq 1} \mathcal{G}^{2n} \, \| Q\|_{I_{n}} \bigg) \, , \\[2mm]
\| F(s)^{-1}\| & \leq (1+\mathcal{G})^{2} \bigg( \| Q^{-1}\| +  2\| Q^{-1}\|^{2}\sum_{n \geq 1} \mathcal{G}^{2n} \, \| Q\|_{I_{n}} \bigg) \, .
\end{align*}
\end{prop}

\noindent \textit{Proof.}
Let us denote $\widetilde{Q}:=\tr_{B}(\rho^{B}Q)$.
Since $Q$ is positive and invertible,
\[ \|Q^{-1}\|^{-1} \, \mathbbm{1} \, \leq \, Q \, \leq \, \| Q\| \, \mathbbm{1}\,. \]
As a consequence, if we apply the (positive and unital) map $P \mapsto \tr_{B}(\rho^{B}P)$ we get
\begin{equation}\label{equa:auxiliarLocality1}
\|Q^{-1}\|^{-1} \, \mathbbm{1} \, \leq \, \widetilde{Q} \, \leq \,  \| Q\| \, \mathbbm{1}\,.
\end{equation} 
Since $P \mapsto \tr_{B}(\rho^{B}P)$ is moreover contractive, we can also get
\begin{equation}\label{equa:auxiliarLocality2}
\| \widetilde{Q}\|_{I_{n}} \, \leq \, \| Q\|_{I_{n}}\,. 
\end{equation}
At this point, note that all factors
\[ F(s) = e^{s H_{AC}} \, \widetilde{Q} \, e^{-s H_{AC}} \]
are supported in $A \cup C$. Making use of the canonical identification $\mathfrak{A}_{A} \otimes \mathfrak{A}_{C} \hookrightarrow \mathfrak{A}_{ABC}$ described in Section \ref{sec:quantum-spin-chains}, we can ``ignore'' $B$ and consider $A$ and $C$ as two adjacent intervals of the form $[1-a,0]$ and $[1,c]$ for some $a,c \in \mathbb{N}$. In this case, the Hamiltonian $H_{AC}$ is made of interactions with the same bounds for their range and strength. Moreover, the interval $I_{n}$ can be identified with $\widetilde{I}_{n}=[1-n,n]$ for each $n \in \mathbb{N}$, so that applying Lemma \ref{lem:localityDynamics} 
\[ \| F(s)\| \, \leq \, \mathcal{G}^{2}\| Q\| + (1+\mathcal{G}^{2}) \sum_{n \geq 1} \mathcal{G}^{2n} \, \| Q\|_{I_{n}}\,. \]
On the other hand, combining \eqref{equa:auxiliarLocality2} with Lemma \ref{lem:localityInverse}, 
\[ \| \widetilde{Q}^{-1} \|_{\widetilde{I}_{n}} \, \leq \, 2 \, \| \widetilde{Q}^{-1}\|^{2} \, \| \widetilde{Q}\|_{\widetilde{I}_{n}} \leq 2\| Q^{-1}\|^{2} \, \| Q\|_{I_{n}}\,, \]
where in the last inequality we have used  that $\big\|\widetilde{Q}^{-1}\big\| \leq \norm{Q^{-1}}$ due to \eqref{equa:auxiliarLocality1}. Thus, using that
\[ F(s)^{-1} = e^{sH_{AC}} \, \widetilde{Q}^{-1} \, e^{-sH_{AC}}\,, \]
we can argue as above, applying Lemma \ref{lem:localityDynamics}  to estimate
\[ \| F(s)^{-1}\| \leq \mathcal{G}^{2}\| Q^{-1}\| + 2(1+\mathcal{G}^{2}) \| Q^{-1}\|^{2} \sum_{n \geq 1} \mathcal{G}^{2n} \, \| Q\|_{I_{n}}\,. \]
This finishes the proof.\qed
\vspace{0.2cm}

The previous proposition allows us to bound the norm of more complicated local observables which we will encounter often in later sections. 
\begin{cor}\label{Coro:BoundingPartialTraceInverses}

Let $I = ABC$ be three adjacent finite intervals (we admit the possibility of some being empty). Then, there is an absolute constant $\mathcal{C}$ depending only on the strength $J$ and range $r$ of the local interactions, such that
\begin{align}
   &  \big\| \tr_{B}(\rho^{B}Q) \big\|, \big\| \tr_{B}(\rho^{B}Q)^{-1} \big\| \, \leq \, \mathcal{C} \, \,  , \quad \quad Q \in \{ E_{B,C}^{\dagger} \, , \, E_{B,C}\, , \, E_{A,B}^{\dagger} \, , \, E_{A,B} \} \, , \label{equa:BoundingPartialTraceInverses1}\\[2mm]
  &   \label{equa:BoundingPartialTraceInverses2} \big\| \tr_{AB}(\rho^{AB}Q) \big\| \, , \, \big\| \tr_{AB}(\rho^{AB}Q)^{-1} \big\| \,   \leq \, \mathcal{C} \, \,  , \quad \quad Q \in \{ E_{A,B}^{\dagger \, -1} \, , \, E_{A,B}^{\, -1} \} \, ,\\[2mm]
  &  \label{equa:BoundingPartialTraceInverses3} \big\| \tr_{B}\big(\rho^{B}E_{A,B}^{\dagger} E_{AB,C}^{\dagger}\big) \big\| \, , \, \big\| \tr_{B}\big(\rho^{B}E_{A,B}^{\dagger} E_{AB,C}^{\dagger}\big)^{-1} \big\| \, \leq \, \mathcal{C} \, .
\end{align}

\end{cor}

\noindent \textit{Proof.}
Let us identify $B$ with the interval $[1,k]$ for some $k\geq 0$, where $k=0$ corresponds to $B$ being empty, and define $I_{n} = [1-n,k+n]$ for each $n \in \mathbb{N}$. We prove \eqref{equa:BoundingPartialTraceInverses1} for the case $Q=E_{A,B}$, since the argument for the other three cases is analogous. Let
\[ F_{A,B}:=\tr_{B}(\rho^{B}E_{A,B}) = e^{-\frac{1}{2}H_{A}} \, \tr_{B}(\rho^{B}Q_{A,B}) \, e^{\frac{1}{2}H_{A}} \, , \]
where
\[ Q_{A, B}:=E_{A,B}^{\dagger}(\tfrac{1}{2})\, E_{A,B}(\tfrac{1}{2})\,. \]
Using Lemma \ref{lem:localityInverse} and Corollary \ref{Coro:mainLocality}, 
\[ \| Q_{A,B}\|_{I_{n}} \, \leq \, 2 \, \| E_{A,B}^{\dagger}(\tfrac{1}{2})\|_{I_{n}} \mathcal{G} + 2 \, \mathcal{G} \|E_{A,B}(\tfrac{1}{2})\|_{I_{n}} \, \leq \, \frac{4 \, \mathcal{G}^{n+1}}{(\lfloor n/r\rfloor + 1)!}\,. \]
The last inequality can be seen by choosing $E_{A\cap I_n, B\cap I_n}(1/2)$ and its adjoint as local approximations. Thus, by Proposition \ref{Theo:auxiliarLocalityPartialTrace}
\[ \| F_{A,B}\| \leq (1 + \mathcal{G})^{2} \bigg(\mathcal{G}^{2} +  \sum_{n \geq 1} \frac{4 \, \mathcal{G}^{3n+1}}{(\lfloor n/r\rfloor + 1)!} \, \bigg) \]
and 
\[ \| F_{A,B}^{-1}\| \leq (1 + \mathcal{G})^{2} \bigg(\mathcal{G}^{2} +  \sum_{n \geq 1} \frac{8 \, \mathcal{G}^{3n+5}}{(\lfloor n/r\rfloor + 1)!} \, \bigg)\,. \]
 Let us next prove \eqref{equa:BoundingPartialTraceInverses2}. We just argue in the case where $Q=E_{A,B}^{-1}$, since the other cases are completely analogous. Let
\[ \widetilde{F}_{A,B}:=\tr_{AB}\big(\rho^{AB} \, E_{A,B}^{-1}\big) =  \tr_{AB}\big(\rho^{AB} \, \widetilde{Q}_{A,B}\big) \, , \]
where
\[ \widetilde{Q}_{A, B}:=E_{A,B}^{\dagger \, -1}\big(\tfrac{1}{2}\big) E_{A,B}^{\, -1}\big(\tfrac{1}{2}\big)\,. \]
Arguing as before we get the desired inequalities.
 Finally, let us prove \eqref{equa:BoundingPartialTraceInverses3}. Let us first rewrite 
\begin{align*} 
F_{A,B,C} := \tr_{B}\big(\rho^{B} \, E_{A,B}^{\dagger} \, E_{AB,C}^{\dagger}\big) \, & = \, \tr_{B}\big(\rho^{B} \, e^{H_{A}+H_{B}+H_{C}} e^{-H_{ABC}}\big)\\[2mm] 
& = e^{\frac{1}{2}H_{A} + \frac{1}{2}H_{C}} \, \tr_{B}\big(\rho^{B} \, Q_{A,B,C}\big) \, e^{-\frac{1}{2}H_{A} - \frac{1}{2}H_{C}} \, ,
\end{align*}
where
\[ Q_{A,B,C}:=E_{A,B}^{\dagger}\big(\tfrac{1}{2}\big) E_{AB,C}^{\dagger}\big(\tfrac{1}{2}\big) E_{AB,C}\big(\tfrac{1}{2}\big) E_{A,B}\big(\tfrac{1}{2}\big) \, .\]
According to Corollary \ref{Coro:mainLocality}, we can bound
\[ \big\| Q_{A, B, C}\big\|, \big\| Q_{A, B,C}^{-1}\big\| \leq \mathcal{G}^{4}\,.\]
Moreover, using Lemma \ref{lem:localityInverse} and Corollary \ref{Coro:mainLocality}
\begin{align*}
\big\| Q_{A,B,C} \big\|_{I_{n}} \, & \leq \, 2 \, \big\| E_{A,B}^{\dagger}\big(\tfrac{1}{2}\big) \, E_{AB, C}^{\dagger}\big(\tfrac{1}{2}\big)\big\|_{I_{n}} \, \mathcal{G}^{2} + 2 \, \mathcal{G}^{2} \, \big\|  E_{AB, C}\big(\tfrac{1}{2}\big) \, E_{A,B}\big(\tfrac{1}{2}\big) \big\|_{I_{n}} \\[2mm]
& \leq \, 4 \, \mathcal{G}^{3} \Big( \, \big\| E_{A,B}^{\dagger}\big(\tfrac{1}{2}\big)\big\|_{I_{n}} \, + \, \big\| E_{AB, C}^{\dagger}\big(\tfrac{1}{2}\big)\big\|_{I_{n}} \, + \,  \big\| E_{AB, C}\big(\tfrac{1}{2}\big)\big\|_{I_{n}} \, + \, \big\| E_{A,B}\big(\tfrac{1}{2}\big) \big\|_{I_{n}}  \, \Big)\\[2mm] 
& \leq \, \frac{16 \, \mathcal{G}^{n+3}}{\big( \lfloor n/r \rfloor + 1 \big)!} \, .
\end{align*}
Thus, applying Proposition \ref{Theo:auxiliarLocalityPartialTrace},
\[
\big\| F_{A,B,C}\big\| \leq (1+ \mathcal{G})^{2} \, \bigg( \,  \mathcal{G}^{4} + \sum_{n \geq 1} \frac{16  \,  \mathcal{G}^{3n+3}}{\big( \lfloor n/r \rfloor + 1 \big)!} \, \bigg)
\]
and 
\[
\big\| F_{A,B,C}^{-1}\big\| \leq (1+ \mathcal{G})^{2}  \, \bigg(  \, \mathcal{G}^{4} + \mathcal{G}^{8} \, \sum_{n \geq 1} \frac{32 \, \mathcal{G}^{3n+3}}{\big( \lfloor n/r \rfloor + 1 \big  )!} \, \bigg)\, .
\] \qed
\vspace{0.2cm}

\section{Approximate factorization of the Gibbs state}\label{sec:approximate_factorization}

The following Theorem is our first main result. It shows that the Gibbs state of a local Hamiltonian on $ABC$ can be approximated by a certain product of its marginals. The error made in the approximation decays superexponentially in the length scale of the system under consideration. Note that by the results of \cite{Bluhm2020}, the two following conditions are equivalent for any quantum channel $\mathcal{T}$ and any strictly positive states $\rho$ and $\sigma$:
\begin{equation*}
    \rho = \sigma \,  \mathcal{T}^* \left( \mathcal{T}(\sigma)^{-1} \mathcal{T}(\rho) \right) \qquad \iff \qquad \widehat D(\rho || \sigma ) = \widehat D(\mathcal{T} (\rho) || \mathcal{T} (\sigma) ) \, ,
\end{equation*}
where the map $ \mathcal{B}_\mathcal{T}^\sigma (\cdot) := \sigma \,  \mathcal{T}^* \left( \mathcal{T}(\sigma)^{-1} (\cdot) \right)$ is called \textit{BS-recovery condition}, in analogy to the Petz recovery map for the analogous equality in the data-processing inequality for the relative entropy. In the particular case of a tripartite space $\hs_{ABC}= \hs_A \otimes \hs_B \otimes \hs_C$, two strictly positive states $\rho_{ABC}$, $\sigma_{ABC} \in \mathfrak{A}_{ABC}$ such that $\sigma_{ABC}= \rho_{AB} \otimes \identity_C / d_C$ and a quantum channel $\mathcal{T}:= \mathbb{E}_A$, we say that $\rho_{ABC}$ is a \textit{BS-recoverable state} if it is a fixed point of the BS-recovery condition composed with the partial trace, i.e., if
\begin{equation*}
     \rho_{ABC} = \rho_{AB} \rho_{B}^{-1} \rho_{BC} \, .
\end{equation*}
Therefore, any quantum Markov chain is a BS-recoverable state but the converse is not true \cite{Hiai2017}. Moreover, note that, by the equivalence stated above, the following clearly holds
\begin{equation*}
    \rho_{ABC} = \rho_{AB} \rho_{B}^{-1} \rho_{BC} \qquad \iff \qquad \widehat D(\rho_{ABC} || \rho_{AB}) = \widehat D(\rho_{BC} || \rho_B)
\end{equation*}
We emphasize that the BS-entropy on the very right is the one on $\mathcal H_{BC}$. The main result of this section thus states that the distance of a quantum Gibbs state from being BS-recoverable decays superexponentially with the size of the middle system. 
This result should be compared to the findings in \cite{Kato2019}, since the equality $ D(\rho_{ABC} || \rho_{AB}) =  D(\rho_{BC} || \rho_B)$ is equivalent to $\rho_{ABC}$ being a quantum Markov chain.

\begin{figure}[h]
\begin{center}

\begin{tikzpicture}[scale=0.7]

\definecolor{frenchblue}{rgb}{0.0, 0.45, 0.73}
\Block[5,frenchblue!50!white,A,1];

\begin{scope}[xshift=5cm]
\Block[5,green!50!white,B,1];
\end{scope}
\begin{scope}[xshift=10cm]
\Block[5,violet!50!white,C,1];
\end{scope}

\node at (15,0.8) {\huge $I$};

\end{tikzpicture}

  \caption{Representation of an interval $I$ split into three subintervals $I=ABC$, where $B$ shields $A$ from $C$. Here we are taking $\ell=5$. }
  \label{fig:1}
      
\end{center}
\end{figure}
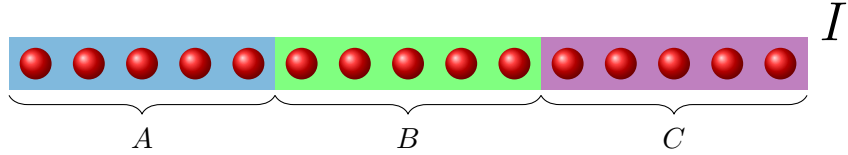

\begin{theo}\label{thm:main_BS_recoverable}
Let $\Phi$ be a finite-range interaction over $\mathbb{Z}$. Then, there exists a positive and decreasing function $\ell \longmapsto \epsilon(\ell)$  with superexponential decay and depending only on the strength $J$ and the range $r$ of the interaction with the following property: 
for every $I \subset \mathbb{Z}$ finite interval split into three subintervals $I=ABC$, where $B$ shields $A$ from $C$ and $|B| \geq \ell$ $($see Figure \ref{fig:1}$)$, and for $\rho = \rho^{I}$ the Gibbs state on $I$,
\[ \big\| \rho_{ABC} - \rho_{AB} \rho_{B}^{-1} \rho_{BC} \big\|_{1}\, \leq \, \epsilon(\ell)\,. \]
\end{theo}

\noindent \textit{Proof.}
 Using Hölder's inequality
\begin{align*}
 \big\| \rho_{ABC} - \rho_{AB} \rho_{B}^{-1} \rho_{BC} \big\|_{1} & \leq \big\| \mathbbm{1} - \rho_{AB} \rho_{B}^{-1} \rho_{BC} \rho_{ABC}^{-1}\big\| \, \| \rho_{ABC}\|_{1}\\[2mm] 
 & = \big\| \mathbbm{1} - \rho_{AB} \rho_{B}^{-1} \rho_{BC} \rho_{ABC}^{-1}\big\| \,. 
 \end{align*}
We are actually going to bound 
\begin{equation}\label{equa:dualInequality} 
\big\| \mathbbm{1} - \rho_{ABC} \rho_{BC}^{-1} \rho_{B} \rho_{AB}^{-1} \big\| \, \leq \, \tilde{\epsilon}(\ell) \,.
\end{equation}
The above result will follow by simply recalling that if an operator $Q$ satisfies\\
\mbox{$\|  \mathbbm{1} - Q\| \leq \tilde{\epsilon} < 1$}, then
\begin{equation}\label{eq:series}
    \big\| \mathbbm{1} - Q^{-1}\big\| = \big\| \mbox{$\sum_{k \geq 1}(\mathbbm{1} - Q)^{k}$} \big\| \leq \tilde{\epsilon}/(1 - \tilde{\epsilon})\,.
\end{equation}
For large $\ell$, the assertion follows from \eqref{eq:series} by taking e.g.\ $\epsilon(\ell)= 2 \tilde{\epsilon}$. For small $\ell$, we can directly bound the LHS of \eqref{equa:dualInequality} by a constant, as we show below. To estimate \eqref{equa:dualInequality}, let us first rewrite 
\begin{align*}
\rho_{ABC} \,  \rho_{BC}^{-1}  & = e^{-H_{ABC}}   \, \operatorname{tr}_{A}\big(e^{-H_{ABC}}\big)^{-1} \, \\[2mm] 
& =   e^{-H_{ABC}} \, e^{H_{A} + H_{BC}} \, e^{-H_{A}} \, \operatorname{tr}_{A}\big( \, e^{-H_{ABC}} e^{H_{A} + H_{BC}} e^{-H_{A}} \big)^{-1} \,  \\[2mm]
& = E_{A,BC} \, \rho^{A} \, \operatorname{tr}_{A}\big(  E_{A, BC} \, \rho^{A} \big)^{-1} \, , \,
\end{align*}
where we are using the notation $\rho^{\Lambda}$ and $E_{\Lambda, \Lambda^{c}}$ introduced in Section \ref{sec:partial-trace}. Analogously 
\begin{align*} 
 \rho_{B} \, \rho_{AB}^{-1} & = \operatorname{tr}_{AC}\big(e^{-H_{ABC}}\big) \, \operatorname{tr}_{C}\big(e^{-H_{ABC}}\big)^{-1} \\[2mm]
& = \operatorname{tr}_{AC}\big(e^{-H_{ABC}}\big) \, \operatorname{tr}_{C}\big(e^{-H_{BC}}\big)^{-1}    \, \left( \,  \operatorname{tr}_{C}\big(e^{-H_{ABC}}\big) \, \operatorname{tr}_{C}\big(e^{-H_{BC}}\big)^{-1}\, \right)^{-1}\\[2mm]
& = \operatorname{tr}_{A}\left( \operatorname{tr}_{C}\big(e^{-H_{ABC}}\big)  \operatorname{tr}_{C}\big(e^{-H_{BC}}\big)^{-1} \right)  \, \left( \, \operatorname{tr}_{C}\big(e^{-H_{ABC}}\big) \, \operatorname{tr}_{C}\big(e^{-H_{BC}}\big)^{-1} \, \right)^{-1}\\[2mm]
& =  \,  \operatorname{tr}_{A}\big(  \widetilde{E}_{A, BC} \, \rho^{A}\big)   \, (\rho^{A})^{-1} \,  \widetilde{E}_{A, BC}^{-1} \, ,
\end{align*}
where 
\[ \widetilde{E}_{A, BC} := \operatorname{tr}_{C}\big(e^{-H_{ABC}}\big)\, \operatorname{tr}_{C}\big(e^{-H_{BC}}\big)^{-1} \, e^{H_{A}}\,. \]
Therefore
\begin{equation}  \label{equa:EstimationsSecondFactor}
\rho_{ABC}\,  \rho_{BC}^{-1} \, \rho_{B} \,  \rho_{AB}^{-1} = E_{A,BC} \, \operatorname{tr}_{A}\big(  E_{A, BC} \, \rho^{A} \big)^{-1} \, \operatorname{tr}_{A}\big(  \widetilde{E}_{A, BC} \, \rho^{A}\big)   \,  \,  \widetilde{E}_{A, BC}^{\, -1}  \, ,
\end{equation}
where the outer $\rho^{A}$'s have cancelled each other. We include now two statements whose proof is postponed by now.\\

\noindent \textbf{Statement 1}: \textit{The factors on the right-hand side of \eqref{equa:EstimationsSecondFactor} and their inverses are uniformly bounded by a constant $\widehat{\mathcal{C}}>0$ depending on $J$ and $r$ of the local interaction $\Phi$ $($but independent of $A,B,C$$)$}.\\

\noindent \textbf{Statement 2}: \textit{There is a superexponentially decaying positive function $\ell \longmapsto \delta(\ell)$ 
depending only on $J$ and $r$ of the local interaction $($but independent of $A,B,C$$)$ such that}
\[ \big\| \widetilde{E}_{A,BC} - E_{A,BC}\big\|\leq \delta(\ell)\,.\\[3mm] \]

\noindent Using that $Q \longmapsto \operatorname{tr}_{A}(Q \, \rho^{A})$ is contractive, we immediately deduce from the second statement that
\[ \|\operatorname{tr}_{A}(\widetilde{E}_{A,BC} \, \rho^{A})  -  \operatorname{tr}_{A}(E_{A,BC} \, \rho^{A})\| \leq \delta(\ell)\,.\\ \]
Therefore, replacing in the right-hand side of \eqref{equa:EstimationsSecondFactor}
\[ E_{A,BC} \leadsto \widetilde{E}_{A,BC} \quad \mbox{ and } \quad \operatorname{tr}_{A}\big(\widetilde{E}_{A,BC}\, \rho^{A}\big) \leadsto \operatorname{tr}_{A}\big(E_{A,BC}\, \rho^{A}\big)\,, \]
we can estimate 
\begin{align*}
\big\| \rho_{ABC} \, \rho_{BC}^{-1} \, \rho_{B} \rho_{AB}^{-1}  - \mathbbm{1} \big\| & \leq  \widehat{\mathcal{C}}^{ \, 3} \, \big\| E_{A, BC} - \widetilde{E}_{A,BC} \big\| + \widehat{\mathcal{C}}^{\, 3} \, \big\|\operatorname{tr}_{A}\big(\widetilde{E}_{A,BC} \, \rho^{A}\big)  -  \operatorname{tr}_{A}\big(E_{A,BC} \, \rho^{A}\big)\big\|\\
& \leq 2 \, \widehat{\mathcal{C}}^{ \, 3} \, \delta(\ell) =: \epsilon(\ell)\,.
\end{align*}
For large $\ell$, the assertion follows from \eqref{eq:series}. For small $\ell$, we can use statement 1 to upper bound $ \big\| \mathbbm{1} - \rho_{AB} \rho_{B}^{-1} \rho_{BC} \rho_{ABC}^{-1} \big\|$ by a constant. This concludes the proof. \qed
\vspace{0.2cm}

It remains to argue that both statements hold.

\vspace{0.2cm}

\noindent \textit{Proof of \textbf{Statement 1}.}
 By Corollaries \ref{Coro:mainLocality} and \ref{Coro:BoundingPartialTraceInverses}, we can bound
\[ \big\| E_{A,BC} \big\| \leq \mathcal{G} \quad \quad \text{and} \quad \quad \big\| \tr_{A}(\rho^{A}E_{A, BC})^{-1} \big\| \leq \mathcal{C}\,. \]
To bound the third and fourth factors, let us first rewrite
\begin{align} 
\widetilde{E}_{A, BC} &= \operatorname{tr}_{C}\big(e^{-H_{ABC}} e^{H_{AB}}\big)\, e^{-H_{AB}} \, e^{H_{A} + H_{B}} \,  \operatorname{tr}_{C}\big(e^{-H_{BC}} e^{H_{B}}\big)^{-1} \nonumber \\[2mm]
& \label{equa:decompEtilde}  = \operatorname{tr}_{C}\big(E_{AB, C} \, \rho^{C}\big) \, E_{A, B} \, \operatorname{tr}_{C}\big(E_{B, C} \,\rho^{C}\big)^{-1}\,.
\end{align}
We can apply once again Corollaries \ref{Coro:mainLocality}  and  \ref{Coro:BoundingPartialTraceInverses} to bound 
\[ \big\| \widetilde{E}_{A,BC}\big\| \, \leq \, \mathcal{G} \, \mathcal{C}^{2}\quad \mbox{ and } \quad \big\| \widetilde{E}_{A,BC}^{\, -1}\big\| \, \leq \, \mathcal{G} \, \mathcal{C}^{2}\,. \]
Finally, since $Q \longmapsto \operatorname{tr}_{A}(Q \, \rho^{A})$ contractive,
\[ \big\| \operatorname{tr}_{A}\big(\widetilde{E}_{A, BC} \, \rho^{A}\big)\big\| \leq \big\| \widetilde{E}_{A,BC}\big\| \leq \mathcal{G} \, \mathcal{C}^{2}\,.  \]
The inverse of this term can be bounded by the same quantity using Jensen's operator inequality  \cite{Hansen2003}. This finishes the proof of the first statement.\qed
\vspace{0.2cm}

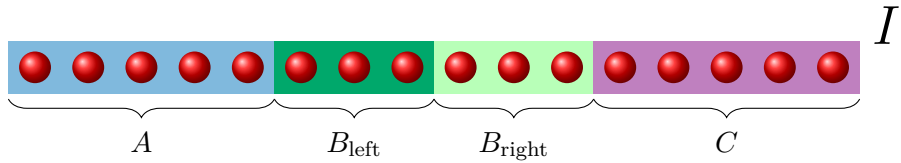
\begin{figure}[h]
\begin{center}

\begin{tikzpicture}[scale=0.7]

\definecolor{frenchblue}{rgb}{0.0, 0.45, 0.73}
\Block[5,white!50!frenchblue,A,1];

\begin{scope}[xshift=5cm]
\definecolor{jade}{rgb}{0.0, 0.66, 0.42}
\Block[3,jade,B_{\text{left}},1];
\end{scope}

\begin{scope}[xshift=8cm]
\definecolor{mintgreen}{rgb}{0.6, 1.0, 0.6}
\Block[3,white!30!mintgreen,B_{\text{right}},1];
\end{scope}

\begin{scope}[xshift=11cm]
\Block[5,violet!50!white,C,1];
\end{scope}

\node at (16,0.8) {\huge $I$};

\end{tikzpicture}

  \caption{Splitting of an interval $I$ into three subintervals $I=ABC$, where $B$ shields $A$ from $C$. Note that $B$ is further split into two 
  parts $B = B_{\mathrm{left}} B_{\mathrm{right}}$ with $|B_{\mathrm{left}}|, |B_{\mathrm{right}}| \geq \lfloor \ell/2 \rfloor$. Here we consider $\ell=6$. }
  \label{fig:2}
  \end{center}
\end{figure}

\vspace{0.2cm}

\noindent \textit{Proof of \textbf{Statement 2}.}
Using formula \eqref{equa:decompEtilde} for $\widetilde{E}_{A, BC}$, we can rewrite the difference $\widetilde{E}_{A, BC}  - E_{A,BC}$ as
\[ \left(  \operatorname{tr}_{C}\big(E_{AB, C} \, \rho^{C}\big) \, E_{A, B} - E_{A,BC} \, \operatorname{tr}_{C}\big(E_{B, C} \,\rho^{C}\big) \, \right) \, \left(\operatorname{tr}_{C}\big(E_{B, C} \,\rho^{C}\big)  \right)^{-1},  \]
and thus, by Corollary \ref{Coro:BoundingPartialTraceInverses}
\[ \big\| \widetilde{E}_{A, BC}  - E_{A,BC} \big\| \, \leq \,  \mathcal{C}  \left\| \operatorname{tr}_{C}\big(E_{AB, C} \, \rho^{C}\big) \, E_{A, B} - E_{A,BC} \, \operatorname{tr}_{C}\big(E_{B, C} \,\rho^{C}\big) \,  \right\|\,. \]
Let us split the set $B$ into two 
parts $B = B_{\mathrm{left}} B_{\mathrm{right}}$ with $|B_{left}|, |B_{right}| \geq \lfloor \ell/2 \rfloor$  (see Figure \ref{fig:2}). Taking into account that $E_{A, B_{\mathrm{left}}}$ and $\operatorname{tr}_{C}\big(E_{B_{\mathrm{right}}, C} \, \rho^{C}\big)$ commute, we can estimate
\begin{align*}
  \big\| \widetilde{E}_{A, BC}  - E_{A,BC} \big\| \,  &  \leq \,  \mathcal{C} \, \left\| \operatorname{tr}_{C}\big(E_{AB, C} \, \rho^{C}\big) \, E_{A, B} -  \operatorname{tr}_{C}\big(E_{B_{\mathrm{right}}, C} \,\rho^{C}\big) \, E_{A, B_{\mathrm{left}}} \right\| \\[2mm]
& \hspace{0.3cm} + \mathcal{C} \left\| E_{A, B_{\mathrm{left}}} \, \operatorname{tr}_{C}\big(E_{B_{\mathrm{right}}, C} \,\rho^{C}\big)  - E_{A,BC} \, \operatorname{tr}_{C}\big(E_{B, C} \,\rho^{C}\big) \,  \right\| \\[2mm]
&  \leq \mathcal{C} \, \left\| \operatorname{tr}_{C}\big(E_{AB, C} \, \rho^{C}\big) \right\| \, \left\| E_{A,B} - E_{A, B_{\mathrm{left}}}\right\|\\[2mm] 
& \hspace{0.3cm} + \mathcal{C} \left\| \operatorname{tr}_{C}\big(E_{AB, C} \, \rho^{C}\big) - \operatorname{tr}_{C}\big( E_{B_{\mathrm{right}}, C} \, \rho^{C}\big) \right\| \, \left\| E_{A,B_{\mathrm{left}}} \right\|\\[2mm]
& \hspace{0.3cm} + \mathcal{C} \left\| E_{A, B_{\mathrm{left}}} \right\| \,  \left\| \operatorname{tr}_{C}\left( E_{B_{\mathrm{right}}, C} \, \rho^{C}\right) - \operatorname{tr}_{C}\left( E_{B,C} \, \rho^{C} \right) \right\|\\[2mm]
& \hspace{0.3cm} + \mathcal{C} \, \left\| E_{A, B_{\mathrm{left}}} - E_{A, BC} \right\| \, \left\| \operatorname{tr}_{C}\big(E_{B,C} \rho^{C}\big)\right\|\\[2mm]
&  \leq \mathcal{C} \, \big\| E_{AB, C}  \big\| \, \big\| E_{A,B} - E_{A, B_{\mathrm{left}}} \big\|\, + \, \mathcal{C} \big\| E_{AB, C} - E_{B_{\mathrm{right}}, C} \big\| \, \big\| E_{A,B_{\mathrm{left}}} \big\|\\[2mm]
& \hspace{0.3cm} + \mathcal{C} \big\| E_{A, B_{\mathrm{left}}} \big\| \,  \big\| E_{B_{\mathrm{right}}, C} - E_{B,C} \big\| \, + \, \mathcal{C} \, \big\| E_{A, B_{\mathrm{left}}} - E_{A, BC} \big\| \, \| E_{B,C}\|\, ,
\end{align*}
where in the last inequality we have used again that the operator $Q \longmapsto \operatorname{tr}_{C}\big(Q \, \rho^{C}\big)$ is  contractive. Finally, applying Corollary \ref{Coro:mainLocality} and Remark \ref{rem:alternative_corollary} we can bound from above the previous expression
\begin{align*}
 \big\| \widetilde{E}_{A, BC}  - E_{A,BC} \big\|  \, \leq \, \frac{4 \, \mathcal{C} \, \mathcal{G}^{ 1 + \lfloor \ell/2 \rfloor}}{\left( \lfloor \, \lfloor\ell/2 \rfloor / r\, \rfloor +1\right)!} \, .
\end{align*}
This easily yields the desired bound.\qed
\vspace{0.2cm}

\section{Correlations in finite intervals vs the infinite chain}\label{sec:infinite-chain}

Thermal states on the infinite chain $\mathbb{Z}$ are characterized in terms of the KMS condition, after which they are also called \emph{KMS states}. In our setting, dealing with a finite-range interaction $\Phi$, it is well-known that, for every finite temperature, the corresponding KMS state exists and is unique \cite[Theorem 6.2.47]{BraRob81}.

For every finite subinterval $I \subset \mathbb{Z}$, let us denote by $\psi_{I}: \mathfrak{A}_{I} \longrightarrow \mathbb{C}$ the local Gibbs state  as
\begin{equation}\label{equa:pointwiselimitKMSstate} 
\psi_{I}(Q) := 
\Tr_{I}(\rho^{I}Q)= \frac{\Tr_{I}(e^{-H_{I}}\, Q)}{\Tr_{I}(e^{-H_{I}})} \,\, , \,\, Q \in \mathfrak{A}_I \, ,
\, 
\end{equation}
and extended to a state on $\mathfrak{A}_{\mathbb Z}$  using the Hahn-Banach Theorem as in \cite[Proposition 2.3.24]{BraRob79}.
We are absorbing the inverse temperature constant $\beta$ in the Hamiltonian to simplify. We denote by $\psi_{\mathbb{Z}}$ the unique KMS state over the infinite chain (at inverse temperature $\beta = 1$). For every increasing and absorbing sequence $I_{n} \nearrow \mathbb{Z}$, the sequence of states $\psi_{I_{n}}$ is weak$^\ast$ convergent to $\psi_{\mathbb Z}$, i.e.,
\begin{equation} \label{equa:pointwiseLimitGibbs}
\psi_{\mathbb{Z}}(Q) = \lim_{n \rightarrow \infty} \psi_{I_{n}}(Q) \, \, , \quad Q \in \mathfrak{A}_{\mathbb{Z}} 
\end{equation}
(see e.g.\ \cite[Proposition 6.2.15 and its preceeding discussion]{BraRob81}).

In his seminal paper, Araki \cite{Araki1969} proved that if $\Phi$ is moreover translation invariant, then it satisfies \emph{exponential decay of correlations}: There exist constants $\mathcal{K}$, $\alpha > 0$ (depending only on the range and the strength of the interaction) such that for every pair $Q_{A} \in \mathfrak{A}_{A}$ and $Q_{C} \in \mathfrak{A}_{C}$ with support in finite intervals $A,C$,
\begin{equation}\label{ArakiCorrelationsExponentialDecay} 
\big|\psi_{\mathbb{Z}}(Q_{A} Q_{C}) - \psi_{\mathbb{Z}}(Q_{A}) \psi_{\mathbb{Z}}(Q_{C}) \big| \, \leq \, \mathcal{K} \, e^{- \alpha \, \operatorname{dist}(A,C)} \, \| Q_{A}\| \, \| Q_{C}\|\,.  
\end{equation}

A natural question is whether these results imply that the interaction $\Phi$ satisfies the \textit{uniform clustering condition} \cite{Brandao2019}, as stated below, with a function $\varepsilon(\ell)$ decaying exponentially fast in $\ell$ and depending only on the range and interaction strength.

\begin{defi}[Uniform clustering] \label{def:unif-clustering}
Let $\Phi$ be a local interaction on $\mathbb{Z}$. We say that it is \emph{uniform clustering} if there is a positive and decreasing function $\ell \mapsto \varepsilon(\ell)$ with the following property: for every finite interval $I \subset \mathbb{Z}$ split into three subintervals $I=ABC$ with $|B| \geq \ell$, 
\begin{equation}\label{eq:uniform_clustering}
    \operatorname{Corr}_{\rho^{I}}(A:C) \leq \varepsilon(\ell)  \, .
\end{equation} 
In particular, we will say that it satisfies \emph{exponential uniform clustering} if $\varepsilon(\ell)$ can be taken to have exponential decay as $\ell$ tends to infinity.
\end{defi}

 In this line, we can actually prove that the exponential decay of correlations proven by Araki implies exponential uniform clustering. That is the content of our next main result.

\begin{theo}\label{Theo:ArakiImpliesUnifCluster} 
Let $\Phi$ be  finite-range interaction over $\mathbb{Z}$. If the infinite-chain thermal state $\psi_{\mathbb{Z}}$ satisfies the exponential decay of correlations condition, then there are constants $\widetilde{\mathcal{K}}$ and $\widetilde{\alpha}$ $($depending only on $\mathcal{K}$, $\alpha$ from \eqref{ArakiCorrelationsExponentialDecay}, and on the range $r$ and strength $J$ of the interaction$)$ with the following property: for every finite interval $I \subset \mathbb{Z}$ split into three subintervals $I=ABC$ with $|B| \geq \ell \geq 0$ and every observable $Q_{A} \in \mathfrak{A}_{A}$ and $Q_{C} \in \mathfrak{A}_{C}$, 
\[ \big| \psi_{ABC}(Q_{A}Q_{C}) \, - \, \psi_{ABC}(Q_{A}) \, \psi_{ABC}(Q_{C})\big| \,\, \leq \,\, \| Q_{A}\| \, \| Q_{C}\| \, \widetilde{\mathcal{K}} \, e^{- \widetilde{\alpha} \ell}\,. \]
\end{theo}

\noindent \textit{Proof.}
We can assume, without loss of generality that $|A|, |B|, |C| \geq  \ell$. Indeed, denoting $\ell' = \lfloor \ell/3 \rfloor$, which has the same scale as $\ell$, we can split $B$ into three subintervals \mbox{$B=B_{1}B_{2}B_{3}$} and redefine $A':=AB_{1}$, $B':=B_{2}$ and $C':=B_{3}C$ satisfying $|A'|, |B'|, |C'| \geq \ell'$. Moreover, we are going to add two adjacent intervals $\widetilde{C}=\widetilde{C}_{1}\widetilde{C}_{2}$ with $|\widetilde{C}_{1}| \geq \ell$ on the right side of  $ABC$, and other two intervals $\widetilde{A}=\widetilde{A}_{2}\widetilde{A}_{1}$ with $|\widetilde{A}_{1}| \geq \ell$ on the left side of $ABC$. Thus we have a larger finite interval
\[ \widetilde{I} \, = \, \widetilde{A} ABC \widetilde{C} \, = \, \widetilde{A}_{2} \widetilde{A}_{1}  A B C \widetilde{C}_{1} \widetilde{C}_{2}\,. \]

\begin{figure}[h]
  
  \begin{center}
     
  \begin{tikzpicture}[scale=0.65]

\Block[2,0,0,0];

\definecolor{frenchblue}{rgb}{0.0, 0.45, 0.73}

\begin{scope}[xshift=2cm]
\Block[4,white!60!frenchblue,\widetilde{A},1];
\Block[2,white!10!frenchblue,\widetilde{A}_2,2];
\end{scope}
\begin{scope}[xshift=4cm]
\Block[2,white!40!frenchblue,\widetilde{A}_1,2];
\end{scope}
\begin{scope}[xshift=6cm]
\Block[4,white!60!frenchblue,A,1];
\end{scope}

\definecolor{caribbeangreen}{rgb}{0.0, 0.8, 0.6}

\begin{scope}[xshift=10cm]
\Block[3,caribbeangreen,B,1];
\end{scope}

\begin{scope}[xshift=13cm]
\Block[4,white!70!violet,C,1];
\end{scope}
\begin{scope}[xshift=17cm]
\Block[4,white!30!violet,\widetilde{C},1];
\Block[2,white!50!violet,\widetilde{C}_1,2];
\end{scope}
\begin{scope}[xshift=19cm]
\Block[2,white!20!violet,\widetilde{C}_2,2];
\end{scope}

\begin{scope}[xshift=21cm]
\Block[2,0,0,0];
\end{scope}

\node at (22,1.5) {\huge $\widetilde{I}$};

\end{tikzpicture}
  
  \caption{Splitting of an interval $I$ into three subintervals $I=ABC$ with $|A|, |B|, |C| \geq \ell$, to which we further append $\widetilde{A}$ and $\widetilde{C}$ at the left and right side, respectively. Subsequently, we split $\widetilde{A}$ and $\widetilde{C}$ into two subintervals each, so that each of them is of size at least $\ell$.} 
  \label{fig:IntervalsGibbs2}
    \end{center}
\end{figure}
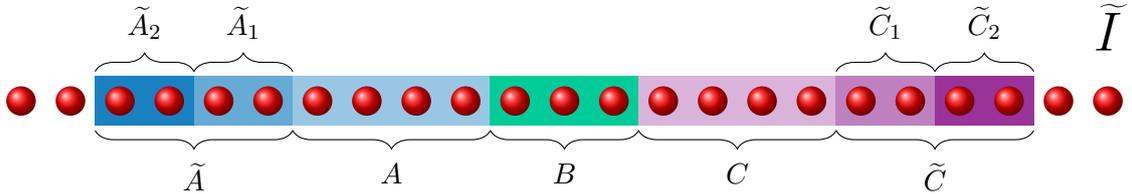

\noindent We will repeatedly use that for scalars $a,a',b,b' \in \mathbb{C}$
\begin{equation}\label{equa:quotientDifferenceAux}
\left| \frac{a}{b} - \frac{a'}{b'} \right| \leq \frac{1}{|b|} |a - a'| + \frac{|a'|}{|b| \, |b'|} \, |b - b'|\,.
\end{equation}

\noindent Let us denote
\[ \Theta_{\widetilde{A}, ABC, \widetilde{C}}  \, := \, e^{\frac{1}{2} H_{\widetilde{A}ABC\widetilde{C}}} \, e^{-\frac{1}{2}H_{\widetilde{A}} - \frac{1}{2} H_{ABC} - \frac{1}{2} H_{\widetilde{C}}} \, = \, \Theta_{\widetilde{A}ABC, \widetilde{C}} \,\Theta_{\widetilde{A}, ABC} \, , \]
where
\[ \Theta_{\widetilde{A}ABC, \widetilde{C}} \, := \, e^{\frac{1}{2} H_{\widetilde{A} ABC \widetilde{C}}} \, e^{-\frac{1}{2} H_{\widetilde{A}ABC} - \frac{1}{2} H_{\widetilde{C}}} \quad , \quad \Theta_{\widetilde{A}, ABC} \, := \, e^{\frac{1}{2}H_{\widetilde{A}ABC}} e^{-\frac{1}{2} H_{\widetilde{A}} \, - \frac{1}{2} H_{ABC}}\, , \]
and in general 
\begin{equation*}
    \Theta_{X, Y} \, := \, e^{\frac{1}{2} H_{XY}} \, e^{-\frac{1}{2} H_{X} - \frac{1}{2} H_{Y}} \,.
\end{equation*}
Note that the norm and locality properties of these observables (and their inverses) are known by Corollary \ref{Coro:mainLocality}. In particular, we can approximate,
\begin{equation}\label{equa:approxFfactors0} 
\Theta_{\widetilde{A}, ABC, \widetilde{C}} \, \approx \, \Theta_{\widetilde{A}_{1}, A} \, \Theta_{C, \widetilde{C}_{1}} 
\end{equation}
using the aforementioned corollary and the conditions $|A|, |\widetilde{A}_{1}|, |B|, |\widetilde{B}_{1}| \geq \ell$. Indeed
\begin{equation}\label{equa:approxFfactors} 
\begin{split}
\big\| \Theta_{\widetilde{A}, ABC, \widetilde{C}} \, - \, \Theta_{\widetilde{A}_{1}, A} \, \Theta_{C, \widetilde{C}_{1}}\big\| & = \big\| \Theta_{\widetilde{A}ABC, \widetilde{C}} \, \Theta_{\widetilde{A}, ABC} \, - \, \Theta_{\widetilde{A}_{1}, A} \, \Theta_{C, \widetilde{C}_{1}} \big\| \\[2mm]
& \leq \mathcal{G} \big\| \Theta_{\widetilde{A}, ABC} - \Theta_{\widetilde{A}_{1}, A} \big\| + \mathcal{G} \, \big\| \Theta_{\widetilde{A}ABC, \widetilde{C}} - \Theta_{C, \widetilde{C}_{1}}\big\| \\[2mm]
& \leq 2 \, \mathcal{G} \, \frac{\mathcal{G}^{\ell}}{(\lfloor \ell / r\rfloor + 1)!}\,.
\end{split}
\end{equation}
Let $Q:=Q_AQ_C$. Next, let us rewrite
\begin{equation}\label{equa:rewrittingFiniteGibbsStateTheoUnifClust}
\begin{split} 
\psi_{ABC}(Q)  \,\, & = \,\, \frac{\mathrm{Tr}_{ABC}\left(e^{-H_{ABC}}\, Q \right)}{\mathrm{Tr}_{ABC} \left(e^{-H_{ABC}}\right)}\\[2mm] 
& = \, \frac{\mathrm{Tr}_{\widetilde{A}ABC\widetilde{C}}\left(e^{-H_{\widetilde{A}} - H_{ABC} - H_{\widetilde{C}}} \, Q \right)}{\mathrm{Tr}_{\widetilde{A}ABC\widetilde{C}}\left(e^{-H_{\widetilde{A}} - H_{ABC} - H_{\widetilde{C}}}\right)}\\[3mm]
& = \, \frac{\mathrm{Tr}_{\widetilde{A}ABC\widetilde{C}}\left(e^{-H_{\widetilde{A}ABC\widetilde{C}}} \, \Theta_{\widetilde{A},ABC,\widetilde{C}} \, Q \,  \Theta_{\widetilde{A},ABC,\widetilde{C}}^{\dagger}\right)}{\mathrm{Tr}_{\widetilde{A}ABC\widetilde{C}}\left(e^{-H_{\widetilde{A}ABC\widetilde{C}}} \,  \Theta_{\widetilde{A},ABC,\widetilde{C}} \, \Theta_{\widetilde{A},ABC,\widetilde{C}}^{\dagger} \right)}\\[3mm]
& = \frac{\psi_{\widetilde{A}ABC \widetilde{C}} \left( \Theta_{\widetilde{A},ABC,\widetilde{C}} \, Q \,  \Theta_{\widetilde{A},ABC,\widetilde{C}}^{\dagger} \right) }{\psi_{\widetilde{A}ABC \widetilde{C}} \left( \Theta_{\widetilde{A},ABC,\widetilde{C}} \, \Theta_{\widetilde{A},ABC,\widetilde{C}}^{\dagger} \right)}\,.
\end{split}
\end{equation}

\noindent This suggests, having \eqref{equa:approxFfactors0} in mind, that we can estimate
\begin{equation}\label{equa:UnifClustTheoApprox1} 
\psi_{ABC}(Q)  \, \approx \, \frac{\psi_{\widetilde{A}ABC \widetilde{C}} \left( \, \Theta_{\widetilde{A}_{1}, A} \, Q_{A} \,  \Theta_{\widetilde{A}_{1}, A}^{\dagger} \,  \Theta_{C,\widetilde{C}_{1}} \, Q_{C} \, \Theta_{C,\widetilde{C}_{1}}^{\dagger} \, \right) }{\psi_{\widetilde{A}ABC \widetilde{C}} \left( \, \Theta_{\widetilde{A}_{1}, A} \, \Theta_{\widetilde{A}_{1}, A}^{\dagger} \,  \Theta_{C,\widetilde{C}_{1}} \, \Theta_{C,\widetilde{C}_{1}}^{\dagger} \, \right)}\,.
\end{equation}
Note that 
by Corollary \ref{Coro:mainLocality},
\begin{equation}\label{equa:UnifClustTheoApprox1.5}
 \mathcal{G}^{-4} \, \leq  \, \psi_{\widetilde{A}ABC \widetilde{C}} \left( \, \Theta_{\widetilde{A}_{1}, A} \, \Theta_{\widetilde{A}_{1}, A}^{\dagger} \,  \Theta_{C,\widetilde{C}_{1}} \Theta_{C,\widetilde{C}_{1}}^{\dagger} \, \right) \, , \, \, \psi_{\widetilde{A}ABC \widetilde{C}} \left( \Theta_{\widetilde{A},ABC,\widetilde{C}} \, \Theta_{\widetilde{A},ABC,\widetilde{C}}^{\dagger} \right) \, \leq \, \mathcal{G}^{4} \,,
\end{equation}  
Combining \eqref{equa:rewrittingFiniteGibbsStateTheoUnifClust}, \eqref{equa:quotientDifferenceAux} and \eqref{equa:UnifClustTheoApprox1.5}, we can bound
\begin{align*}
    & \left| \psi_{ABC}(Q)  \, - \, \frac{\psi_{\widetilde{A}ABC \widetilde{C}} \left( \, \Theta_{\widetilde{A}_{1}, A} \, Q_{A} \, \Theta_{\widetilde{A}_{1}, A}^{\dagger} \,  \Theta_{C,\widetilde{C}_{1}} \, Q_{C} \, \Theta_{C,\widetilde{C}_{1}}^{\dagger} \, \right) }{\psi_{\widetilde{A}ABC \widetilde{C}} \left( \, \Theta_{\widetilde{A}_{1}, A} \, \Theta_{\widetilde{A}_{1}, A}^{\dagger} \,  \Theta_{C,\widetilde{C}_{1}} \, \Theta_{C,\widetilde{C}_{1}}^{\dagger} \, \right)} \right| \,  \\[2mm]
    & \hspace{2.5cm} \leq \, \mathcal{G}^{4} \, \left\| \Theta_{\widetilde{A},ABC,\widetilde{C}} \,  Q \, \Theta_{\widetilde{A},ABC,\widetilde{C}}^{\dagger} - \Theta_{\widetilde{A}_{1}, A} \, Q_{A} \, \Theta_{\widetilde{A}_{1}, A}^{\dagger} \,  \Theta_{C,\widetilde{C}_{1}} \, Q_{C} \, \Theta_{C,\widetilde{C}_{1}}^{\dagger} \right\| \, + \, \\[2mm]
    & \hspace{3.4cm} + \, \mathcal{G}^{12} \, \| Q_{A}\| \, \| Q_{C}\| \, \left\|\Theta_{\widetilde{A},ABC,\widetilde{C}}  \, \Theta_{\widetilde{A},ABC,\widetilde{C}}^{\dagger} \, - \,  \Theta_{\widetilde{A}_{1}, A} \, \Theta_{\widetilde{A}_{1}, A}^{\dagger} \,  \Theta_{C,\widetilde{C}_{1}} \, \Theta_{C,\widetilde{C}_{1}}^{\dagger} \right\|\\[2mm]
     & \hspace{2.5cm} \leq \, 2 \, \mathcal{G}^{6} \, \| Q_{A}\| \, \| Q_{C}\| \, \left\| \Theta_{\widetilde{A},ABC,\widetilde{C}}  - \Theta_{\widetilde{A}_{1}, A}  \,  \Theta_{C,\widetilde{C}_{1}}  \right\| \, + \, \\[2mm]
      & \hspace{3.4cm} + \, 2 \, \mathcal{G}^{14} \, \| Q_{A}\| \, \| Q_{C}\| \, \left\| \Theta_{\widetilde{A},ABC,\widetilde{C}}   \, - \,  \Theta_{\widetilde{A}_{1}, A}   \,  \Theta_{C,\widetilde{C}_{1}}  \right\|\\[2mm]
      & \hspace{2.5cm} \leq 8 \, \mathcal{G}^{15} \, \| Q_{A}\| \, \| Q_{C}\| \, \frac{\mathcal{G}^{\ell}}{(\lfloor \ell / r\rfloor + 1)!}\,.
\end{align*} 
Note that in the above approximation, the inner observables $Q_{A}, Q_{C}, \Theta_{\widetilde{A}_{1}, A}, \Theta_{C,\widetilde{C}_{1}}$ are independent of the size of $\widetilde{A}_{2}$ and $\widetilde{C}_{2}$. Hence, we can take limit when $|\tilde{A}_{2}|, |\widetilde{C}_{2}|$ tend to infinity  so that
\begin{equation}\label{equa:UnifClustTheoApprox2}
\left| \psi_{ABC}(Q) - \frac{\psi_{\mathbb{Z}} \left( \Theta_{\widetilde{A}_{1}, A} \, Q_{A} \, \Theta_{\widetilde{A}_{1}, A}^{\dagger} \,  \Theta_{C,\widetilde{C}_{1}} \, Q_{C} \, \Theta_{C,\widetilde{C}_{1}}^{\dagger}  \right) }{\psi_{\mathbb{Z}} \left(  \Theta_{\widetilde{A}_{1}, A} \, \Theta_{\widetilde{A}_{1}, A}^{\dagger} \,  \Theta_{C,\widetilde{C}_{1}} \, \Theta_{C,\widetilde{C}_{1}}^{\dagger}  \right)} \right| \, \leq \, 8 \, \mathcal{G}^{15} \, \| Q_{A}\| \, \| Q_{C}\| \, \frac{\mathcal{G}^{\ell}}{(\lfloor \ell / r\rfloor + 1)!}\,. 
\end{equation}
We can next use the exponential decaying condition on $\psi_{\mathbb{Z}}$ and \eqref{equa:quotientDifferenceAux} to estimate
\begin{align*} 
& \left| \frac{\psi_{\mathbb{Z}} \left(  \Theta_{\widetilde{A}_{1}, A} \, Q_{A} \, \Theta_{\widetilde{A}_{1}, A}^{\dagger} \,  \Theta_{C,\widetilde{C}_{1}} \, Q_{C} \, \Theta_{C,\widetilde{C}_{1}}^{\dagger}  \right) }{\psi_{\mathbb{Z}} \left(  \Theta_{\widetilde{A}_{1}, A} \, \Theta_{\widetilde{A}_{1}, A}^{\dagger} \,  \Theta_{C,\widetilde{C}_{1}} \, \Theta_{C,\widetilde{C}_{1}}^{\dagger} \right)} \, - \, 
\frac{\psi_{\mathbb{Z}} \left(  \Theta_{\widetilde{A}_{1}, A} \, Q_{A} \, \Theta_{\widetilde{A}_{1}, A}^{\dagger}   \right) }{\psi_{\mathbb{Z}} \left(   \Theta_{\widetilde{A}_{1},A}\, \Theta_{\widetilde{A}_{1},A}^{\dagger}  \right)} \,
\frac{\psi_{\mathbb{Z}} \left(   \Theta_{C,\widetilde{C}_{1}} \, Q_{C} \, \Theta_{C,\widetilde{C}_{1}}^{\dagger}  \right) }{\psi_{\mathbb{Z}} \left(   \Theta_{C,\widetilde{C}_{1}} \, \Theta_{C,\widetilde{C}_{1}}^{\dagger}  \right)} \right| \, \\[2mm]
& \hspace{1.5cm} \leq \, \mathcal{G}^{4} \, \left| \psi_{\mathbb{Z}} \left(  \Theta_{\widetilde{A}_{1}, A} \,  Q_{A} \, \Theta_{\widetilde{A}_{1}, A}^{\dagger} \,  \Theta_{C,\widetilde{C}_{1}} \, Q_{C}  \, \Theta_{C,\widetilde{C}_{1}}^{\dagger}  \right) \right. \\[2mm]
& \hspace{3cm} \left.- \psi_{\mathbb{Z}} \left(  \Theta_{\widetilde{A}_{1}, A} \, Q_{A} \, \Theta_{\widetilde{A}_{1}, A}^{\dagger}  \right) \, \psi_{\mathbb{Z}} \left(   \Theta_{C,\widetilde{C}_{1}} \, Q_{C} \, \Theta_{C,\widetilde{C}_{1}}^{\dagger}  \right) \right| \\[2mm]
& \hspace{2.5cm} +  \mathcal{G}^{12} \, \| Q_{A}\| \, \| Q_{C}\| \, \left| \psi_{\mathbb{Z}} \left(  \Theta_{\widetilde{A}_{1}, A} \, \Theta_{\widetilde{A}_{1}, A}^{\dagger} \,  \Theta_{C,\widetilde{C}_{1}} \,  \Theta_{C,\widetilde{C}_{1}}^{\dagger} \right)\right. \\[2mm]
& \hspace{6.2cm} \left.- \psi_{\mathbb{Z}} \left( \Theta_{\widetilde{A}_{1}, A} \, \Theta_{\widetilde{A}_{1}, A}^{\dagger}   \right) \, \psi_{\mathbb{Z}} \left(   \Theta_{C,\widetilde{C}_{1}} \, \Theta_{C,\widetilde{C}_{1}}^{\dagger}  \right) \right| \\[2mm]
& \hspace{1.5cm} \leq \mathcal{G}^{4} \, \mathcal{K} \, e^{- \alpha \, \ell} \, \left\| \Theta_{\widetilde{A}_{1}, A} \, Q_{A} \, \Theta_{\widetilde{A}_{1}, A}^{\dagger} \right\| \, \left\|  \Theta_{C,\widetilde{C}_{1}} \, Q_{C} \, \Theta_{C,\widetilde{C}_{1}}^{\dagger} \right\| \,  \\[2mm]
& \hspace{2.5cm} + \mathcal{G}^{12} \, \| Q_{A}\| \, \| Q_{C}\| \, \mathcal{K} \, e^{- \alpha \ell} \, \left\| \Theta_{\widetilde{A}_{1}, A} \, \Theta_{\widetilde{A}_{1}, A}^{\dagger} \right\| \, \left\|  \Theta_{C,\widetilde{C}_{1}} \, \Theta_{C,\widetilde{C}_{1}}^{\dagger} \right\|\\[2mm]
& \hspace{1.5cm} \leq \, 2 \, \mathcal{G}^{16} \, \| Q_{A}\| \, \| Q_{C}\| \, \mathcal{K} \, e^{- \alpha \, \ell} \, .
\end{align*}
Therefore, by combining the two inequalities
\begin{multline}\label{equa:UnifClustTheoApprox3}
\left| \psi_{ABC}(Q) - \frac{\psi_{\mathbb{Z}} \left( \, \Theta_{\widetilde{A}_{1}, A} \, Q_{A}  \, \Theta_{\widetilde{A}_{1}, A}^{\dagger}  \, \right) }{\psi_{\mathbb{Z}} \left( \,  \Theta_{\widetilde{A}_{1},A} \, \Theta_{\widetilde{A}_{1},A}^{\dagger} \, \right)} \,
\frac{\psi_{\mathbb{Z}} \left( \,  \Theta_{C,\widetilde{C}_{1}} \, Q_{C} \, \Theta_{C,\widetilde{C}_{1}}^{\dagger} \, \right) }{\psi_{\mathbb{Z}} \left( \,  \Theta_{C,\widetilde{C}_{1}} \, \Theta_{C,\widetilde{C}_{1}}^{\dagger} \, \right)} \right|   \\
 \leq \, 8 \, \mathcal{G}^{16}  \, \| Q_{A}\| \, \| Q_{C}\| \, \left(\, \frac{\mathcal{G}^{\ell}}{(\lfloor \ell / r\rfloor + 1)!} + \mathcal{K}e^{- \alpha \ell} \,\right) \, .
\end{multline}
In particular, applying \eqref{equa:UnifClustTheoApprox3} with $Q_{C} = \mathbbm{1}_{C}$ we deduce
\begin{equation}\label{equa:UnifClustTheoApprox4} 
\left| \psi_{ABC}(Q_{A}) - \frac{\psi_{\mathbb{Z}} \left( \, \Theta_{\widetilde{A}_{1}, A}  \, Q_{A} \, \Theta_{\widetilde{A}_{1}, A}^{\dagger}  \, \right) }{\psi_{\mathbb{Z}} \left( \,  \Theta_{\widetilde{A}_{1},A} \, \Theta_{\widetilde{A}_{1},A}^{\dagger} \, \right)} \right| \, \leq 8 \, \mathcal{G}^{16} \, \| Q_{A}\| \, \left(\, \frac{\mathcal{G}^{\ell}}{(\lfloor \ell / r\rfloor + 1)!} + \mathcal{K}e^{- \alpha \ell} \,\right) \,, 
\end{equation}
and analogously 
\begin{equation}\label{equa:UnifClustTheoApprox5} 
\left| \psi_{ABC}(Q_{C}) - \frac{\psi_{\mathbb{Z}} \left( \,  \Theta_{C,\widetilde{C}_{1}} \, Q_{C} \, \Theta_{C,\widetilde{C}_{1}}^{\dagger} \, \right) }{\psi_{\mathbb{Z}} \left( \,  \Theta_{C,\widetilde{C}_{1}} \, \Theta_{C,\widetilde{C}_{1}}^{\dagger} \, \right)} \right| \, \leq 8 \, \mathcal{G}^{16} \, \| Q_{C}\| \, \left(\, \frac{\mathcal{G}^{\ell}}{(\lfloor \ell / r\rfloor + 1)!} + \mathcal{K}e^{- \alpha \ell} \,\right) \,. 
\end{equation}
Finally, if we combine \eqref{equa:UnifClustTheoApprox3}, \eqref{equa:UnifClustTheoApprox4} and \eqref{equa:UnifClustTheoApprox5}, then we conclude 
\begin{align*}
\left| \psi_{ABC}(Q) - \psi_{ABC}(Q_{A}) \psi_{ABC}(Q_{C})\right| & \leq \,  24 \, \mathcal{G}^{16} \| Q_{A}\| \, \| Q_{C}\| \,\left(\, \frac{\mathcal{G}^{\ell}}{(\lfloor \ell / r\rfloor + 1)!} + \mathcal{K}e^{- \alpha \ell} \,\right)\,.  
\end{align*}   
The proof of the theorem is now complete.\qed
\vspace{0.2cm}

\vspace{0.2cm}

\section{Local indistinguishability and clustering}\label{sec:clustering}

In order to prove our third main result in Section \ref{sec:decay-of-mutual-info}, we 
first need to show some conditions on the Hamiltonian under study. More specifically, we need to prove that, whenever exponential uniform clustering holds (see Definition \ref{def:unif-clustering}), the expectation values of local observables do not distinguish between Gibbs states on different intervals in some sense if the intervals are large enough.

The next result is proved for general lattices in \cite[Theorem 5]{Brandao2019} by means of the Quantum Belief Propagation estimates for perturbed Hamiltonians \cite{Hastings2007}. It shows that uniform clustering implies another property of the Gibbs state under study, namely local indistinguishability. For the sake of completeness, we include a proof based on Araki's expansionals from Proposition \ref{Theo:mainLocality} and Corollary \ref{Coro:mainLocality}.

\begin{prop}[Local indistinguishability]\label{thm:clust-implies-indist}
Let $\Phi$ be a finite-range interaction on $\mathbb{Z}$. If $\Phi$ satisfies uniform clustering with decay $\varepsilon(\ell)$ in \eqref{eq:uniform_clustering}, then  for every finite interval $I \subset \mathbb{Z}$ split into three subintervals $I=ABC$ with $|B| \geq 2 \ell \geq 0$ and every pair of observables $Q_{A} \in \mathfrak{A}_{A}$ and $Q_{C} \in \mathfrak{A}_{C}$ we have
\begin{align*} 
& \left| \Tr_{ABC}(\rho^{ABC} \, Q_{A}) - \Tr_{AB}(\rho^{AB} \, Q_{A}) \right| \, \leq \, \| Q_{A}\| \, \frac{4 \, \mathcal{G}^{3+ \ell}}{(\lfloor \ell/r \rfloor + 1)!} + \mathcal{G}^{4} \, \| Q_{A}\| \, \varepsilon(\ell) \,, \\[2mm]  
& \left| \Tr_{ABC}(\rho^{ABC} \, Q_{C}) - \Tr_{BC}(\rho^{BC} \, Q_{C}) \right| \, \leq \, \| Q_{C}\| \, \frac{4 \, \mathcal{G}^{3+ \ell}}{(\lfloor \ell/r \rfloor + 1)!} + \mathcal{G}^{4} \, \| Q_{C}\| \, \varepsilon(\ell) \,.
\end{align*}
\end{prop}

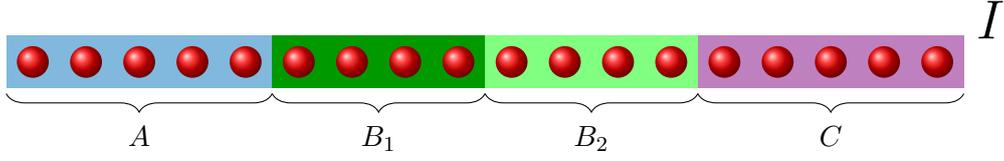
\begin{figure}[h]
\begin{center}

  \begin{tikzpicture}[scale=0.7]

\definecolor{frenchblue}{rgb}{0.0, 0.45, 0.73}

\Block[5,frenchblue!50!white,A,1];
\begin{scope}[xshift=5cm]
\Block[4,green!60!black,B_{1},1];
\end{scope}
\begin{scope}[xshift=9cm]
\Block[4,green!50!white,B_{2},1];
\end{scope}
\begin{scope}[xshift=13cm]
\Block[5,violet!50!white,C,1];
\end{scope}

\node at (18,0.8) {\huge $I$};

\end{tikzpicture}

  \caption{Splitting of an interval $I$ into four subintervals $I=AB_1B_2C$ with $|B_1|, |B_2| \geq \ell$. Here we take $\ell=4$.}
  \label{fig:3}
  \end{center}
\end{figure}

\noindent \textit{Proof.} We illustrate the proof of the first inequality, the other one is completely analogous. We can split $B$ into two subintervals $B=B_{1}B_{2}$ such that $|B_{1}|, |B_{2}| \geq \ell$ (see Figure \ref{fig:3}). Let us denote, following the nomenclature from Corollary \ref{Coro:mainLocality},
\[ \Xi_{X,Y}:=E^{\dagger}_{X,Y}(-\tfrac{1}{2}) = e^{-\frac{1}{2}H_{X} - \frac{1}{2} H_{Y}} \, e^{\frac{1}{2} H_{XY}}\,. \]
Note that $ \Xi_{X,Y}= \Theta_{X,Y}^\dagger$ in the previous section. As a consequence of the aforementioned result, as well as Remark \ref{rem:alternative_corollary}, we have  
\begin{equation}\label{equa:clusteringUndistingueAux1} 
\left\| \, \Xi_{AB, C} \, \right\| \, , \, \,  \| \,  \Xi_{AB, C}^{-1} \, \| \, \leq \, \mathcal{G} \, \, , \quad \| \, \Xi_{AB, C} - \Xi_{B_{2},C} \, \| \leq \frac{\mathcal{G}^{\ell}}{(\lfloor \ell/r \rfloor + 1)!}\,. 
\end{equation}
Note that
\[ e^{-H_{AB} - H_{C}} \, = \, \Xi_{AB,C} \, e^{-H_{ABC}} \, \Xi_{AB,C}^{\dagger}\,.  \]
Hence, we can rewrite
\begin{align*}
\Tr_{AB}(\rho^{AB} \, Q_{A}) 
& = \frac{\Tr_{ABC} \left(e^{-H_{AB}-H_{C}} \,Q_{A} \right)}{\Tr_{ABC}\left(e^{-H_{AB} - H_{C}}\right)} \\
& = \frac{\Tr_{ABC} \left(e^{-H_{ABC}} \, \Xi_{AB,C}^{\dagger} \, Q_{A} \, \Xi_{AB,C} \right)}{\Tr_{ABC} \left(e^{-H_{ABC}} \, \Xi_{AB,C}^{\dagger} \, \Xi_{AB,C} \right)}\\[2mm] 
&  =  \frac{\Tr_{ABC} \left(\rho^{ABC} \, \Xi_{AB,C}^{\dagger} \,  Q_{A} \, \Xi_{AB,C} \right)}{\Tr_{ABC} \left(\rho^{ABC} \, \Xi_{AB,C}^{\dagger} \, \Xi_{AB,C} \right)} \, . \end{align*}
The denominator is controlled by \eqref{equa:clusteringUndistingueAux1}, 
\[ \mathcal{G}^{-2} \, \leq \, \Tr_{ABC} \left(\rho^{ABC} \, \Xi_{AB,C}^{\dagger} \, \Xi_{AB,C} \right) \, \leq \, \mathcal{G}^{2} \, , \]
so that
\begin{align*} 
& \left| \Tr_{AB}(\rho^{AB} \, Q_{A})   - \Tr_{ABC}(\rho^{ABC} \, Q_{A}) \right| \\[2mm]
&  \leq \, \left| \frac{\Tr_{ABC} \left(\rho^{ABC} \, \Xi_{AB,C}^{\dagger} \, Q_{A}  \, \Xi_{AB,C} \right)}{\Tr_{ABC} \left(\rho^{ABC} \, \Xi_{AB,C}^{\dagger} \, \Xi_{AB,C} \right)} - \Tr_{ABC}(\rho^{ABC} Q_{A}) \right| \\[2mm]
&  \leq \mathcal{G}^{2} \, \left| \Tr_{ABC}\left(\rho^{ABC} \, \Xi_{AB,C}^{\dagger} \, Q_{A} \, \Xi_{AB,C} \right) - \Tr_{ABC} \left(\rho^{ABC} \, \Xi_{AB,C}^{\dagger} \, \Xi_{AB,C} \right) \Tr_{ABC} \left(\rho^{ABC} \, Q_{A} \right) \right| \\[2mm]
& \leq \mathcal{G}^{2} \left| \Tr_{ABC} \left(\rho^{ABC} \, \Xi_{AB,C}^{\dagger} \, Q_{A} \, \Xi_{AB,C}\right) -  \Tr_{ABC} \left(\rho^{ABC} \, \Xi_{B_{2},C}^{\dagger} \,  Q_{A} \, \Xi_{B_{2},C} \right) \right|  \\[2mm]
&  \; \; + \mathcal{G}^{2} \left| \Tr_{ABC} \left(\rho^{ABC} \, \Xi_{B_{2},C}^{\dagger} \, Q_{A} \, \Xi_{B_{2},C}\right) -  \Tr_{ABC} \left(\rho^{ABC} \, Q_{A} \right) \Tr_{ABC}\left(\rho^{ABC}  \, \Xi_{B_{2},C}^{\dagger} \, \Xi_{B_{2},C} \right)  \right|  \\[2mm]
& \; \; + \mathcal{G}^{2} \left| \Tr_{ABC} \left(\rho^{ABC} \, Q_{A} \right) \right| \, \left| \Tr_{ABC} \left(\rho^{ABC} \, \Xi_{B_{2},C}^{\dagger}  \, \Xi_{B_{2},C} \right) -  \Tr_{ABC} \left(\rho^{ABC} \, \Xi_{AB,C}^{\dagger} \, \Xi_{AB,C} \right)  \right| \,.
\end{align*}
The first and third summands in the last expression can be bounded similarly: Using \eqref{equa:clusteringUndistingueAux1}, it follows that
\begin{align*}
& \left| \Tr_{ABC} \left(\rho^{ABC} \, \Xi_{AB,C}^{\dagger}  \, Q_{A} \, \Xi_{AB,C} \right) -  \Tr_{ABC} \left(\rho^{ABC} \, \Xi_{B_{2},C}^{\dagger} \, Q_{A} \, F_{B_{2},C} \right) \right|\\[2mm]
& \quad \quad \quad \quad \quad \quad \leq \,  \left\| \Xi_{AB,C}^{\dagger} \, Q_{A} \, \Xi_{AB,C} - \Xi_{B_{2},C}^{\dagger} \, Q_{A} \, \Xi_{B_{2},C}  \right\| \, \\
&  \quad \quad \quad \quad \quad \quad\leq \, \| Q_{A}\| \, \frac{2 \mathcal{G}^{\ell+1}}{(\lfloor \ell/r\rfloor + 1)!} \, ,
\end{align*}
and
\begin{align*}
& \left| \Tr_{ABC}(\rho^{ABC} \, Q_{A}) \right| \, \left| \Tr_{ABC} \left(\rho^{ABC} \, \Xi_{AB,C}^{\dagger} \, \Xi_{AB,C} \right) -  \Tr_{ABC} \left(\rho^{ABC} \, \Xi_{B_{2},C}^{\dagger} \, \Xi_{B_{2},C} \right) \right|\\[2mm]
& \quad \quad \quad \quad \quad \quad \quad \quad \quad \quad  \leq \, \| Q_{A}\| \, \left\| \Xi_{AB,C}^{\dagger} \, \Xi_{AB,C} - \Xi_{B_{2},C}^{\dagger}  \, \Xi_{B_{2},C} \right\| \, \\
& \quad \quad \quad \quad \quad \quad  \quad \quad  \quad \quad  \leq \, \| Q_{A}\| \, \frac{2 \mathcal{G}^{\ell+1}}{(\lfloor \ell/r\rfloor + 1)!} \, .
\end{align*}
On the second summand we apply the uniform clustering condition. Note that $Q_{A}$ and $F_{B_{2}, C}$ commute, and that $Q_{A}$ and $F_{B_{2}, C}^{\dagger} F_{B_{2}, C}$ have supports separated by $B_{1}$, thus
\begin{align*}
& \left| \Tr_{ABC}\left(\rho^{ABC} \, \Xi_{B_{2},C}^{\dagger} \, Q_{A}  \, \Xi_{B_{2},C} \right) -  \Tr_{ABC} \left(\rho^{ABC} \, Q_{A} \right)  \Tr_{ABC}\left(\rho^{ABC} \, \Xi_{B_{2},C}^{\dagger}\, \Xi_{B_{2},C} \right)  \right| \\
& \phantom{asdasdadsasdasdasdadasdasdasasdasdasddasd} \leq  \,\varepsilon(\ell) \, \| Q_{A}\| \, \left\| \Xi_{B_{2}, C}^{\dagger} \, \Xi_{B_{2}, C}  \right\| \\[2mm]
& \phantom{asdasdadsasdasdasdadasdasdasasdasdasddasd} \leq \, \mathcal{G}^{2} \, \varepsilon(\ell) \, \| Q_{A}\|\,.
\end{align*}
Combining all these estimates we conclude the result.\qed
\vspace{0.2cm}

\begin{remark} 
Whenever $\varepsilon(\ell)$ is exponentially decaying with $\ell$, the second summand in each of the upper bounds shown in Proposition \ref{thm:clust-implies-indist} is more restrictive than the previous one $($which exhibits superexponential decay$)$. Therefore, in such a case, the first inequality $($and analogously for the second one$)$ in the previous theorem can be simplified to:
\begin{equation*}
    \left| \Tr_{ABC}(\rho^{ABC}Q_{A}) - \Tr_{AB}(\rho^{AB}Q_{A}) \right| \, \leq \, \| Q_{A}\| \, \widetilde{\mathcal{C}}\, \operatorname{e}^{-\gamma \ell} \, ,
\end{equation*}

\noindent for certain positive constants $\widetilde{\mathcal{C}}$, $\gamma >0$.
\end{remark}

\vspace{0.2cm}

\section{Decay of mutual information for Gibbs states of local Hamiltonians}\label{sec:decay-of-mutual-info}
In this section, we show that the BS-mutual information $\widehat I_\rho(A:C)$ decays exponentially fast in the distance of the two regions $A$, $C$, if $\rho$ is the Gibbs state of a local, finite-range, translation-invariant Hamiltonian on an interval $I=ABC$. In the proof, we make use of the fact that the interactions satisfy uniform clustering as in Definition \ref{def:unif-clustering} with an exponential decay. We proved this fact in Theorem \ref{Theo:ArakiImpliesUnifCluster}. 

More specifically, given a local interaction $\Phi$, by virtue of  Theorem \ref{Theo:ArakiImpliesUnifCluster} we know that there exist absolute positive constants $\widetilde{\mathcal{K}}$ and $\widetilde{\alpha}$  (depending only on $r$ and $J$) such that $\Phi$ satisfies uniform clustering as in  Definition \ref{def:unif-clustering} with
\begin{equation}\label{eq:exp_uniform_clustering}
\epsilon(\ell) \, = \,  \widetilde{\mathcal{K}} \, e^{- \widetilde{\alpha} \, \ell} \,,
\end{equation}
Now, we are ready to prove the main technical result of this section, which will be used subsequently to prove an exponential decay of the BS-mutual information:

\begin{prop}\label{Theo:MIdecay}
Let $\Phi$ be a finite-range interaction over $\mathbb{Z}$ that satisfies exponential uniform clustering. Then, there are absolute constants $\widehat{\mathcal{K}}$ and $\widehat{\alpha}$  $($depending only on the range $r$ and strength $J$ of the interaction, the local dimension $d$ and the constants $\widetilde{\mathcal{K}}$ and $\widetilde{\alpha}$ appearing in \eqref{eq:exp_uniform_clustering}$)$ with the following property: For each finite interval $I=ABC \subset \mathbb{Z}$ split into three adjacent subintervals $A, B$ and $C$ with $|B| \geq 3 \ell \geq 0$, and for $\rho=\rho^I$ the Gibbs state on $I$, 
\[ \big\| \rho_{A}^{-1} \rho_{C}^{-1} \rho_{AC} - \mathbbm{1} \big\| \,\, \leq \,\, \widehat{\mathcal{K}} \, e^{- \widehat{\alpha} \, \ell} \, . \]
\end{prop}

\noindent Before we give the proof of Proposition \ref{Theo:MIdecay}, we will use it to prove the main result of this section:

\begin{theo}\label{Theo:MIdecay-main-result}
 Let $\Phi$ be a finite-range interaction over $\mathbb{Z}$ that satisfies exponential uniform clustering. Then, there are absolute constants $\widehat{\mathcal{K}}$ and $\widehat{\alpha}$  $($depending only on the range $r$ and strength $J$ of the interaction, the local dimension $d$ and the constants $\widetilde{\mathcal{K}}$ and $\widetilde{\alpha}$ appearing in \eqref{eq:exp_uniform_clustering}$)$ with the following property: For each finite interval $I=ABC \subset \mathbb{Z}$ split into three adjacent subintervals $A, B$ and $C$ with $|B| \geq 3 \ell \geq 0$ and for $\rho=\rho^I$ the Gibbs state on $I$, 
\begin{equation*}
    I_{\rho}(A:C) \leq \widehat I_\rho(A:C) \leq \widehat{I}^q_\rho(A:C) \leq \widehat{\mathcal K} e^{- \widehat{\alpha} \ell} \, ,
\end{equation*}
for every $q \geq 1$. 
\end{theo}
\noindent \textit{Proof.}
It is a direct consequence of Proposition \ref{Theo:MIdecay} and Lemma \ref{lem:boundBSrelativeEntropy}.\qed
\vspace{0.2cm}

In particular, the previous Theorem implies exponential decay of the BS-mutual information in the setting studied by Araki \cite{Araki1969}:
\begin{cor}
Let $\Phi$ be a finite-range and translation-invariant interaction over $\mathbb{Z}$. Then, there are absolute constants $\mathcal{K}^\prime$ and $\alpha^\prime$  $($depending only on the range $r$ and strength $J$ of the interaction, the local dimension $d$ and the constants $\mathcal{K}$ and $\alpha$ appearing in \eqref{ArakiCorrelationsExponentialDecay}$)$ with the following property: For each finite interval $I=ABC \subset \mathbb{Z}$ split into three adjacent subintervals $A, B$ and $C$ with $|B| \geq 3 \ell$ and for $\rho=\rho^I$ the Gibbs state on $I$, 
\begin{equation*}
    \widehat I_\rho(A:C) \leq \mathcal K^\prime e^{- \alpha^\prime \ell} \, .
\end{equation*}
\end{cor}
\noindent \textit{Proof.}
Due to \cite{Araki1969}, \eqref{ArakiCorrelationsExponentialDecay} holds true. As we have seen, this implies exponential uniform clustering by Theorem \ref{Theo:ArakiImpliesUnifCluster}. Thus, the assertion follows from Theorem \ref{Theo:MIdecay-main-result}.\qed
\vspace{0.2cm}

\noindent Now we come to the proof of Proposition \ref{Theo:MIdecay}. Since its proof is rather technical, we have split it into several steps that will be proved in subsequent subsections.

The first step consists of rewriting the product $\rho_{A}^{-1} \rho_{C}^{-1} \rho_{AC}$ using the notation for $E_{X,Y}$ that was introduced in Corollary \ref{Coro:mainLocality}.

\begin{step}\label{step:1}
We can rewrite
\begin{equation} 
\rho_{A}^{-1}\, \rho_{C}^{-1} \, \rho_{AC} =  \tr_{BC}\big(\rho^{BC} \, E_{A, BC}^{\dagger}\big)^{-1} \, \tr_{AB}\big(\rho^{AB}E_{AB,C}^{\dagger}\big)^{-1} \, \tr_{B}\big(\rho^{B} E_{A,B}^{\dagger} \, E_{AB,C}^{\dagger}  \big) \, \lambda_{ABC} \, , 
\end{equation}
where 
\begin{equation}
 \lambda_{ABC} \,\, = \,\, \Tr_{ABC}\big(\rho^{ABC} E_{A, BC}^{\dagger \, -1}\big)^{-1} \, \Tr_{AB}\big(\rho^{AB} \, E_{A,B}^{\dagger \, -1}\big) \, .
\end{equation}
\end{step}

\vspace{2mm}
The formulation given in Step \ref{step:1} has the advantage that all the involved factors and their inverses are, by Corollary \ref{Coro:BoundingPartialTraceInverses}, (uniformly) bounded by a constant $\mathcal C$ depending only on the strength and range of the local interaction, which proves the first part of the theorem.

\vspace{2mm}

 Next, we need two estimates to conclude the proof of the theorem.

\vspace{2mm}

\begin{step}\label{step:2}
There are absolute constants $\mathcal{K}'', \alpha'' >0$ satisfying
\begin{equation*} 
|\lambda_{ABC} - 1| \,\, \leq \,\, \mathcal{K}'' \, e^{- \alpha'' \, \ell} \, . 
\end{equation*}
\end{step}

\begin{step}\label{step:3}
There are absolute constants $\mathcal{K}''', \alpha''' >0$ satisfying
\begin{equation*} 
\left\| \tr_{B}(\rho^{B} E_{A,B}^{\dagger} \, E_{AB,C}^{\dagger}  ) \, - \, \tr_{BC}(\rho^{BC} \, E_{A, BC}^{\dagger}) \, \tr_{AB}(\rho^{AB}E_{AB,C}^{\dagger}) \right\| \,  \leq \mathcal{K}''' \, e^{- \alpha''' \, \ell} \, .
\end{equation*}
\end{step}

\vspace{2mm}

Before proceeding into the proofs of the three steps, let us show how the aforementioned theorem can be proved from them.

\noindent \textit{Proof of Proposition \ref{Theo:MIdecay}.}
As a combination of Corollary \ref{Coro:BoundingPartialTraceInverses} with Step \ref{step:1}, we have
\begin{align*}
 &\left\| \rho_{A}^{-1}\, \rho_{C}^{-1} \, \rho_{AC} \, - \, \mathbbm{1} \right\|\,\,  \\[2mm]
 & \quad  \leq \,\,  \left\| \rho_{A}^{-1}\, \rho_{C}^{-1} \,  \rho_{AC} \, - \, \tr_{BC}\big(\rho^{BC} \, E_{A, BC}^{\dagger}\big)^{-1} \, \tr_{AB}\big(\rho^{AB}E_{AB,C}^{\dagger}\big)^{-1} \, \tr_{B}\big(\rho^{B} E_{A,B}^{\dagger} \, E_{AB,C}^{\dagger}  \big) \right\| \\[2mm]
 & \quad   \quad  \quad +  \left\| \mathbbm{1} \, - \, \tr_{BC}\big(\rho^{BC} \, E_{A, BC}^{\dagger}\big)^{-1} \, \tr_{AB}\big(\rho^{AB}E_{AB,C}^{\dagger}\big)^{-1} \, \tr_{B}\big(\rho^{B} E_{A,B}^{\dagger} \, E_{AB,C}^{\dagger}  \big) \right\| \\[2mm] 
 & \quad  \leq \,\, \norm{\tr_{BC}\big(\rho^{BC} \, E_{A, BC}^{\dagger}\big)^{-1} } \, \norm{\tr_{AB}\big(\rho^{AB}E_{AB,C}^{\dagger}\big)^{-1}} \, \norm{\tr_{B}\big(\rho^{B} E_{A,B}^{\dagger} \, E_{AB,C}^{\dagger}  \big)}  \, \left| \lambda_{ABC} - \mathbbm{1} \right| \,  \\[2mm]
  & \quad   \quad  \quad + \, \norm{\tr_{BC}\big(\rho^{BC} \, E_{A, BC}^{\dagger}\big)^{-1}} \, \norm{\tr_{AB}\big(\rho^{AB}E_{AB,C}^{\dagger}\big)^{-1}} \, \\[2mm]
 & \quad   \quad  \quad  \; \cdot \, \; \left\| \, \tr_{BC}\big(\rho^{BC} \, E_{A, BC}^{\dagger}\big) \, \tr_{AB}\big(\rho^{AB}E_{AB,C}^{\dagger}\big) \, - \, \tr_{B}\big(\rho^{B} E_{A,B}^{\dagger} \, E_{AB,C}^{\dagger}  \big) \right\|  \\[2mm]
  & \quad \leq \, \mathcal{C}^3 \, \left| \lambda_{ABC} - \mathbbm{1} \right|  + \, \mathcal{C}^{2} \, \left\| \, \tr_{BC}\big(\rho^{BC} \, E_{A, BC}^{\dagger}\big) \, \tr_{AB}\big(\rho^{AB}E_{AB,C}^{\dagger}\big) \, - \, \tr_{B}\big(\rho^{B} E_{A,B}^{\dagger} \, E_{AB,C}^{\dagger}  \big) \right\|  \, .
\end{align*}
Therefore, applying Step \ref{step:2} and Step \ref{step:3},
\[ \left\| \rho_{A}^{-1}\, \rho_{C}^{-1} \, \rho_{AC} \, - \, \mathbbm{1} \right\|\,\, \leq \,\, \mathcal{C}^3  \, \mathcal{K}'' \, e^{- \alpha'' \, \ell} + \mathcal{C}^{2} \, \mathcal{K}''' \, e^{- \alpha''' \, \ell} \, .\]
\qed
\vspace{0.2cm}

We devote the rest of the section to proving the three steps stated above, as well as some auxiliary ones that will be introduced afterwards for clarity. Each step is proven below in a different subsection.

\subsection{Proof of Step \ref{step:1}}

First, we introduce some auxiliary terms that we combine with the initial ones to obtain a more tractable expression. 
\begin{align*}
& \rho_{A}^{-1}\, \rho_{C}^{-1} \, \rho_{AC} \, \\
& \hspace{0.2cm}= \, \tr_{BC}\big(e^{-H_{ABC}}\big)^{-1} \, \tr_{AB}\big(e^{-H_{ABC}}\big)^{-1}\, \tr_{B}\big(e^{-H_{ABC}}\big) \, \Tr_{ABC}\big(e^{-H_{ABC}}\big)\\[2mm]
& \hspace{0.2cm}= \, \tr_{BC}\big(e^{-H_{ABC}}\big)^{-1} \, \tr_{AB}\big(e^{-H_{ABC}}\big)^{-1} \, e^{-H_{A}-H_{C}} \, e^{H_{A}+H_{C}}  \, \tr_{B}\big(e^{-H_{ABC}}\big) \, \Tr_{ABC}\big(e^{-H_{ABC}}\big)\\[2mm]
& \hspace{0.2cm}= \, \tr_{BC}\big(e^{H_{A}} \, e^{-H_{ABC}}\big)^{-1} \, \tr_{AB}\big(e^{H_{C}} \, e^{-H_{ABC}}\big)^{-1}\, \, \tr_{B}\big(e^{H_{A} + H_{C}} \, e^{-H_{ABC}}\big) \, \Tr_{ABC}\big(e^{-H_{ABC}}\big)\\[2mm]
& \hspace{0.2cm} = \tr_{BC}\left( e^{-H_{BC}} \, e^{H_{A} + H_{BC}} \, e^{-H_{ABC}} \right)^{-1} \, \tr_{AB}\left( e^{-H_{AB}} \, e^{H_{AB} + H_{C}} \, e^{- H_{ABC}}\right)^{-1} \\[2mm] 
& \hspace{0.8cm} \cdot \tr_{B}\left( e^{-H_{B}} \, e^{H_{A} + H_{B} +H_{C}} \, e^{-H_{ABC}} \right) \, \Tr_{ABC}\big(e^{-H_{ABC}}\big) \, . 
\end{align*}
Next, we multiply by $\Tr_{AB}(e^{-H_{AB}}) \, , \, \Tr_{BC}(e^{-H_{BC}}) \, , \, \Tr_{B}(e^{-H_{B}})$ and their inverses, so that after rearranging terms we arrive at
\begin{equation*}
\begin{split}
\rho_{A}^{-1}\, \rho_{C}^{-1} \, \rho_{AC} &  = \tr_{BC}\left( \rho^{BC} \, e^{H_{A} + H_{BC}} \, e^{-H_{ABC}} \right)^{-1} \, \tr_{AB}\left( \rho^{AB} \, e^{H_{AB} + H_{C}} \, e^{- H_{ABC}}\right)^{-1} \\[2mm] 
& \hspace{6.5cm}  \cdot \tr_{B}\left( \rho^{B} \, e^{H_{A} + H_{B} +H_{C}} \, e^{-H_{ABC}} \right) \, \lambda_{ABC} \, ,
\end{split}
\end{equation*}

\noindent where $\lambda_{ABC}$ is the scalar 
\begin{align*} 
\lambda_{ABC} & := \,\, \Tr_{ABC}\left( e^{-H_{ABC}} \right) \, \Tr_{BC}\left( e^{-H_{BC}} \right)^{-1} \, \Tr_{AB}\left( e^{-H_{AB}} \right)^{-1} \, \Tr_{B}\big(e^{-H_{B}}\big)\\[2mm] 
& = \,\, \Tr_{ABC}\big(\rho^{ABC} \, e^{H_{ABC}} \, e^{-H_{A} -H_{BC}}\big)^{-1} \, \operatorname{Tr}_{AB}\left( \rho^{AB} \, e^{H_{AB}} \, e^{-H_{A} - H_{B}}\right)\,. 
\end{align*}

\noindent Recall the notation $E_{\Lambda, \Lambda^{c}}$ introduced in Corollary \ref{Coro:mainLocality}. We can then rewrite
\begin{equation*} 
\rho_{A}^{-1}\, \rho_{C}^{-1} \, \rho_{AC} \, = \, \tr_{BC}\big(\rho^{BC} \, E_{A, BC}^{\dagger}\big)^{-1} \, \tr_{AB}\big(\rho^{AB}E_{AB,C}^{\dagger}\big)^{-1} \, \tr_{B}\big(\rho^{B} E_{A,B}^{\dagger} \, E_{AB,C}^{\dagger}  \big) \, \lambda_{ABC}  
\end{equation*}
and
\begin{equation*}
 \lambda_{ABC} \,\, = \,\, \Tr_{ABC}\big(\rho^{ABC} E_{A, BC}^{\dagger \, -1}\big)^{-1} \, \Tr_{AB}\big(\rho^{AB} \, E_{A,B}^{\dagger \, -1}\big) \, .
\end{equation*}
\qed

\subsection{Proof of Step \ref{step:2}} 

Note that we can bound:
\begin{align*}
      \abs{\lambda_{ABC}-1} & = \abs{\Tr_{ABC}\big(\rho^{ABC} E_{A, BC}^{\dagger \, -1}\big)^{-1} \, \Tr_{AB}\big(\rho^{AB} \, E_{A,B}^{\dagger \, -1}\big) - 1} \\
    & \leq \abs{\Tr_{ABC}\big(\rho^{ABC} E_{A, BC}^{\dagger \, -1}\big)^{-1} } \abs{\Tr_{AB}\big(\rho^{AB} \, E_{A,B}^{\dagger \, -1}\big) - \Tr_{ABC}\big(\rho^{ABC} E_{A, BC}^{\dagger \, -1}\big) } \\
    & \leq \mathcal{C}\abs{\Tr_{AB}\big(\rho^{AB} \, E_{A,B}^{\dagger \, -1}\big) - \Tr_{ABC}\big(\rho^{ABC} E_{A, BC}^{\dagger \, -1}\big) }  \, ,
\end{align*}
where the last inequality comes from Corollary \ref{Coro:BoundingPartialTraceInverses}. Since $|B| \geq 3 \ell$, we can split the interval $B$ into two subintervals $B=B_{1}B_{2}$, such that $|B_{1}|, |B_{2}| \geq \ell$. Firstly, we add and subtract some intermediate terms, which allow us to bound
\begin{align*}
 &\left| \Tr_{AB}\big(\rho^{AB} E_{A,B}^{\dagger \,-1}\big) \, - \, \Tr_{ABC}\big(\rho^{ABC} E_{A,BC}^{\dagger \,-1}\big) \right| \\
 & \hspace{5cm} \leq \,\, \left| \Tr_{AB}\big(\rho^{AB} E_{A,B}^{\dagger \,-1}\big) \, - \, \Tr_{AB}\big(\rho^{AB} E_{A,B_{1}}^{\dagger \,-1}\big) \right| \, \\[2mm]
& \hspace{5cm} \quad \quad + \left| \Tr_{AB}\big(\rho^{AB} E_{A,B_{1}}^{\dagger \,-1}\big) \, - \, \Tr_{ABC}\big(\rho^{ABC} E_{A,B_{1}}^{\dagger \,-1}\big) \right|  \\[2mm]
& \hspace{5cm} \quad  \quad + \left| \Tr_{ABC}\big(\rho^{ABC} E_{A,B_{1}}^{\dagger \,-1}\big) \, - \, \Tr_{ABC}\big(\rho^{ABC} E_{A,BC}^{\dagger \,-1}\big) \right|\,.
\end{align*}

\noindent Note that we can enlarge $A$ and $C$ trivially to ensure $|A|$, $|C| > \ell$ if needed (see Remark \ref{rem:alternative_corollary}). Using that $Q \mapsto \Tr_{\Lambda}(\rho^{\Lambda}Q)$ is contractive, and estimates from Corollary \ref{Coro:mainLocality}, we deduce that the first and third summands can be bounded 
\begin{align*}
\left| \Tr_{AB}\big(\rho^{AB} E_{A,B}^{\dagger \,-1}\big) \, - \, \Tr_{AB}\big(\rho^{AB} E_{A,B_{1}}^{\dagger \,-1}\big) \right| \,\, & \leq \,\, \left\| E_{A,B}^{\dagger \,-1} \, - \, E_{A,B_{1}}^{\dagger \,-1}  \right\| \,\, \leq \,\, \frac{\mathcal{G}^{\ell}}{(\lfloor \ell/r \rfloor + 1)!} \, , \\[2mm]
\left| \Tr_{ABC}\big(\rho^{ABC} E_{A,B_{1}}^{\dagger \,-1}\big) \, - \, \Tr_{ABC}(\rho^{ABC} E_{A,BC}^{\dagger \,-1}) \right| \,\, & \leq \,\, \left\| E_{A,B_{1}}^{\dagger \,-1} \, - \, E_{A,BC}^{\dagger \,-1}  \right\| \,\, \leq \,\, \frac{\mathcal{G}^{\ell}}{(\lfloor \ell/r \rfloor + 1)!} \, .
\end{align*}
To bound the second summand, we can then use Proposition \ref{thm:clust-implies-indist}, since $|B_{2}| \geq \ell$, which yields that
\begin{align*}
\left| \Tr_{AB}\big(\rho^{AB} E_{A,B_{1}}^{\dagger \,-1}\big) \, - \, \Tr_{ABC}\big(\rho^{ABC} E_{A,B_{1}}^{\dagger \,-1}\big) \right| \,\, & \leq \,\, \| E_{A,B_{1}}^{\dagger \,-1} \| \, \widetilde{\mathcal{C}} \, e^{- \gamma \, \ell}  \,\, \leq \,\, \mathcal{G} \, \widetilde{\mathcal{C}} \, e^{-  \gamma \, \ell} \, .
\end{align*} 
\qed

\subsection{Proof of Step \ref{step:3}} 
Since $|B| \geq 3 \ell$, we can split $B$ into three subintervals $B=B_{1} B_{2} B_{3}$ such that $|B_{j}| \geq \ell$  for each $j=1,2,3$, as in the next figure: 

\begin{figure}[h]
 \begin{center}
     
 \begin{tikzpicture}[scale=0.7]

\definecolor{frenchblue}{rgb}{0.0, 0.45, 0.73}

\Block[5,frenchblue!50!white,A,1];

\definecolor{junglegreen}{rgb}{0.16, 0.67, 0.53}

\begin{scope}[xshift=5cm]
\Block[3,junglegreen,B_{1},1];
\end{scope}

\begin{scope}[xshift=8cm]
\Block[3,green!50!white,B_{3},1];
\end{scope}

\definecolor{celadon}{rgb}{0.67, 0.88, 0.69}

\begin{scope}[xshift=11cm]
\Block[3,celadon,B_{2},1];
\end{scope}
\begin{scope}[xshift=14cm]
\Block[5,violet!60!white,C,1];
\end{scope}

\node at (19,0.8) {\huge $I$};

\end{tikzpicture}
 
  \caption{An interval $I$ split into five subintervals $I=AB_1B_2B_3C$ with $|B_1|, |B_2|, |B_3| \geq \ell$. Here we take $\ell=3$.}
  \label{fig:4}
  \end{center}
\end{figure}
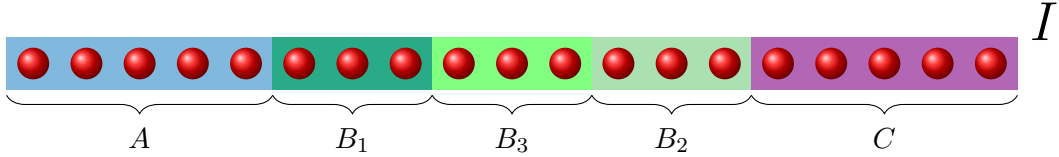

Let us first add and subtract some extra terms, and apply the triangle inequality to get four more tractable summands to estimate:
\begin{align*}
& \left\| \tr_{B}\big(\rho^{B} E_{A,B}^{\dagger} \, E_{AB,C}^{\dagger}  \big) \, - \, \tr_{BC}\big(\rho^{BC} \, E_{A, BC}^{\dagger}\big) \, \tr_{AB}\big(\rho^{AB}E_{AB,C}^{\dagger}\big) \right\| \, \, \\[2mm]
& \hspace{0.7cm} \leq \,\, \left\| \tr_{B}\big(\rho^{B} E_{A,B}^{\dagger} \, E_{AB,C}^{\dagger}  \big) \, - \, \tr_{B}\big(\rho^{B} E_{A,B_{1}}^{\dagger} \, E_{B_{3},C}^{\dagger}  \big) \right\|  \\[2mm]
& \hspace{1cm} + \left\| \tr_{B}\big(\rho^{B} E_{A,B_{1}}^{\dagger} \, E_{B_{3},C}^{\dagger}  \big)  \, - \, \tr_{B}\big(\rho^{B} \, E_{A, B_{1}}^{\dagger}\big) \, \tr_{B}\big(\rho^{B}E_{B_{3},C}^{\dagger}\big) \right\| \\[2mm]
& \hspace{1cm} + \left\| \tr_{B}\big(\rho^{B} \, E_{A, B_{1}}^{\dagger}\big) \, \tr_{B}\big(\rho^{B}E_{B_{3},C}^{\dagger}\big)  \, - \, \tr_{BC}\big(\rho^{BC} \, E_{A, B_{1}}^{\dagger}\big) \, \tr_{AB}\big(\rho^{AB}E_{B_{3},C}^{\dagger}\big) \right\| \\[2mm]
& \hspace{1cm} + \left\| \tr_{BC}\big(\rho^{BC} \, E_{A, B_{1}}^{\dagger}\big) \, \tr_{AB}\big(\rho^{AB}E_{B_{3},C}^{\dagger}\big) \, - \, \tr_{BC}\big(\rho^{BC} \, E_{A, BC}^{\dagger}\big) \, \tr_{AB}\big(\rho^{AB}E_{AB,C}^{\dagger}\big) \right\| \, .
\end{align*}
The first and fourth summands can be easily handled by using the fact that $Q \mapsto \operatorname{tr}_{\Lambda}(\rho^{\Lambda}Q)$ is contractive and estimates from Corollary \ref{Coro:mainLocality} and Remark \ref{rem:alternative_corollary}. Indeed,
\begin{align*}
& \left\| \tr_{B}\big(\rho^{B} E_{A,B}^{\dagger} \, E_{AB,C}^{\dagger}  \big) \, - \, \tr_{B}\big(\rho^{B} E_{A,B_{1}}^{\dagger} \, E_{B_{3},C}^{\dagger}  \big) \right\| \,\, \\[2mm]
& \hspace{2cm} \leq \,\, \left\| E_{A,B}^{\dagger} \, E_{AB,C}^{\dagger}  \, - \, E_{A,B_{1}}^{\dagger} \, E_{B_{3},C}^{\dagger}  \right\|\\[2mm]
& \hspace{2cm} \leq \,\,  \left\| E_{A, B}^{\dagger} - E_{A, B_{1}}^{\dagger}\right\| \, \left\| E_{AB, C}^{\dagger} \right\| \, + \, \left\| E_{A, B_{1}}^{\dagger} \right\| \, \left\| E_{AB, C}^{\dagger} - E_{B_{3}, C}^{\dagger} \right\|  \\[2mm]
& \hspace{2cm} \leq \,\, 2 \, \frac{\mathcal{G}^{\ell+1}}{(\lfloor \ell/r \rfloor + 1)!} \, ,
\end{align*}
and analogously
\begin{align*}
& \left\| \tr_{BC}\big(\rho^{BC} \, E_{A, B_{1}}^{\dagger}\big) \, \tr_{AB}\big(\rho^{AB}E_{B_{3},C}^{\dagger}\big) \, - \, \tr_{BC}\big(\rho^{BC} \, E_{A, BC}^{\dagger}\big) \, \tr_{AB}\big(\rho^{AB}E_{AB,C}^{\dagger}\big) \right\| \,\,  \\[2mm]
& \hspace{2cm}  \leq \,\, \left\| \tr_{BC}\big(\rho^{BC} \, E_{A, B_{1}}^{\dagger}\big) \, - \, \tr_{BC}\big(\rho^{BC} \, E_{A, BC}^{\dagger}\big)\right\| \, \left\| \tr_{AB}\big(\rho^{AB} \, E_{B_{3}, C}^{\dagger}\big)\right\|\\[2mm] 
& \hspace{3cm}+ \, \left\| \tr_{BC}\big(\rho^{BC} \, E_{A, B_{1}}^{\dagger}\big)\| \, \| \tr_{AB}\big(\rho^{AB} \, E_{B_{3}, C}^{\dagger}\big) - \tr_{AB}\big(\rho^{AB} \, E_{AB, C}^{\dagger}\big) \right\|  \\[2mm]
& \hspace{2cm}  \leq \,\, \left\| \, E_{A, B_{1}}^{\dagger} \, -  \, E_{A, BC}^{\dagger} \right\| \, \left\| \, E_{B_{3}, C}^{\dagger} \right\| + \, \left\| \, E_{A, B_{1}}^{\dagger} \right\| \, \left\| \, E_{B_{3}, C}^{\dagger} - \, E_{AB, C}^{\dagger} \right\|  \\[2mm]
& \hspace{2cm} \leq \,\, 2 \, \frac{\mathcal{G}^{\ell+1}}{(\lfloor \ell/r \rfloor + 1)!} \, .
\end{align*}

\noindent Bounding the second and third summands requires some extra work, so we are going to do it in two more steps.\\ 

\begin{step}[Auxiliary for Step \ref{step:3}]
\label{step:4}
There are absolute constants $\mathcal{K}^{*}$ and $\alpha^{*}$ such that
\[  \left\| \tr_{B}\big(\rho^{B} \, E_{A, B_{1}}^{\dagger}\big) \, \tr_{B}\big(\rho^{B} \, E_{B_{3},C}^{\dagger}\big)  \, - \, \tr_{BC}\big(\rho^{BC} \, E_{A, B_{1}}^{\dagger}\big) \, \tr_{AB}\big(\rho^{AB} \, E_{B_{3},C}^{\dagger}\big) \right\|  \, \leq \mathcal{K}^{*} \, e^{- \alpha^{*} \, \ell} \, . \]
\end{step}

\begin{step}[Auxiliary for Step \ref{step:3}]
\label{step:5}
There are absolute constants $\mathcal{K}^{\sharp}$ and $\alpha^{\sharp}$ such that
\[ \left\| \tr_{B}\big(\rho^{B} \, E_{A,B_{1}}^{\dagger} \, E_{B_{3},C}^{\dagger}  \big)  \, - \, \tr_{B}\big(\rho^{B} \, E_{A, B_{1}}^{\dagger}\big) \, \tr_{B}\big(\rho^{B} \, E_{B_{3},C}^{\dagger}\big) \right\|  \, \leq \mathcal{K}^{\sharp} \, e^{- \alpha^{\sharp} \, \ell} \, . \]
\end{step}

The proofs for these steps are provided in the next subsections.\qed

\vspace{3mm}

\subsection{Proof of Step \ref{step:4}}

We need two auxiliary results:

\begin{lem}\label{Lemm:SchmidtDecompEstimate1}
Let $\widetilde{A}ABC\widetilde{C}$ be a finite subsystem of $\mathbb{Z}$ and let us fix $m$, $\ell \in \mathbb{N}$ such that $|\widetilde A|$, $|A| \leq m$, $|\widetilde C|$, $|C| \leq m$ and $|B| \geq \ell$. Then, for all observables $Q_{\widetilde{A} A} \in \mathfrak{A} _{\widetilde{A} A}$ and $Q_{C\widetilde{C}} \in \mathfrak{A}_{C\widetilde{C}}$,
\begin{align*}  
& \left\| \tr_{ABC}\big(\rho^{ABC} Q_{\widetilde{A}A}\big) \,  - \, \tr_{AB}\big(\rho^{AB} Q_{\widetilde{A} A}\big) \right\| \,\,  \leq \,\, \, \| Q_{\widetilde{A}A}\| \, d^{2m} \, \mathcal{K}_1\, e^{- \alpha_1 \, \ell}\,.\\[2mm]
&  \left\| \tr_{ABC}\big(\rho^{ABC} Q_{C\widetilde{C}}\big) \,  - \, \tr_{BC}\big(\rho^{BC} Q_{C\widetilde{C} }\big) \right\| \,\,  \leq \,\, \, \| Q_{C\widetilde{C}}\| \, d^{2m} \, \mathcal{K}_2\, e^{- \alpha_2 \, \ell}\,.
\end{align*}
\end{lem}

\noindent \textit{Proof.}
We only prove the first inequality, the other is analogous. Let $Q_{A \widetilde{A}}$ be an observable with support in a subsystem $A \widetilde{A}$ of $\mathbb{Z}$. Considering $\mathfrak{A}_{\widetilde{A}A} = \mathfrak{A}_{\widetilde{A}} \otimes \mathfrak{A}_{A}$ as the tensor product of two Hilbert spaces ($\mathfrak{A}_{A}$ of dimension less than or equal to $d^{2m}$) endowed with the Hilbert-Schmidt scalar product, we infer that $Q_{A \widetilde{A}}$ can be written using the operator Schmidt decomposition as
\begin{equation}\label{equa:SchmidtDecompOperators} 
Q_{A \widetilde{A}} = \sum_{j=1}^{d^{2m}} \, \sqrt{\lambda_{j}} \,\, Q_{A}^{(j)} \otimes Q_{\widetilde{A}}^{(j)}  \quad , \quad Q_{A}^{(j)} \in \mathfrak{A}_{A} \quad , \quad Q_{\widetilde{A}}^{(j)} \in \mathfrak{A}_{\widetilde{A}} \, ,
\end{equation}
where the factors $Q_{A}^{(j)}$, resp.  $Q_{\widetilde{A}}^{(j)}$, are orthonormal, i.e.,
\begin{equation} 
\Tr(Q_{A}^{(j) \, \dagger} \, Q_{A}^{(k)}) \,\, = \,\, \Tr(Q_{\widetilde{A}}^{(j) \, \dagger} \, Q_{\widetilde{A}}^{(k)}) \,\, = \,\, \delta_{j k}  \quad \text{for every } j,k \, ,
\end{equation}
and the Schmidt coefficients $\lambda_{j}$ satisfy the bound
\begin{equation} 
\sum_j |\lambda_{j}| \,\, \leq \,\, \| Q_{A \widetilde{A}}\|_2^2 \,\, \leq \,\, d^{2m}\| Q_{A \widetilde{A}}\|^{2}\,. 
\end{equation}
We can finally estimate
\begin{align*}
& \left\| \tr_{ABC}(\rho^{ABC} Q_{\widetilde{A}A}) \,  - \, \tr_{AB}(\rho^{AB} Q_{\widetilde{A} A}) \right\| \,\, \\[2mm] 
& \hspace{1cm} \leq \,\, \sum_{j=1}^{d^{2m}} \, \sqrt{\lambda_{j}} \, \big\| Q_{\widetilde{A}}^{(j)}\big\| \cdot \big\| \tr_{ABC}(\rho^{ABC} Q_{A}^{(j)}) \, - \, \tr_{AB}(\rho^{AB}Q_{A}^{(j)})\big\| \\[2mm]
& \hspace{1cm} \leq \,\, \, \sum_{j=1}^{d^{2m}} \sqrt{\lambda_j}\, \big\| Q_{\widetilde{A}}^{(j)}\big\| \, \big\| Q_{A}^{(j)}\big\| \, \widetilde{\mathcal{C}} \, e^{- \gamma \, \ell}\\[2mm]
& \hspace{1cm} \leq d^{2m} \, \| Q_{\widetilde{A}A}\| \, \widetilde{\mathcal{C}}\, e^{- \gamma \, \ell}\, ,
\end{align*}
where in the second inequality we are using Proposition \ref{thm:clust-implies-indist} and, in the last line, the Cauchy-Schwarz inequality. Note that $\| Q^{(j)}_{ A}\|$, $\| Q_{\widetilde{A}}^{(j)}\|\leq 1$ since the factors are orthonormal with respect to the Hilbert-Schmidt inner product.\qed
\vspace{0.2cm}

\begin{prop}\label{Prop:SchmidtDecompEstimate1}
Let $AB_{1}B_{2}B_{3}C$ be a subsystem of $\mathbb{Z}$ and let us fix $\ell \in \mathbb{N}$ satisfying $|B_{1}|$, $|B_{3}| \geq \ell$.  Then
\begin{align*} 
& \left\| \tr_{B}\big(\rho^{B} E_{A, B_{1}}^{\dagger}\big) - \operatorname{tr}_{BC}\big(\rho^{BC} E_{A, B_{1}}^{\dagger}\big) \right\| \,\, \leq \,\,  \widetilde{\mathcal K}_2 \, e^{- \alpha_2 \, \ell} \, , \\[2mm]
& \left\| \tr_{B}\big(\rho^{B} E_{B_{3}, C}^{\dagger}\big) - \operatorname{tr}_{AB}\big(\rho^{AB} E_{B_{3}, C}^{\dagger}\big) \right\| \,\, \leq \,\, \widetilde{\mathcal K}_1 \, e^{- \alpha_1 \, \ell} \, .
\end{align*}
\end{prop}

\noindent \textit{Proof.}
We just prove the first inequality, the other is analogous. We can assume that $A$ corresponds to the interval $A=[1-a,0]$ and $B_1=[1,b]$ for some $a,b \in \mathbb{N}$. Then, using Proposition \ref{Theo:mainLocality} (see also Corollary \ref{Coro:mainLocality}) we can rewrite it as a finite series
\[  E_{A,B_{1}}^{\dagger} \,\, = \,\, \sum_{m=1}^{\infty} \widetilde{E}_{m} \, , \]
where each $\widetilde{E}_{m}$ has support in $[1-m, m] \cap A B_{1}$ and $\| \widetilde{E}_{1}\| \leq \mathcal G$, 
\[ \| \widetilde{E}_{m}\| \leq \frac{\mathcal{G}^{m-1}}{( \lfloor (m-1)/r\rfloor + 1 )!}  \qquad \forall m \geq 2 \, .\]
We can then use Lemma \ref{Lemm:SchmidtDecompEstimate1} and Corollary \ref{Coro:mainLocality} to bound
\begin{align*}
\left\| \tr_{B}\big(\rho^{B} E_{A,B_{1}}^{\dagger \,-1}\big) \, - \, \tr_{BC}\big(\rho^{BC} E_{A,B_{1}}^{\dagger \,-1}\big) \right\| \,\, & \leq \,\, \sum_{m=1}^{\infty} \, \left\| \tr_{B}\big(\rho^{B} \widetilde{E}_{m}\big) \, - \, \tr_{BC}\big(\rho^{BC} \widetilde{E}_{m}\big) \right\| \\[2mm]
& \leq \,\, \sum_{m=1}^{\infty} \, \| \widetilde{E}_{m} \| \, d^{2m}\, \mathcal{K}_2 \, e^{- \alpha_2 \, \ell }\\[2mm]
& \leq \,\, d^{2}\, \mathcal{K}_2 \, e^{- \alpha_2 \, \ell } \mathcal{G} + d^2\sum_{m=1}^{\infty} \, \frac{(\mathcal{G}d^{2})^{m}}{([m/r] + 1)!} \, \mathcal{K}_2 \, e^{- \alpha_2 \, \ell } \, .
\end{align*}  
It is easy to see that the series converges to a finite constant. We conclude the proof by denoting
\begin{equation*}
    \widetilde{\mathcal{K}}_2 := d^2 \mathcal{K}_2 \left( \mathcal{G} + \sum_{m=1}^{\infty} \, \frac{(\mathcal{G}d^{2})^{m}}{([m/r] + 1)!} \right) \, .
\end{equation*}
\qed
\vspace{0.2cm}

\noindent We can finally prove the step.

\vspace{0.2cm}

\noindent \textit{Proof of Step \ref{step:4}.}
\noindent Using Proposition \ref{Prop:SchmidtDecompEstimate1} and estimates from Corollary \ref{Coro:mainLocality}

\begin{align*}
& \left\| \tr_{B}\big(\rho^{B} \, E_{A, B_{1}}^{\dagger}\big) \, \tr_{B}\big(\rho^{B}E_{B_{3},C}^{\dagger}\big)  \, - \, \tr_{BC}\big(\rho^{BC} \, E_{A, B_{1}}^{\dagger}\big) \, \tr_{AB}\big(\rho^{AB}E_{B_{3},C}^{\dagger}\big) \right\| \,\,  \\[2mm]
& \hspace{3cm} \leq \,\, \left\|  \tr_{B}\big(\rho^{B} \, E_{A, B_{1}}^{\dagger}\big) \, - \,  \tr_{BC}\big(\rho^{BC} \, E_{A, B_{1}}^{\dagger}\big) \right\| \, \left\| \tr_{B}\big(\rho^{B}E_{B_{3},C}^{\dagger}\big) \right\|  \\[2mm]
& \hspace{3cm} + \left\| \tr_{BC}\big(\rho^{BC}E_{A,B_1}^{\dagger}\big) \right\| \,  \left\|  \tr_{B}\big(\rho^{B} \, E_{B_{3}, C}^{\dagger}\big) \, - \,  \tr_{AB}\big(\rho^{AB} \, E_{B_{3}, C}^{\dagger}\big) \right\| \,\,  \\[2mm]
& \hspace{3cm}  \leq \,\, \, \widetilde{\mathcal K }_1 \, e^{- \alpha_1 \, \ell} \, \| E_{B_{3}, C}^{\dagger}\| \, + \, \| E_{A, B_{1}}^{\dagger} \| \, \widetilde{\mathcal K}_2 \, e^{-\alpha_2 \, \ell} \, \\[2mm]
& \hspace{3cm} \leq \,\, 2 \, \mathcal{G} \, {\mathcal K}^{**} \, e^{- {\alpha^{**}} \, \ell}\,.
\end{align*}
\qed
\vspace{0.2cm}

\subsection{Proof of Step \ref{step:5}}

Again we need an auxiliary lemma in the same spirit of Lemma \ref{Lemm:SchmidtDecompEstimate1}.

\begin{lem}\label{Lemm:SchmidtDecompEstimate2}
Let $AB_{1}B_{2}B_{3}C$ be a subsystem of $\mathbb{Z}$ with $|A|$, $|B_{1}| \leq n$, $|B_{2}| \geq \ell$ and \mbox{$|C|$, $|B_{3}| \leq m$} for some $n,m, \ell \in \mathbb{N}$. Then, for every pair of observables $Q_{A B_{1}} \in \mathfrak{A}_{A B_{1}}$ and $Q_{B_{3}C} \in \mathfrak{A}_{B_{3}C}$ we deduce that
\[
\big\| \tr_{B}\big(\rho^{B}Q_{A B_{1}} Q_{B_{3}C}\big) \, - \, \tr_B\big(\rho^{B} Q_{A B_{1}}\big) \, \tr_{B}\big(\rho^{B} Q_{B_{3}C}\big)\big\| \, \leq  d^{2n + 2m} \,\, \| Q_{AB_{1}}\| \,\, \| Q_{B_{3}C}\| \,\, \breve{\mathcal{K}} \, e^{- \breve{\alpha} \ell} \, .
\]
\end{lem}

\noindent \textit{Proof.}
Using the operator Schmidt decomposition, we can write
\begin{align*}
Q_{AB_{1}} \, = \, \sum_{a=1}^{d^{2n}} \, \sqrt{\lambda_{a}} \, Q_{A}^{(a)} \otimes Q_{B_{1}}^{(a)} \quad \quad , \quad \quad Q_{B_{3}C} \, = \, \sum_{c=1}^{d^{2m}} \, \sqrt{\mu_{c}} \,\, Q_{B_{3}}^{(c)} \otimes Q_{C}^{(c)} \, ,
\end{align*}
where for every $a$
\[ Q_{A}^{(a)} \in \mathfrak{A}_{A} \; \; , \quad \| Q_{A}^{(a)}\| \, \leq \, 1 \; \; , \quad Q_{B_{1}}^{(a)} \in \mathfrak{A}_{B_{1}} \; \; , \quad \| Q_{B_1}^{(a)}\| \, \leq \, 1 \; \; , \quad \sum_{a = 1}^{d^{2n}}|\lambda_{a}| \leq d^{2n}\| Q_{A B_{1}}\|^2  \, , \]
and for every $c$
\[ Q_{B_{3}}^{(c)} \in \mathfrak{A}_{B_{3}} \; \; , \quad \| Q_{B_{3}}^{(c)}\| \, \leq \, 1 \; \; , \quad Q_{C}^{(c)} \in \mathfrak{A}_{C} \; \; , \quad \| Q_{C}^{(c)}\| \, \leq \, 1 \; \; , \quad \sum_{c = 1}^{d^{2m}} |\mu_{c}| \leq d^{2m} \| Q_{B_{3}C}\|^2  \, . \] 
Therefore
\begin{align*}
& \big\| \tr_{B}\big(\rho^{B}Q_{A B_{1}} Q_{B_{3}C}\big) \, - \, \tr\big(\rho^{B} Q_{A B_{1}}\big) \, \tr_{B}\big(\rho^{B} Q_{B_{3}C}\big)\big\| \,\,  \\[2mm]
& \hspace{2cm}\hspace{2cm} \leq \,\, \sum_{a=1}^{d^{2n}} \, \sum_{c=1}^{d^{2m}} \, \sqrt{\lambda_{a}} \,\, \sqrt{\mu_{c}} \,\, \| Q_{A}^{(a)}\| \, \| Q_{C}^{(c)}\| \\[2mm] 
& \hspace{2cm}\hspace{4cm} \cdot \big\|  \tr_{B}\big(\rho^{B} Q_{B_{1}}^{(a)} Q_{B_{3}}^{(c)}\big) \, - \, \tr_{B}\big(\rho^{B} Q_{B_{1}}^{(a)}\big) \cdot \tr_{B}\big(\rho^{B} Q_{B_{3}}^{(c)}\big)\big\| \\[2mm]
& \hspace{2cm}\hspace{2cm} \leq \,\, \sum_{a=1}^{d^{2n}} \, \sum_{c=1}^{d^{2m}} \, \sqrt{\lambda_{a}} \,\, \sqrt{\mu_{c}} \,\, \big\| Q_{A}^{(a)}\big\| \, \big\| Q_{C}^{(c)}\big\| \, \big\| Q_{B_{1}}^{(a)}\big\| \, \big\| Q_{B_{3}}^{(c)}\big\|   \,\, \widetilde{\mathcal{K}} \, e^{- \widetilde{\alpha} \ell}\\[2mm]
& \hspace{2cm}\hspace{2cm} \leq \,\, d^{2n + 2m} \,\, \| Q_{AB_{1}}\| \,\, \| Q_{B_{3}C}\| \,\, \widetilde{\mathcal{K}} \, e^{- \widetilde{\alpha} \ell}\,,
\end{align*}
where we have used the uniform exponential decay of correlations and the Cauchy-Schwarz inequality.\qed
\vspace{0.2cm}

\begin{prop}\label{Prop:ExponentialDecayPartialTrace}
Let $AB_{1}B_{2}B_{3}C$ be a system with $|B_{2}| \geq \ell$ for some $\ell \in \mathbb{N}$. Then,
\[ \left\| \Tr_{B}\big(\rho^{B} E_{A, B_{1}} E_{B_{3}, C}\big) \, - \, \Tr_B \big(\rho^{B} E_{A, B_{1}}\big) \cdot \Tr_B \big(\rho^{B} E_{B_{3}, C}\big) \right\|  \,\, \leq \,\, \breve{\mathcal{K}}^\prime \, e^{-  \breve{\alpha}^\prime \ell}\,. \] 
\end{prop}

\noindent \textit{Proof.}
Let us identify $A$ and $B_{1}$ with $[1-a,0]$ and $[1,b_{1}]$ for some $a,b_{1} \in \mathbb{N}$. We can find, by Proposition \ref{Theo:mainLocality} (see also Corollary \ref{Coro:mainLocality}), a decomposition
\[ E_{A, B_{1}} = \sum_{n=1}^{\infty} \widetilde{E}_{n}  \]  where 
\[\widetilde{E}_{n} \, \in \, \mathfrak{A}_{[1-n,n] \cap AB_{1}}  \, ,  \; \; \|  \widetilde{E}_{n} \| \,\, \leq \,\, \frac{\mathcal{G}^{n-1}}{(\lfloor (n-1)/r \rfloor + 1)!}\, \, \forall n \geq 2\,, \; \; \|  \widetilde{E}_{1} \| \leq \mathcal G. \]
 Analogously we can identify $B_{3}$ and $C$ with intervals $[1-b_{3},0]$ and $[1,c]$ for certain $b_{3}, c \in \mathbb{N}$ and find a decomposition
\[ E_{A, B_{1}} = \sum_{m=1}^{\infty} \widetilde{F}_{m} \] 
where
\[\widetilde{F}_{m} \, \in \, \mathfrak{A}_{[1-m,m] \cap B_{3}C}  \,,  \; \; \|  \widetilde{F}_{m} \| \,\, \leq \,\, \frac{\mathcal{G}^{m-1}}{(\lfloor (m-1)/r \rfloor + 1)!}\, \, \forall m \geq 2\,, \; \; \|  \widetilde{F}_{1} \| \leq \mathcal G. \]
 Note that whenever $1 \leq n,m \leq \ell$, then $\widetilde{E}_{n}$ and $\widetilde{F}_{m}$ have support in $AB_{1}$ and $B_{3}C$, respectively, where we can enlarge $A$ and $C$ if necessary. Let us assume $\mathcal G \geq 1$, making the constant possibly larger. Since both supporting intervals are separated by an interval $B_{2}$ with $|B_{2}| \geq \ell$, we can apply Lemma \ref{Lemm:SchmidtDecompEstimate2} to bound, for $1 \leq n,m \leq \ell$ 
\begin{align*}
\left\| \tr_{B}(\rho^{B}\widetilde{E}_{n} \widetilde{F}_{m}) \, - \, \tr_{B}(\rho^{B} \widetilde{E}_{n}) \cdot \tr_{B}(\rho^{B} \widetilde{F}_{m}) \right\| \,\, & \leq \,\, d^{2n + 2m} \, \| \widetilde{E}_{n}\| \, \| \widetilde{F}_{m}\| \, \breve{\mathcal{K}} \, e^{- \breve{\alpha} \ell} \\[2mm]
& \hspace{-0.7cm} \leq \,\, \breve{\mathcal{K}} \, e^{- \breve{\alpha} \, \ell} \, \frac{(\mathcal{G}d^{2})^{n}}{(\lfloor (n-1)/r \rfloor + 1)!} \, \frac{(\mathcal{G}d^{2})^{m}}{(\lfloor (m-1)/r \rfloor + 1)!} \, .
\end{align*}
If $n>\ell$ or $m>\ell$, we will estimate in a rudimentary way
\begin{align*}
\left\| \tr_{B}(\rho^{B}\widetilde{E}_{n} \widetilde{F}_{m}) \, - \, \tr_{B}(\rho^{B} \widetilde{E}_{n}) \cdot \tr_{B}(\rho^{B} \widetilde{F}_{m}) \right\| \,\, & \leq \,\, 2 \, \| \widetilde{E}_{n}\| \, \| \widetilde{F}_{m}\| \\[2mm]
& \leq \,\, 2 \, \frac{\mathcal{G}^{n-1}}{(\lfloor (n-1)/r \rfloor + 1)!} \,\, \frac{\mathcal{G}^{m-1}}{(\lfloor (m-1)/r \rfloor + 1)!}  \, .
\end{align*}
Combining these bounds
\begin{align*}
&  \left\| \tr_{B}(\rho^{B} E_{A, B_{1}} E_{B_{3}, C}) \, - \, \tr_B(\rho^{B} E_{A, B_{1}}) \cdot \tr_B(\rho^{B} E_{B_{3}, C}) \right\|  \,\,  \\[2mm]
 & \hspace{4cm} \leq \, \sum_{n=1}^{\infty} \, \sum_{m=1}^{\infty} \, \left\|\tr_{B}(\rho^{B} \widetilde{E}_{n} \, \widetilde{F}_{m}) \, - \, \tr_B(\rho^{B} \widetilde{E}_{n}) \cdot \tr_B(\rho^{B} \widetilde{F}_{m})  \right\|\\[2mm]
 & \hspace{4cm} \leq \, \breve{\mathcal{K}} \, e^{-\breve{\alpha} \, \ell} \, \left( \sum_{n=1}^{\ell} \, \frac{(\mathcal{G}d^{2})^{n}}{(\lfloor (n-1)/r \rfloor + 1)!} \right)^{2}\\[2mm]
 &\hspace{5cm}  + 4 \, \left( \sum_{n=1}^{\infty} \, \frac{\mathcal{G}^{n}}{(\lfloor (n-1)/r \rfloor + 1)!} \right) \left( \sum_{n=\ell +1}^{\infty} \, \frac{\mathcal{G}^{n}}{(\lfloor
(n-1)/r \rfloor + 1)!} \right) \\[2mm]
 & \hspace{4cm} \leq \, \breve{\mathcal K} \, \left( \, e^{-\breve{\alpha} \, \ell}  + \sum_{n=\ell +1}^{\infty} \, \frac{\mathcal{G}^{n}}{(\lfloor n/r \rfloor + 1)!} \right).
\end{align*} 
The fact that the series decays superexponentially in $\ell$ can be seen by considering the remainder term of the Taylor expansion of the exponential function.\qed
\vspace{0.2cm}

\noindent \textit{Proof of Step \ref{step:5}.}
This follows directly from an application of Proposition \ref{Prop:ExponentialDecayPartialTrace}.\qed
\vspace{0.2cm}

\section{Final remarks and Conclusions}\label{sec:conclusions}

In this paper, we have proven some new results related to the decay of correlations between spatially separated regions for Gibbs states of local, finite-range, translation-invariant Hamiltonians in 1D. The main tools employed in their proofs have been derived from Araki's expansionals and the BS-entropy.

First, we have shown that the distance of a Gibbs state in the previous setting from being BS-recoverable between three adjacent intervals decays superexponentially with the size of the middle interval. This result can be compared to \cite[Theorem 4]{Kato2019}, where the distance from a Gibbs state in the same context to a recovery map is shown to decay subexponentially instead. 

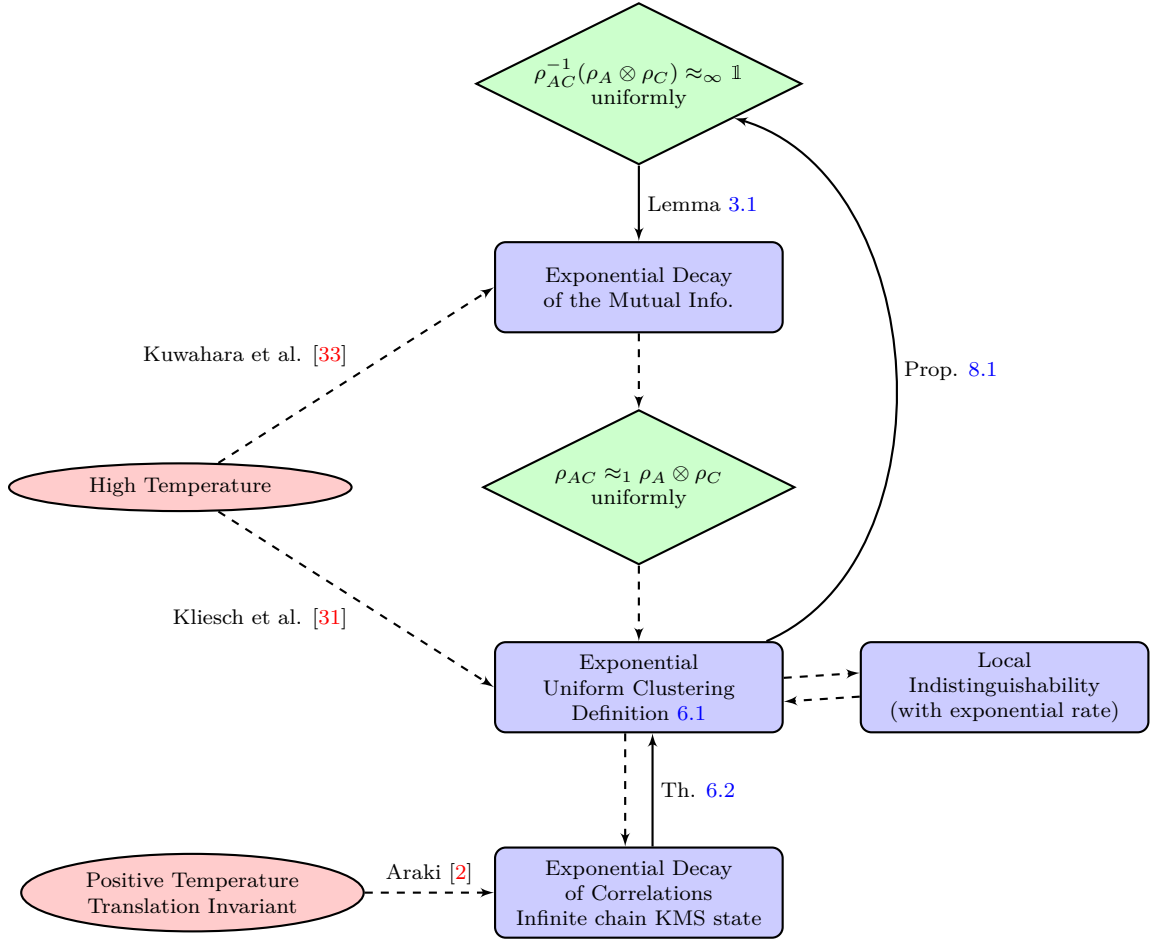
\begin{figure}[ht]
    \centering

 \tikzstyle{greenBlock} = [diamond, aspect=2, draw, fill=green!20, 
     text width=10em, text badly centered, node distance=3cm, inner sep=0pt]
 \tikzstyle{block} = [rectangle, aspect=2, draw, fill=blue!20, 
     text width=12em, text centered, rounded corners, minimum height=4em]
\tikzstyle{line} = [draw, -latex']
 \tikzstyle{cloud} = [draw, ellipse, aspect=2 ,fill=red!20, text width=10em, text badly centered, node distance=3cm,
     minimum height=1.5em]
    
\begin{tikzpicture}[node distance = 2cm, auto, thick, scale=0.6]
    \scriptsize
    
    
     \node [block] (UnifCluster) {Exponential \\ Uniform Clustering \\ Definition \ref{def:unif-clustering}};    
    

     \node [greenBlock, above=1cm of UnifCluster] (ApproxNormOne) {$\rho_{AC} \approx_{1} \rho_{A} \otimes \rho_{C}$ \\ uniformly};  
    
  \path [line, dashed] (ApproxNormOne.south) -- (UnifCluster.north); 
  
  
       \node [block, above=1cm of ApproxNormOne] (ExpDecayMI) {Exponential Decay \\ of the Mutual Info.};      
  
   \path [line, dashed] (ExpDecayMI.south) -- (ApproxNormOne.north);

  
 \node [cloud, left=1.7cm of ApproxNormOne] (Ndim) {High Temperature};

   \path [line, dashed] (Ndim) --  (UnifCluster.west) node[midway,below left] {Kliesch  et al. \cite{Kliesch2014}};

   \path [line, dashed] (Ndim) --  (ExpDecayMI.west) node[midway] {Kuwahara  et al. \cite{Kuwahara2019}};

   
       \node [greenBlock, above=1cm of ExpDecayMI] (ApproxNormInfinity) {$\rho_{AC}^{-1}(\rho_{A} \otimes \rho_{C}) \approx_{\infty} \mathbbm{1}$ \\ uniformly};  
 
   \path [line] (ApproxNormInfinity) --  (ExpDecayMI) node[midway]  {Lemma \ref{lem:boundBSrelativeEntropy}};

  
       \node [block, right=1cm of UnifCluster] (LocalUndistinguishable) {Local \\  Indistinguishability \\ (with exponential rate) };    
  
   \path [line, dashed] ([yshift=0.8cm]UnifCluster) -- ([yshift=0.8cm]LocalUndistinguishable) node[midway] {};
 
    \path [line, dashed] ([yshift=-0.8cm]LocalUndistinguishable) -- ([yshift=-0.8cm]UnifCluster) node[midway] {};

  \path [line] ([xshift=5cm]UnifCluster)  edge [bend right=70] node[midway, right] {Prop. \ref{Theo:MIdecay}} (ApproxNormInfinity) ; 
 
  
     \node [block, below=1.5cm of UnifCluster] (ExpDecay) {Exponential Decay\\ of Correlations \\ Infinite chain KMS state};      
    
  \node [cloud, left = 1.7cm of ExpDecay] (OneDim) {Positive Temperature \\ Translation Invariant};  
    
   \path [line, dashed] (OneDim) -- (ExpDecay) node[midway] {Araki \cite{Araki1969}};      
    
    
     \path [line, dashed] ([xshift=-0.3cm]UnifCluster.south) -- ([xshift=-0.3cm]ExpDecay.north) ;
    
     \path [line] ([xshift=0.3cm]ExpDecay.north) -- ([xshift=0.3cm]UnifCluster.south) node[midway,xshift=1.2cm] {Th. \ref{Theo:ArakiImpliesUnifCluster}};


 \end{tikzpicture}    
   
    \caption{Here we show the connections between several properties and results in 1D that have appeared in the paper. We represent by red ellipsoids some assumptions on the temperature and interactions on the system under study, by blue rounded rectangles some results on decay of correlations and by green rhombi some conditions on approximation of observables. The dashed arrows stand for results that were previously known (or easily obtained from previously-known results), and the solid ones, for original outcomes of this paper.}
    \label{fig:scheme}
 \end{figure}

Furthermore, we have proven that such Gibbs states exhibit exponential decay of correlations over any finite interval, as opposed to Araki's original result, which concerns exponential decay of correlations over an infinite chain. This result has subsequently allowed us to show exponential decay of the mutual information for any Gibbs state of a local, finite-range, translation-invariant Hamiltonian in 1D, for any temperature. These results allow us to close a cycle by obtaining the equivalence between several well-known properties in the context of decay of correlations (see Figure \ref{fig:scheme}), namely:
\begin{itemize}
    \item[(i)] Exponential decay of correlations for thermal states on the infinite chain.
    \item[(ii)] Exponential uniform clustering.
    \item[(iii)] Exponential decay of the mutual information.
\end{itemize}

It might be interesting to remark here the relation between exponential uniform clustering and local indinguishability as it was introduced by Brand\~{a}o and Kastoryano in \cite{Brandao2019} as a consequence of the exponential uniform clustering condition. Particularizing \cite[Theorem 5]{Brandao2019} to 1D, we have that for every finite subsets $A \subset X \subset Y$ with $\operatorname{dist}(A, Y \setminus X) \geq \ell$ and every observable $Q$ with support in $A$
\begin{equation}\label{equa:localindistCondition}
 | \operatorname{Tr}_{X}(\rho^{X} Q) \, - \, \operatorname{Tr}_{Y}(\rho^{Y} Q) | \, \leq \, |\partial X|  \, \mathcal{K} \, e^{- \alpha \ell} \, \| Q\|   \end{equation}
where $\partial X$ is the set of nodes in $X$ with a neighbouring site in $X^{c}$, and where $\mathcal{K}, \alpha$ are constants independent of the supporting sets. This local indistinguishability property is more general than the one we consider in Proposition \ref{thm:clust-implies-indist} (the supporting sets are not necessarily intervals, but arbitrary finite subsets). Indeed, it can be easily shown that this condition implies exponential uniform clustering, hence both conditions are equivalent. For that, consider a finite interval $I \subset \mathbb{Z}$ split into  $I=ABC$, where $B$ shields $A$ from $C$, and for which $|B| > 3 \ell$ for some $\ell \in \mathbb{N}$ with $\ell \geq r$. We can then split $B = B_{1} B_{2} B_{3}$, where $B_{2}$ shields $B_{1}$ from $B_{3}$ and $|B_{j}| \geq \ell$ for $j=1,2,3$. Let $\widetilde{A}=AB_{1}$, $\widetilde{C} = B_{3}C$ and $X= \widetilde{A} \cup \widetilde{C}$. Then, for each $Q_{A} \in \mathfrak{A}_{A}$ and $Q_{C} \in \mathfrak{A}_{C}$,
\begin{equation}\label{equa:localindistCondition2}
 | \operatorname{Tr}_{I}(\rho^{I} Q_{A} Q_{C}) \, - \, \operatorname{Tr}_{X}(\rho^{X} Q_{A} Q_{C}) | \, \leq \, 4  \, \mathcal{K} \, e^{- \alpha \ell} \, \| Q_{A}\| \, \| Q_{C}\|  \,. 
 \end{equation}
We observe that $H_{X} = H_{\widetilde{A}} + H_{\widetilde{C}} $ since the distance between $\widetilde{A}$ and $\widetilde{C}$ is larger than $\ell \geq r$, so that $\rho^{X} = \rho^{\widetilde{A}} \rho^{\widetilde{C}}$  and hence
\begin{equation}\label{equa:localindistCondition3}
\operatorname{Tr}_{X}(\rho^{X} Q_{A} Q_{C})  = \operatorname{Tr}_{\widetilde{A}}(\rho^{\widetilde{A}} Q_{A}) \, \operatorname{Tr}_{\widetilde{C}}(\rho^{\widetilde{C}} Q_{C})   \, .
\end{equation}
Moreover, using again the local indistinguishability condition \eqref{equa:localindistCondition}, we can estimate
\begin{equation}\label{equa:localindistCondition4}
\big| \operatorname{Tr}_{\widetilde{A}}(\rho^{\widetilde{A}} Q_{A} ) - \operatorname{Tr}_{I}(\rho^{I} Q_{A})   \big| \, \leq \, 2 \mathcal{K} e^{- \alpha \ell} \, \| Q_{A}\| \,\, , \,\, \big| \operatorname{Tr}_{\widetilde{C}}(\rho^{\widetilde{C}} Q_{C} ) - \operatorname{Tr}_{I}(\rho^{I} Q_{C})   \big| \, \leq \, 2 \mathcal{K} e^{- \alpha \ell} \, \| Q_{C}\| \, .
\end{equation}
Combining \eqref{equa:localindistCondition2},\eqref{equa:localindistCondition3}, and \eqref{equa:localindistCondition4} we conclude that
\[ \left| \operatorname{Tr}_{I}(\rho^{I}Q_{A}Q_{C}) - \operatorname{Tr}_{I}(\rho^I Q_{A}) \, \operatorname{Tr}_{I}(\rho^I Q_{C})  \right| \, \leq \, 8 \mathcal{K} e^{- \alpha \ell} \, \| Q_{A}\| \, \| Q_{C}\|\,.  \]
\noindent This finishes the proof of the statement.
  
 Regarding the decay of mutual information for Gibbs states, it was proven in \cite{Kastoryano2013} that systems satisfying the property of rapid mixing exhibit exponential decay of correlations. In particular, Theorem 12 in that reference shows that, under the assumption of a positive \emph{modified logarithmic Sobolev constant} (MLSI constant) for a certain dissipative system, the mutual information between spatially separated regions on a Gibbs state decays exponentially with the distance between those regions, with a mutiplicative factor depending polynomially on the system size. This result, together with the main findings of \cite{CapelRouzeStilckFranca-MLSIcommuting-2020} (in which it is shown that for local, finite-range, commuting Hamiltonians in 1D or with nearest-neighbour interactions, at high enough temperature, there is a certain evolution, satisfying the conditions of \cite[Theorem 12]{Kastoryano2013}, with 
 the Gibbs state of such a Hamiltonian as a fixed point, that has a positive MLSI constant) yield an exponential decay for the mutual information on Gibbs states in their setting. However, the main drawback of this result is the polynomial dependence on the system size. Our Theorem \ref{Theo:MIdecay-main-result} is a clear improvement to the latter in 1D, since we have no dependence with the system size and our result holds for non-commuting Hamiltonians and for any temperature.

One motivation to find a decay rate for the mutual information on Gibbs states of certain systems is that of employing them to prove area laws. Indeed, it might be possible to extend Theorem \ref{Theo:MIdecay-main-result}  to larger dimensions for high enough temperature by using \cite[Theorem 5]{Brandao2019}, which is based on Quantum Belief Propagation \cite{Hastings2007}, instead of our Proposition \ref{thm:clust-implies-indist} based on Araki's expansionals. Recall that the former presents the advantage with respect to the latter that it holds for any dimension, and the drawback that it only works for high-enough temperature. The possible extension of our result might then yield in particular exponential decay of the mutual information on Gibbs states of local, finite-range, non-commuting Hamiltonians at high-enough temperature, a setting already explored in \cite{Kuwahara2019} for the conditional mutual information. Note that the authors of \cite{Kuwahara2019} point out that their results extend to the mutual information. However, as we would actually be estimating geometric Rényi divergences, it might also be possible to use it to obtain an area law for the mutual information that arises from such quantities. This direction has recently been explored in \cite{Scalet2021}.

To conclude, we expect that the results and techniques developed in the current manuscript find applications in the fields of quantum information theory and quantum many-body systems beyond those already mentioned here. For instance, to illustrate the potential of Theorem \ref{Theo:MIdecay-main-result}, it can actually be used in the context of quantum functional inequalities to prove examples of positivity of a modified logarithmic Sobolev inequality (MLSI). Indeed, the norm that is upper bounded in Proposition \ref{Theo:MIdecay} can be interpreted as a mixing condition for a Gibbs state, as stated in \cite[Assumption 1]{BardetCapelLuciaPerezGarciaRouze-HeatBath1DMLSI-2019}. In that paper, the authors show that such a condition needs to be assumed in order to prove that the heat-bath dynamics in 1D has a positive MLSI constant. Our Proposition  \ref{Theo:MIdecay} can then be used to show that such an assumption holds for any Gibbs state of a local, finite-range, translation-invariant Hamiltonian in 1D, for any temperature.  Jointly with some other recent techniques, it yields that Davies generators in 1D converging to such a Gibbs state, at any temperature, have a positive MLSI constant, and hence exhibit rapid mixing \cite{BardetCapelGaoLuciaPerezGarciaRouze-Davies1DMLSIlong-2021,BardetCapelGaoLuciaPerezGarciaRouze-Davies1DMLSIshort-2021}.

\vspace{\baselineskip}
\textbf{Acknowledgements:} The authors would like to thank {\'A}lvaro M.\ Alhambra, in particular for suggesting the use of the geometric Rényi divergences that improved the presentation of Section \ref{subsec:mutualinfo}; Yoshiko Ogata, whose comments inspired the proof of Theorem \ref{Theo:ArakiImpliesUnifCluster}; and also Kohtaro Kato, Angelo Lucia and David P{\'e}rez-Garc{\'i}a for insightful discussions. AB acknowledges support from the VILLUM FONDEN via the QMATH Centre of Excellence (Grant no.  10059) and from the QuantERA ERA-NET Cofund in Quantum Technologies implemented within the European Union’s Horizon 2020 Programme (QuantAlgo project) via the Innovation Fund Denmark. AC is partially supported by a MCQST Distinguished PostDoc fellowship, by the Seed Funding Program of the MCQST and by the Deutsche Forschungsgemeinschaft (DFG, German Research Foundation) under Germany's Excellence Strategy EXC-2111 390814868. APH is partially supported by the grants “Juan de la Cierva Formación” (FJC2018-036519-I) and 2021-MAT11 (ETSI Industriales, UNED), as well as by the Spanish Ministerio de Ciencia e Innovación project PID2020-113523GB-I00, by Comunidad de Madrid project QUITEMAD-CM P2018/TCS4342, and by the European Research Council (ERC) under the European Union’s Horizon 2020 research and innovation programme (grant agreement No 648913).

\vspace{0.5cm}

\bibliographystyle{plainnat}
\bibliography{lit}

\vspace{0.5cm}

\end{document}